\documentclass[aps,prd,reprint,nofootinbib,showkeys,showpacs,superscriptaddress,floatfix]{revtex4-1}

\pdfoutput=1

\usepackage{amsmath}
\usepackage{amssymb}
\usepackage[american]{babel}
\usepackage{booktabs}
\usepackage{graphicx}
\usepackage{dcolumn}  % Align table columns on decimal point
\usepackage[T1]{fontenc}  % Font encoding that enables acents in the PDF
\usepackage{graphicx}  % Include figure files
\usepackage[]{hyperref}  % Add hypertext capabilities
\usepackage[utf8]{inputenc}  % Accept the Western European characters set
\usepackage{url}
\usepackage[dvipsnames]{xcolor}
\usepackage{float}

\DeclareUnicodeCharacter{2212}{-}

\usepackage{tikz}
\usepackage[normalem]{ulem}

\usetikzlibrary{arrows,shapes,positioning,shadows,trees}

\tikzset{
	basic/.style  = {draw, text width=2cm, drop shadow, font=\sffamily, rectangle},
	root/.style   = {basic, rounded corners=2pt, thin, align=center,
		fill=green!30},
	level 2/.style = {basic, rounded corners=6pt, thin,align=center, fill=green!60,
		text width=8em},
	level 3/.style = {basic, thin, align=left, fill=pink!60, text width=6.5em}
}

\usetikzlibrary{arrows,decorations.pathmorphing,backgrounds,fit,positioning,shapes.symbols,chains}

\usepackage{hyperref}
\hypersetup{
    colorlinks=true, % false: boxed links; true: colored links
    linkcolor=royalblue, % color of internal links (change box color with linkbordercolor)
    citecolor=magenta}

\newcolumntype{d}[1]{D{.}{.}{#1}}
\newcolumntype{v}[1]{D{,}{,\ }{#1}}

\makeatletter

\newcommand{\Rmnum}[1]{\expandafter\@slowromancap\romannumeral #1@}

\definecolor{greenW}{rgb}{0.0, 0.55, 0.1}

\makeatother

\hypersetup{
	citecolor = red,
	linkcolor = blue,
	urlcolor = magenta
}

\begin{document}

\title{In search of an interaction in the dark sector through Gaussian Process and ANN approaches}

\author{Mazaharul Abedin}
\email{mazaharul.rs@presiuniv.ac.in}
\affiliation{Department of Mathematics, Presidency University, 86/1 College Street,  Kolkata 700073, India}

\author{Guo-Jian Wang}
\email{gjwang2018@gmail.com}
\affiliation{Department of Physics, Stellenbosch University, Matieland 7602, South Africa}
\affiliation{National Institute for Theoretical and Computational Sciences (NITheCS), South Africa}

\author{Yin-Zhe Ma}
\email{mayinzhe@sun.ac.za}
\affiliation{Department of Physics, Stellenbosch University, Matieland 7602, South Africa}
\affiliation{National Institute for Theoretical and Computational Sciences (NITheCS), South Africa}

\author{Supriya Pan}
\email{supriya.maths@presiuniv.ac.in}
\affiliation{Department of Mathematics, Presidency University, 86/1 College Street,  Kolkata 700073, India}
\affiliation{Institute of Systems Science, Durban University of Technology, PO Box 1334, Durban 4000, Republic of South Africa}

\begin{abstract}

Whether the current observational data indicate any evidence of interaction between the dark sector is a matter of supreme interest at the present moment. This article searched for an interaction in the dark sector between a pressure-less dark matter and a dark energy fluid with constant equation of state, $w_{\rm DE}$. For this purpose, two non-parametric approaches, namely, the Gaussian Process (GP) and the Artificial Neural Networks (ANN) have been employed and using the Hubble data from Cosmic Chronometers (CC), Pantheon+ from Supernovae Type Ia and their combination we have reconstructed the interaction function. We find that for $w_{\rm DE} =-1$, the interaction in the dark sector is not prominent while for $w_{\rm DE} \neq -1$, evidence of interaction is found depending on the value of $w_{\rm DE}$.
In particularly, we find that if we start deviating from $w_{\rm DE} = -1$ either in the quintessence ($w_{\rm DE} > -1$) or phantom ($w_{\rm DE} < -1$) direction, an emergence of dark interaction is observed from both GP and ANN reconstructions. We further note that ANN which is applied for the first time in this context seems to play a very efficient role compared to GP.

\end{abstract}
%----------------------------------------------
\maketitle
%------------------------------------------------

\section{Introduction}\label{sec1}

Late-time accelerating expansion of the universe is one of the landmark discoveries in modern cosmology. To describe this accelerating expansion, usually two well known approaches are used. The first and the simplest approach is to include a dark energy (DE) component \cite{Copeland:2006wr} within the framework of Einstein's General Relativity (GR). 
The second approach relies on the modifications of GR in various ways, henceforth, they are classified as modified gravity models, which result in a DE-like fluid arising due to the gravitational (effectively geometric) corrections (also known as geometrical DE)~ \cite{Nojiri:2006ri,Sotiriou:2008rp,DeFelice:2010aj,Clifton:2011jh,Cai:2015emx, Nojiri:2017ncd,Bahamonde:2021gfp}. On the quantitative direction, according to the up-to-date observational evidences,  nearly 68\% of the total energy budget of the universe is occupied by DE or geometrical DE and about 28\% of the total energy budget is occupied by non-luminous dark matter (DM), the key source of the observed formation of structures of the universe.  These information eventually imply that the dark sector of the universe is occupied by DM and DE or geometric DE. Despite many astronomical surveys and their data release, the fundamental nature of DE or modified gravity models is still unclear and  currently we have a large number of such models which try to explain the dynamics of the universe in the realm of the accelerating expansion of the universe.

One of the appealing cosmological models in the literature is the interacting or coupled DE models where DM and DE interact with each other through an exchange of energy-momentum between them.
The possibility of an interaction 
in the dark sector was proposed by Amendola \cite{Amendola:1999er} and it was observed that an interaction in the dark sector could alleviate the cosmic coincidence problem \cite{Amendola:1999er,Cai:2004dk,Pavon:2005yx,Huey:2004qv,delCampo:2008sr,delCampo:2008jx}. Subsequently, it was observed that such cosmological models have many fascinating consequences such as, crossing the phantom divide line~\cite{Das:2005yj,Wang:2005jx,Sadjadi:2006qb}, alleviation of the cosmological tensions \cite{Kumar:2016zpg,Pourtsidou:2016ico,An:2017crg,DiValentino:2017iww,Yang:2018euj,Kumar:2019wfs,Pan:2019gop,Shah:2024rme,Giare:2024smz} (see also the section on interacting dark energy in Ref. \cite{DiValentino:2021izs}). For a specific functional form describing the interaction rate between DM and DE, using the gravitational equations, one can in principle determine the dynamics of the universe. However, if the information about the interaction between these dark sectors is directly obtained from the observational data without assuming any specific model in the background, then this could offer deeper insights on this particular field. At present two known data driven (also known as non-parametric) approaches are getting attention in the cosmology community, one is the Gaussian process  and the other one is Artificial Neural Network.

The Gaussian process (GP) is a generalization of Gaussian distributions to function space \cite{Seikel:2012uu}. It is a Bayesian approach  describing a distribution over functions and it is a completely  non-parametric in the sense that it does not assume any model or parametrization. In cosmology GP has been extensively used over the years, such as the reconstruction of the DE equation of state (EoS) \cite{Seikel:2012uu,Seikel:2012cs,Yahya:2013xma}, reconstructing the expansion history of the universe \cite{Li:2015nta}, investigating the curvature of the universe \cite{Cai:2015pia,Cai:2016vmn,Wei:2016xti,Yu:2017iju,Wang:2017lri,Mukherjee:2022ujw}, estimating the Hubble constant \cite{Busti:2014dua,Gomez-Valent:2018hwc}, 
reconstructing the cosmic growth and matter perturbations \cite{Shafieloo:2012ms,Gonzalez:2017fra}, examining the distance duality relation \cite{Santos-da-Costa:2015kmv,Li:2017zrx,Mukherjee:2021kcu}, reconstructing quintom DE and modified gravity~\cite{Yang:2024kdo,Yang:2025kgc,Yang:2025mws}, and reconstructing the interaction in the dark sector \cite{Yang:2015tzc,Cai:2017yww,Aljaf:2020eqh,Mukherjee:2021ggf,Bonilla:2021dql,Escamilla:2023shf}.

On the other hand, Artificial Neural Network (ANN) is a machine learning approach which has caught significant attention in cosmology. The use of  ANN in order to reconstruct the cosmological parameters as well as to understand the expansion history of the universe has been accelerated in recent times, see for instance \cite{Wang:2019vxv,Wang:2020dbt,Wang:2020hmn,Dialektopoulos:2021wde,Wang:2022qta,Wang:2023vej,Qi:2023oxv,Giare:2024syw}.   
This machine learning approach has also been used to reconstruct various modified gravity theories~\cite{Mukherjee:2022yyq,Dialektopoulos:2023jam}.  According to the existing records in the literature, a large number of papers have been written so far \cite{Kim:2014nba,Cheng:2018nhz,Cheng:2020gec,
Choudhury:2019vat,Choudhury:2020azd,Zhang:2022caa,Pal:2022hpi,Sikder:2022hzk,Gomez-Vargas:2021zyl,
Ran:2023jmh,Liu:2024gne,Fortunato:2024hfm,
Qi:2024acx} which clearly indicates its emergence in the field of cosmology and astrophysics.

Thus, considering both GP and ANN, and using two model independent datasets, namely, the Hubble parameter measurements from Cosmic Chronometers and Pantheon+ sample from Supernovae Type Ia, in this article we have reconstructed the interaction between these dark components. Although GP has been used earlier to reconstruct the interaction between DM and DE \cite{Yang:2015tzc, Bonilla:2021dql,Mukherjee:2021ggf,Escamilla:2023shf}, but the ANN approach has not been considered yet in this context. Thus, the use of ANN  in reconstructing the interaction between DM and DE, is a new ingredient in this article. According to the existing literature on the reconstruction of the DE-DM interaction using GP, evidence of the interaction depends on many factors, the EoS of DE being one of them~\cite{Yang:2015tzc, Bonilla:2021dql,Mukherjee:2021ggf,Escamilla:2023shf}, because, as argued in \cite{Yang:2015tzc}, if the DE EoS deviates from $-1$, then an evidence of interaction in the dark sector can be found.

The article has been organized as follows. In section~\ref{sec2-set-up} we describe the basic gravitational equations of an interacting DE-DM model and the interaction function that we wish to reconstruct with the data. 
In section~\ref{sec3-non-parametric-approaches} we describe two non-parametric approaches, namely, GP and ANN that have been considered in this work. In \ref{sec-data} we describe the observational data used to reconstruct the interaction function using these non-parametric approaches. Then in section~\ref{sec-results} we present the results. Finally, in section~\ref{sec-summary} we conclude the present article by summarizing the main findings.

\section{Interacting Dark energy: Theoretical set-up}
\label{sec2-set-up}

We consider that our universe is homogeneous and isotropic in the large scale and its geometrical configuration is well described by the spatially flat Friedmann-Lema\^{i}tre-Robertson-Walker (FLRW) line element 
$ds^2 = -dt^2 + a^2 (t) \left[dr^2 + r^2 \left(d\theta^2 + \sin^2 \theta d\phi^2 \right) \right]$
where $a(t)$ is the expansion scale factor of the universe.  We assume that the gravitational sector of the universe is well described by GR and the matter sector of the universe is comprised of a pressure-less DM and DE which are interacting with each other. The gravitational equations can be described as 

\begin{eqnarray}
    \rho_{\rm DM} + \rho_{\rm DE}  &=& \frac{3}{\kappa^2}  H^2, \label{Friedmann-1}\\
   p_{\rm DE} &=&  -\frac{1}{\kappa^2} \left(2 \dot{H} + 3 H^2 \right),\label{Friedmann-2}
\end{eqnarray}
where $\kappa^2 = 8 \pi G$ is the Einstein's gravitational constant ($G$ denotes the Newton's gravitational constant),  an overhead dot represents the derivative with respect to the cosmic time; $H \equiv \dot{a}/a$ is the Hubble rate of the FLRW universe. As the fluids are interacting with each other, hence, their conservation equations will be modified as 

\begin{eqnarray}
&& \dot{\rho}_{\rm DM} + 3 H \rho_{\rm DM} = - Q (t),\label{cons-DM}\\
&& \dot{\rho}_{\rm DE} + 3 H (1 + w_{\rm DE}) \rho_{\rm DE} =  Q (t), \label{cons-DE}
\end{eqnarray}
where $w_{\rm DE} = p_{\rm DE}/\rho_{\rm DE}$ is the barotropic EoS of DE and it is a constant; $Q (t)$ denotes the interaction function which describes a transfer of energy-momentum between these dark sectors. For $Q (t)> 0$, energy-momentum transfer takes place from DM to DE and $Q (t) < 0$ describes the opposite situation, that means the energy-momentum transfer occurs from DE to DM.   The conservation equations (\ref{cons-DM})
and (\ref{cons-DE}) can also be written in terms of the effective EoS parameters for DM and DE as follows 
\begin{eqnarray}
&& \dot{\rho}_{\rm DM} + 3 H \left(1+ w_{\rm DM}^{\rm eff} \right) \rho_{\rm DM} =0, \label{cons-DM-1}\\
&& \dot{\rho}_{\rm DE} + 3 H (1 + w_{\rm DE}^{\rm eff}) \rho_{\rm DE} =0, \label{cons-DE-1}
\end{eqnarray}
where 
\begin{eqnarray}\label{eff-eos}
&& w_{\rm DM}^{\rm eff} = \frac{Q}{3H \rho_{\rm DM}}, \quad w_{\rm DE}^{\rm eff} = w_{\rm DE}  - \frac{Q}{3H \rho_{\rm DE}},
\end{eqnarray}
are respectively the effective EoS parameter for DM and DE.  Now, looking at Eqs. (\ref{cons-DM-1})  and  (\ref{cons-DE-1}), one can conclude that the present interacting scenario between a pressure-less DM and DE with constant EoS can be viewed as a non-interacting DM-DE system in which the EoS parameters of the dark fluids are dynamical.  Note that the nature of the effective EoS parameters are influenced by the interaction function and they could have some far reaching consequences. For example, even if $w_{\rm DE} > -1$, a positive interaction function can help $w_{\rm DE}$ to cross the phantom divide line $w_{\rm DE} =-1$.  The reverse scenario, i.e. from $w_{\rm DE} < -1$ to $w_{\rm DE} > -1$ is also possible if $Q <0$.   Additionally, for $Q <0$, the effective EoS for DM, $w_{\rm DM}^{\rm eff}$ could be negative. 
Now, using the gravitational Eqs. (\ref{Friedmann-1}), (\ref{Friedmann-2}) together with the conservation Eqs. (\ref{cons-DM}) and (\ref{cons-DE}), one arrives at the following equation \cite{Yang:2015tzc,Bonilla:2021dql}

\begin{align}\label{eqn:Q}
w_{\rm DE} \frac{\kappa^2 Q}{H_0^3} = -\bigg[2 E^2 E'' + 2 E E'^2 \bigg](1+z)^2 - 9 E^3 (1+w_{\rm DE}) \nonumber\\+ \bigg[ 4 E^2 E' + 6 E^2 E' (1+w_{\rm DE})\bigg] (1+z), 
\end{align}
where prime denotes the derivative with respect to the redshift $z$\footnote{The cosmological redshift $z$ is related to the scale factor $a$ as, $1+z  = \frac{a_0}{a}$ in which $a_0$ stands for the present day value of the scale factor. } and $E = H/H_0$ is a dimensionless Hubble rate in which $H_0$ refers to the present value of the Hubble parameter.  
There exists an alternate relation  of $Q$ in terms of the  dimensionless co-moving distance $D(z)$.  Now, using the relation \cite{Hogg:1999ad} $D(z)=\int_0^z \frac{d\tilde{z}}{E(\tilde{z})}$ in Eq. (\ref{eqn:Q})
we obtain

\begin{align}\label{geneqn:Q-D_form}
w_{\rm DE} \frac{\kappa^2 Q}{H_0^3} &= -\frac{6D''^2-2D'D'''}{D'^5} (1+z)^2 \nonumber\\
& -\frac{9}{D'^3} (1+w_{\rm DE}) -\frac{10D'' + 6D'' w_{\rm DE}}{D'^4} (1+z). 
\end{align}
Thus, having the information on $D$, one can also reconstruct the interaction between the dark sectors. Although one can reconstruct $Q$ of Eq. (\ref{eqn:Q}), however, in the present article we shall work with the redefined interaction function as  $\widetilde{Q} (z) = \kappa^2 Q/(1+z)^6/H_0^3$.  
This $(1+z)^6$ has been used for the scaling purpose and there is no physics associated with it.

Now, using the expression of $\widetilde{Q} (z)$ in terms of $E$ and its derivatives, the effective EoS parameters can also be expressed as

\begin{subequations}
\begin{align}\label{eff-eos-v1}
   w_{\rm DM}^{\rm eff}& = \frac{-(2EE''+2E'^2)(1+z)^2-9E^2(1+w_{\rm DE})}{3  \bigl[3 (1+w_{\rm DE}) E^2 - 2 (1+z) E E'\bigr]}
   \nonumber\\
    &+\frac{(4E'+6E'(1+w_{\rm DE}))(1+z)}{3  \bigl[3 (1+w_{\rm DE}) E - 2 (1+z) E'\bigr]}~, \\[10pt]
     w_{\rm DE}^{\rm eff} &= -1 - \frac{(2EE'' + 2E'^2) (1+z)^2}{3 (3E^2- 2 (1+z) EE')}\nonumber\\ &\quad + \frac{4E' (1+z)}{3 (3E - 2 (1+z) E')}, \label{eff-eos-v1-2}
\end{align}
\end{subequations}
which in terms of the co-moving distance $D$, read
\begin{subequations}
\begin{align}\label{eff-eos-v2}
    w_{\rm DM}^{\rm eff} &= \frac{-(6D''^2-2D'D''')(1+z)^2-9D'^2(1+w_{\rm DE})}{3 \bigl[3 (1+w_{\rm DE}) D'^2 + 2 (1+z) D' D''\bigr]} \nonumber\\
    &\quad -\frac{(4D''+6D''(1+w_{\rm DE}))(1+z)}{3  \bigl[3 (1+w_{\rm DE}) D' + 2 (1+z)  D''\bigr]}~,\\[10pt]
    w_{\rm DE}^{\rm eff}  &= -1 - \frac{(6D''^2 - 2 D' D''') (1+z)^2}{3 (3D'^2 + 2 (1+z) D' D'')} \nonumber\\
    &\quad - \frac{4D'' (1+z)}{3 (3D' + 2 (1+z)D'')}. \label{eff-eos-v2-2}
\end{align}
\end{subequations}

Note that in the above expressions, effective EoS for DE does not include $w_{\rm DE}$ explicitly while the effective EoS for DM does.

\section{Non-parametric approaches}
\label{sec3-non-parametric-approaches}

In this section we describe the Gaussian process  and the Artificial Neural Network that have been considered to reconstruct the interaction function using the observational datasets. The publicly available packages used in this article are {\texttt{GaPP}\footnote{\url  {https://github.com/astrobengaly/GaPP}}} (Gaussian Processes in Python) and  \href{https://github.com/Guo-Jian-Wang/refann}{\texttt{ReFANN}\footnote{\url{https://github.com/Guo-Jian-Wang/refann}}} 
 (Reconstruct Functions with Artificial Neural Network), which is based on \href{https://pytorch.org/docs/stable/index.html}{\texttt{PyTorch}\footnote{\url{https://pytorch.org/docs/stable/index.html}}}.  In the following we describe each methodology in detail. 

\subsection{Gaussian Process}
\label{sec3.1-GP}

In this paper, we utilize observational data to reconstruct cosmological parameters without any background cosmological model through the application of Gaussian Process Regression (GPR) \cite{books/lib/RasmussenW06,2012JCAP...06..036S,Mukherjee:2021ggf}. We consider vectors $Z$ (representing redshift), $Y$ (which includes the Hubble parameter $H(z)$ and the luminosity distance $d_L(z)$), and their corresponding errors encapsulated in the covariance matrix $C$. This matrix may be correlated or uncorrelated with the given $n$ observational data points. Inherently, in GPR, $f(z)$ is a random function at $z$ as it follows a Gaussian distribution with a mean function $m(z)$ and a covariance function (Kernel) $K(z,\Tilde{z})$ \cite{2012JCAP...06..036S,Mukherjee:2021ggf}: 
 \begin{eqnarray}
 m(z) &=& \mathbb{E}[f(z)]~,\label{mean_f}\\
k(z,\Tilde{z}) &=& \mathbb{E}[(f(z)-m(z))(f(\Tilde{z})-m(\Tilde{z}))]~,\label{kernal_f}\\  
\text{Var}(z) &=& k(z,z). \label{var_f}
\end{eqnarray} 
Applying GPR to $n$ observational data at the location $Z$, we produce $n^*$ number of prediction functions ${ f^*}$ at the location $Z^*$. These function ${ f^*}$  adhere to the properties of the Gaussian distribution.~\cite{2012JCAP...06..036S,Mukherjee:2021ggf}
\begin{equation}\label{eqn_predf}
f^* \sim \mathcal{GP}\left( \boldsymbol{m}^*, K(Z^*, Z^*)
\right) \;, 
\end{equation} 
where a mean at the location $Z^*$ is defined as $\boldsymbol{m}^*$.
The data errors $C$ are Gaussian since the observational data $y\in Y$ fluctuate arbitrarily with the function $f(z)$ \cite{2012JCAP...06..036S,Mukherjee:2021ggf} 
\begin{equation}\label{eqn_obs_y}
y \sim \mathcal{GP}\left( \boldsymbol{m}, K(Z, Z)+ C
\right),
\end{equation}
where $\boldsymbol{m}$ denotes the mean at the location $Z$.
We obtain the joint probability distribution function by combining  Eqs. (\ref{eqn_predf}) and (\ref{eqn_obs_y}). Since $y$ is known, we can find $f^*$ using a conditional distribution over $y$ \cite{2012JCAP...06..036S,Mukherjee:2021ggf}
\begin{equation}\label{conditional}
{f^*}| Z^*, Z, y \sim \mathcal{GP} \left(
\overline{f^*}, \text{cov}({f^*})
\right). 
\end{equation}
Also the mean and covariance are
\begin{eqnarray}
\overline{f^*} =\boldsymbol{m}^* + K(Z^*, Z)\left[K( Z,Z) +
  C\right]^{-1} ({y} - \boldsymbol{m}),\label{gp-mean_f}
\end{eqnarray}
\begin{eqnarray}
\text{cov}({f^*}) =  K(Z^*,Z^*) - K(Z^*,Z)\nonumber\\ \times \left[K(Z,Z) + C\right]^{-1} K(Z,Z^*). \label{gp-covf*}
\end{eqnarray}
There are different types of covariance functions used to reconstruct the function in Gaussian Process. The different  types of covariance functions are \cite{2012JCAP...06..036S,books/lib/RasmussenW06,e6e4c2b74de0412ab4feae906164e191,MacKay2003}: 
\begin{enumerate}

\item { \bf Squared Exponential:} In this case the kernel takes the form 
    
\begin{equation}
K(z,\tilde{z}) =
\sigma_f^2 \exp\left( -\frac{(z - \tilde{z})^2}{2l^2} \right). \label{eqn_squared_Exp}
\end{equation}

\item { \bf Mat\'{e}rn:} In this case the form of the kernel takes 

\begin{eqnarray}
\label{cov_Matern}
 K_{\nu=p+\frac{1}{2}}(z,\tilde{z}) = \sigma_f^2 \exp \left( \frac{-\sqrt{2p+1}}{l} \vert z - \tilde{z} \vert \right) \frac{p!}{(2p)!} \nonumber\\ 
 \times~ \sum_{i=0}^{p} \frac{(p+i)!}{i!(p-i)!} \left( \frac{2\sqrt{2p+1}}{l} \vert z - \tilde{z} \vert \right)^{p-i}.
\end{eqnarray}
For each value of $p$,  there is a distinct covariance function of Mat\'{e}rn, which are Mat\'{e}rn $9/2$ (for $p=4$), Mat\'{e}rn $7/2$ ($p= 3$), Mat\'{e}rn $5/2$ ($p=2$), and Mat\'{e}rn $3/2$ ($p=1$). Note that the  Mat\'{e}rn covariance functions are $r$-times mean square differentiable if $r<\nu$ where $\nu = p+ 1/2$  as given in  Eq. \eqref{cov_Matern}.

\item {\bf Cauchy:} The kernel is given by 

\begin{equation}\label{Cauchy_cov}
   K(z, \tilde{z}) = \sigma_f^2 \left(\frac{l}{(z - \tilde{z})^2 + l^2}\right). 
\end{equation}

\item {\bf Rational Quadratic:} The kernel assumes the following form 

\begin{equation}\label{Rational_cov}
   K(z, \tilde{z}) = \sigma_f^2 \left(1 + \frac{(z - \tilde{z})^2}{2\alpha l^2}\right)^{-\alpha}.  
\end{equation}  
\end{enumerate}

In the descriptions of the kernels presented above, $\sigma_f$, $\alpha$ and $l$ are defined as the hyperparameters. The posterior probability distribution function, denoted as Eq. (\ref{conditional}), is determined by the Bayes theorem, which states that it is dependent on both the likelihood and the prior. By performing marginalisation on the posterior probability distribution function, we can derive the marginal likelihood. Subsequently, the hyperparameters are then trained and optimized in order to maximize the log marginal likelihood~  \cite{2012JCAP...06..036S,Mukherjee:2021ggf} 

\begin{equation}\label{log-marginal-p}
\begin{split}
    \ln \mathcal{L} =&\ln \mathbf{p}(y|Z, \sigma_f, l) \\ 
                    =& -\frac{1}{2}(y - \boldsymbol{m})^T \left[K( Z,Z) + C\right]^{-1} (y- \boldsymbol{m}) \\
                    & \hspace{1.7cm} - \frac{1}{2}\ln\left|K( Z,Z)+C\right|- \frac{n}{2}\ln 2\pi  \;.
\end{split}
\end{equation}
If we choose Rational Quadratic kernel, the form of  log marginal likelihood is given below:
\begin{equation}\label{log-marginal-p_Rational}
\begin{split}
    \ln \mathcal{L} =&\ln \mathbf{p}(y|Z, \sigma_f,\alpha, l) \\ 
                    =& -\frac{1}{2}(y - \boldsymbol{m})^T \left[K( Z,Z) + C\right]^{-1} (y- \boldsymbol{m}) \\
                    & \hspace{1.7cm} - \frac{1}{2}\ln\left|K( Z,Z)+C\right|- \frac{n}{2}\ln 2\pi  \;.
\end{split}
\end{equation}
In GPR, the reconstruction of the  first, second, and third derivatives of the function $f(z)$ also follows the Gaussian Property. We ascertain the covariance between the function and the derivatives following the Gaussian property \cite{2012JCAP...06..036S,Mukherjee:2021ggf}
\begin{align}
f'(z) &\sim \mathcal{GP} \left(m'(z), \frac{\partial^2 K(z, \tilde{z})}{\partial z \partial \tilde{z}}\right), \label{ch1:f1z_gp}\\
f''(z) &\sim \mathcal{GP} \left(m''(z), \frac{\partial^4 K(z, \tilde{z})}{\partial z^2 \partial \tilde{z}^2}\right), \label{ch1:f2z_gp}
\textbf{}\end{align}
\begin{align}
f'''(z) &\sim \mathcal{GP} \left(m'''(z), \frac{\partial^6
K(z, \tilde{z})}{\partial z^3 \partial \tilde{z}^3}\right). \label{ch1:f3z_gp}
\end{align}
Next, we apply the same procedure to the derivative functions, similar to the method used for $f(z)$. Eventually, we obtain all the reconstructed functions, namely $f(z)$, $f'(z)$, $f''(z)$, and $f'''(z)$. With these reconstructed functions, we can now rebuild the function $h(f(z),f'(z),f''(z),f'''(z))$, that is $\widetilde{Q} (z)$ in our case.  

\subsection{Artificial Neural Network}
\label{sec3.2-ANN}

An ANN functions similarly to the animal brains. In this article, we employ ANN to reconstruct the interaction between DM and DE, eliminating the need for background cosmological models.

\begin{figure}
    \centering
    \includegraphics[scale=1.2]{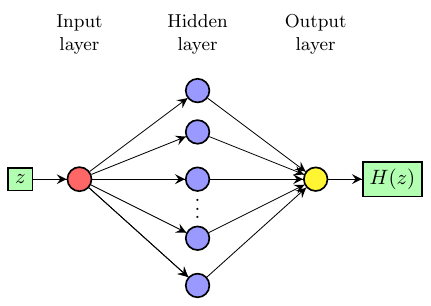}
     \caption{The structure of an ANN model with given redshift $z$ to reconstruct the Hubble function $H(z)$.  }
    \label{fig:ann}
    \end{figure}

An ANN is made up by three things, namely, input, one or more hidden layers, and an output. A Neural Network is characterized as an interconnected network of layers, each consisting of multiple elements, known as neurons or nodes. The output, which includes the Hubble parameter $H(z)$, is not initially known to us. 
We can formulate the output functions using redshift $z$ as the sole input parameter, as shown in Fig. \ref{fig:ann}. The procedure to generate output $H(z)$ and its error $\sigma_H(z)$ 
as depicted in Fig. \ref{fig:ann}, is given below:
\begin{enumerate}
    \item First, we consider $n$ observational data points for $H(z)$ and their associated errors, $\sigma_H(z)$. Now following the Gaussian distribution  $\mathcal{N}( H(z),\sigma_H(z))$, we create 1000 realizations of a data-like sample of  $n$~$ H(z)$ measurements. 
    
    \item After training the ANN model on each sample of $H(z)$, we reconstruct $H(z)$. We  recreate $1000$ reconstructed $H(z)$ by repeating this process for every sample using the trained ANN model. 
    
    \item On the completion of step 2, one can compute the covariance between two Hubble parameters at different redshifts. Using the following formula for the $1000$ reconstructed of $H(z)$ \cite{Wang:2019vxv}
    \begin{align}\label{equ:cov}
 	\nonumber &{\rm\bf Cov}(H(z_i), H(z_j))\\
 	&=\frac{1}{N}\sum_{k=1}^N[(H(z_i)_k - \bar{H}(z_i))(H(z_j)_k - \bar{H}(z_j))]~,
 	\end{align}
  where $N=1000$; $z_i$, $z_j$ represent different redshifts, $\bar{H}(z)$ denotes the mean function of $H(z)$, we get the covariance matrix. 
  \item In the case of $H'(z)$, we use the central differentiation approach, which is 
  \begin{equation}\label{rec_H'(z)}
      H'(z_i)\simeq \frac{H(z_{i+1})-H(z_{i-1})}{z_{i+1}-z_{i-1}}~.
  \end{equation}
By using $1000$ reconstructed $ H(z)$ at each redshift $z$ (which is obtained by completing Step 2) in Eq. \eqref{rec_H'(z)}, we obtain $1000$ reconstructed $H'(z)$ realizations  and the corresponding covariance can be calculated using the same procedure as Step 3. The second and third derivatives of $H(z)$ can also be calculated using the same procedure.
\end{enumerate}
The input layer propagates a signal to the next layer through a weighted sum of inputs and non linear activation function. Let the output  row vector of the $i$-th layer is denoted by $x_i$  and the linear weights and biases are represented by $w_{i+1}$ and $b_{i+1}$, respectively, which are learned. The form is provided below ~\cite{Wang:2019vxv}
\begin{equation}\label{weight_sum}
    z_{i+1}= x_iw_{i+1}+b_{i+1},
\end{equation}
where $z_{i+1}$ represents the weighted sum of the input vector of the $(i+1)$-th layer.  Additionally, a nonlinear activation function is used to transform the $(i+1)$-th layer to $(i+2)$-th layer, that means \cite{Wang:2019vxv} 
\begin{equation}\label{activation_f}
    x_{i+1}=f(z_{i+1})~,
\end{equation}
where $x_{i+1}$ is the output of $(i+1)$-th layer. The topology of neural networks allows for the parallel operation of a batch of data. Let $X$ be an ${l\times n}$ real matrix, where $l$ be the number of row vectors $1\times n$ of the matrix $X$ in which $l$ rows of the matrix $X$ are independent to each other. Now the Eqs. \eqref{weight_sum} and \eqref{activation_f} are written in the generalised way as  ~\cite{Wang:2019vxv}
\begin{equation}\label{eq_matrixdef1}
Z_{i + 1} = X_{i}W_{i+1}+B_{i + 1},
\end{equation}
\begin{equation}\label{eq_matrixdef2}
X_{i+1} = f(Z_{i+1}),
\end{equation}
Thus, in Eq. \eqref{weight_sum}, $B_{i+1}$ is the vertically duplicated matrix of $b_{i+1}$.
Using the similar procedure on the $(i+2)$-th layer, we can obtain the next $(i+3)$-th layer value and continue this process until we get the output value in the output layer. After training the neural networks, we get the predicted output $\Tilde{Y}$, which is $l\times v$  real matrix. Now since we already have an actual output $l\times v$ real matrix $Y$ using the data in the output layer, therefore, the difference between the actual output and the predicted output can be measured using a loss function. Here we take the SmoothL1 as the loss function, which is defined as~\cite{2015arXiv150408083G}

\begin{equation}\label{loss_fun}
\mathcal{L} =
\begin{cases}
    0.5(\widetilde{Y} - Y)^2 / (lv\beta) ~, & |\widetilde{Y} - Y| < \beta \\
    \bigskip \\
    (|\widetilde{Y} - Y| - 0.5\beta) / (lv) ~, & \text{otherwise},
\end{cases}
\end{equation}
where $\beta$ is a non-negative hyperparameter. Without any loss of generality, here we fixed the value of $\beta$ to be unity.
If the loss function $\mathcal L$ is very small, then we get the output and also optimize the value of the hyperparamters like weights and biases. When the loss function $\mathcal L$ is not small, 
and the ANN's hyperparameters are not optimized.  Specifically, the hyperparameters $W$ and $B$ are updated and optimized during the training process using a gradient descent method. Here we adopt the Adam optimizer \cite{2014arXiv1412.6980K} to optimize the hyperparameters by minimizing the loss function. In addition, we use L2 weight decay~\cite{Loshchilov:2017bsp}
on the parameters to prevent over-fitting.

Finally, concerning the activation functions to work with the ANN,  there are several activation functions such as the Softplus activation function having the form
~\cite{7280459} 

\begin{equation}
    f(x)=\frac{\log(1+\exp(\theta x))}{\beta},
\end{equation}
where $\theta$ is the parameter of this activation function and we set it to be unity,  and the LogSigmoid activation function ~\cite{inproceedings}
\begin{equation}
    f(x)=\log\left(\frac{1}{1+\exp(-x)}\right). 
\end{equation}
Notice that this activation function does not have any free parameter unlike the Softplus activation function. 
Having these activation functions, one can proceed to reconstruct the interaction function $\widetilde{Q} (z)$ using the methodology of ANN and the observational datasets.   In the next sections we shall describe the observational datasets, and the results.

\begin{table}
\caption{34 CC $H(z)$ measurements obtained from the differential age method. The $H(z)$ obtained from \citet{Ratsimbazafy:2017vga} is $89 \pm 23~(\rm stat) \pm 44~(\rm syst) ~\rm km\ s^{-1}\ Mpc^{-1}$, here we consider their total error $89 \pm 49.6~(\rm tot) ~\rm km\ s^{-1}\ Mpc^{-1}$ in our analysis. The $H(z)$ obtained from \citet{Jiao:2022aep} is $113.1 \pm 15.1~(\rm stat)^{+29.1}_{-11.3}~(\rm syst) ~\rm km\ s^{-1}\ Mpc^{-1}$,
   here we consider their total error $113.1 \pm 25.22~(\rm tot) ~\rm km\ s^{-1}\ Mpc^{-1}$
   by taking 20.2 $(= (29.1+11.3)/2)$ as the systematic error to avoid lower estimations of errors. The total error can be calculated via $\sigma_{\rm tot} = \sqrt{\sigma_{\rm stat}^2 + \sigma_{\rm syst}^2}$. }
    \label{tab:Hz}
    \begin{center}
    \resizebox{0.4\textwidth}{!}{
	\begin{tabular}{cccc}
		\hline
		\hline
		$z$  & $H(z) \pm \sigma_{H(z)}~\rm km\ s^{-1}\ Mpc^{-1}$ & References \\
		\hline
		0.09 	&   69 $\pm$  12	&  \citet{Jimenez:2003iv} \\
		\hline
		0.17	&   83   $\pm$   8  &	\\
		0.27	&	77   $\pm$  14	&	\\
		0.4		&	95   $\pm$  17	&	\\
		0.9		&	117  $\pm$  23	&   \citet{Simon:2004tf} \\
		1.3		&	168  $\pm$  17	&	\\
		1.43	&	177  $\pm$  18	&	\\
		1.53	&	140  $\pm$  14	&	\\
		1.75	&	202  $\pm$  40	&	\\
		\hline
		0.48	&	97   $\pm$  62	&   \citet{Stern:2009ep} \\
		0.88	&	90  $\pm$  40	&	\\
		\hline
		0.1791	&	75   $\pm$  4	&	\\
		0.1993	&	75  $\pm$  5	&	\\
		0.3519	&	83   $\pm$  14	&	\\
		0.5929	&	104 $\pm$  13	&	\citet{Moresco:2012by} \\
		0.6797	&	92  $\pm$  8	&	\\
		0.7812	&	105  $\pm$  12	&	\\
		0.8754	&	125  $\pm$  17	&	\\
		1.037	&	154  $\pm$  20	&	\\
		\hline
		0.07	&	69   $\pm$ 19.6	&	\\
		0.12	&	68.6 $\pm$ 26.2	&	\citet{Zhang:2012mp} \\
		0.2		&	72.9 $\pm$ 29.6	&	\\
		0.28	&	88.8 $\pm$ 36.6	&	\\
		\hline
		1.363	&	160  $\pm$ 33.6	&	\citet{Moresco:2015cya} \\
		1.965	&	186.5 $\pm$  50.4	&	\\
		\hline
		0.3802	&	83   $\pm$ 13.5	&	\\
		0.4004	&	77   $\pm$ 10.2	&	\\
		0.4247	&	87.1 $\pm$  11.2	&	\citet{Moresco:2016mzx} \\
		0.44497	&	92.8 $\pm$  12.9	&	\\
		0.4783	&	80.9 $\pm$ 9	&	\\
		\hline
		0.47    &   89   $\pm$ 49.6 &  \citet{Ratsimbazafy:2017vga} \\
		\hline
		0.75    &   98.8 $\pm$ 33.6 &  \citet{Borghi:2021rft} \\
		\hline
		0.80    &  113.1 $\pm$ 25.22 &  \citet{Jiao:2022aep} \\
		\hline
		1.26    &  135 $\pm$  65 &  \citet{Tomasetti:2023kek} \\  
		\hline\hline
	\end{tabular}}
    \end{center}
\end{table}

\section{Observational data and Methodology}
\label{sec-data}

In order to reconstruct the dimensionless interaction function, $\widetilde{Q} (z)$, we used two different sources of model-independent datasets, namely the Hubble parameter measurements from Cosmic Chronometers and Pantheon+ sample of Supernovae Type Ia. In what follows we describe the datasets:

\begin{enumerate}
    \item {\bf Cosmic Chronometers (CC):} The Hubble Parameter $H(z)$ quantification describes the rate at which the universe is expanding.  We exclusively address the model independent method measurements in this paper. The model independent value of $H(z)$ can be found for the passively developing galaxies at varying redshifts $z$ with regard to 
    time $t$. This method is known as Cosmic Chronometers(CC). The formula of this  method is given by
\begin{equation}\label{CC}
    H(z)\simeq-\frac{1}{1+z}\frac{\Delta z}{\Delta t}~.
\end{equation}
In this work, we make use of 34 compilation of the CC data, which are displayed in Table~\ref{tab:Hz} with their corresponding references. 

\item {\bf Pantheon$+$ sample of Supernovae Type Ia:}
Given that in Pantheon+ sample of Supernovae Type Ia (SNIa)~\cite{Brout:2022vxf}, there are total 1701 light curves in the given range of $0.01 \leq z \leq 2.26$ with SH0ES included.
In this paper, we use 1590 SNIa data points with some SH0ES data in this redshift range of $0.01016 \leq z \leq 2.26137$ with the condition $z > 0.01$ \cite{Chan-GyungPark:2024mlx} which 
allows us to reduce the effects of the unusual velocity correction. 
The distance modulus $ \mu$ and corresponding covariance matrix of the given Pantheon+ data are publicly available\footnote{\url  {https://github.com/PantheonPlusSH0ES/DataRelease}}.
Here, the relation between the distance modulus $\mu$ and the luminosity distance $d_{\rm L} (z)$ 
is given by

\begin{equation}
    \mu = 5\log_{10} \frac{d_{\rm L} (z)}{\rm Mpc} + 25 ~.\label{SN_dl}
\end{equation}
The dimensionless comoving distance is

\begin{equation} \label{ch1:D_from_DL}
D(z) \equiv \frac{H_0 d_{\rm L}(z)}{c(1+z)}, 
\end{equation}
where $c$ is the speed of light.
Now, since the covariance matrix of $\mu$ is known, we use the error propagation rule to get the covariance matrix of $D(z)$~\cite{Mukherjee:2021kcu}. In this procedure, we create the 1590 data points of Pantheon+ sample.  Given that Eq. \eqref{ch1:D_from_DL} contains an $H_0$, one needs to provide a value of $H_0$ while reconstructing the cosmological parameters, here the interaction in the dark sector. Since the debate on the many values of $H_0$ is still not over \cite{DiValentino:2021izs,Perivolaropoulos:2021jda, Schoneberg:2021qvd, Abdalla:2022yfr}, therefore, in this work, while reconstructing with the Pantheon+ sample, we consider two different values of $H_0$, namely,  $H_0 = 67.36 \pm 0.54$ $\rm km\ s^{-1}\ Mpc^{-1}$ at 68\% CL~\cite{Planck:2018vyg} and $H_0 = 73.04 \pm 1.04$ $\rm km\ s^{-1}\ Mpc^{-1}$ at 68\% CL~\cite{Riess:2021jrx}.

 \end{enumerate}
For the combined dataset CC+Pantheon+, firstly, we convert CC data points in the form of $D(z)$ and combine the data points of Pantheon+ sample.  For the case of combining the covariance matrix of CC and Pantheon+ data, we use the concept of block diagonal matrix.

\begin{figure*}
 \centering 
     \includegraphics[width=0.35\textwidth]{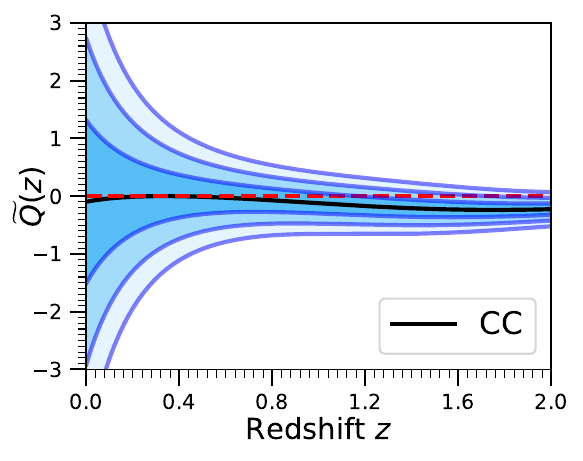}\\
\includegraphics[width=0.35\textwidth]{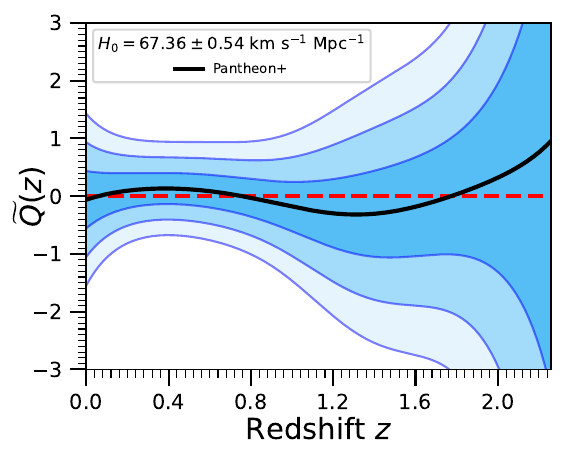}
\includegraphics[width=0.35\textwidth]{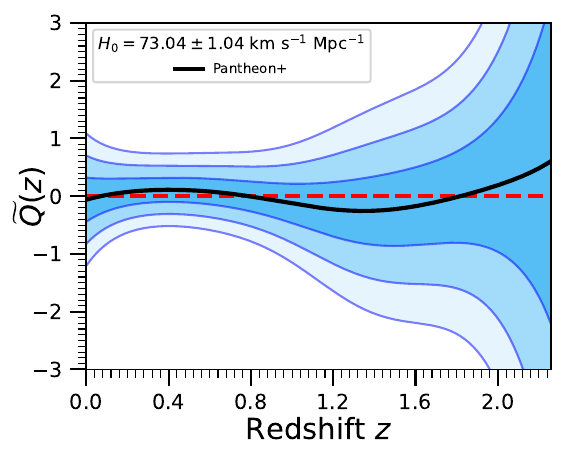}
\includegraphics[width=0.35\textwidth]{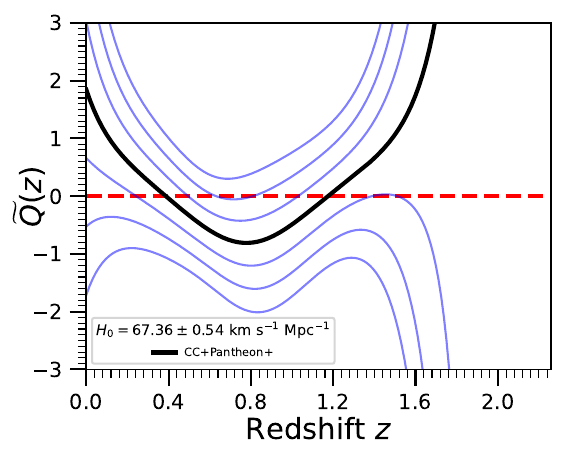}
\includegraphics[width=0.35\textwidth]{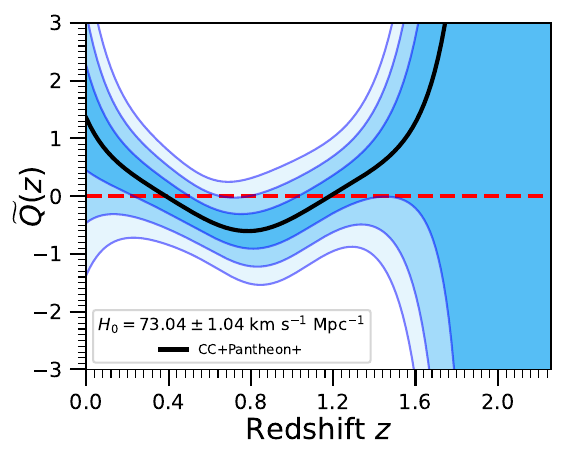}
\caption{Reconstructed interaction function $\widetilde{Q} (z)$ using the Gaussian process considering 34 CC $H(z)$, Pantheon+, and CC+Pantheon+ data for $w_{\rm DE} =-1$. The Hubble constant $H_{0}=67.36 \pm 0.54$~$\rm km\ s^{-1}\ Mpc^{-1}$ at 68\% CL \protect\cite{Planck:2018vyg} 
and $H_{0}=73.04 \pm 1.04$~$\rm km\ s^{-1}\ Mpc^{-1}$ at 68\% CL 
\protect\cite{Riess:2021jrx} 
are used in the reconstruction of $D(z)$. The horizontal dashed line (red) in each plot corresponds to $\widetilde{Q} (z) =0$, i.e. no-interaction between the dark sectors and
the solid curve (black) stands for the mean curve for each case. }
    \label{fig:squared-GP}
\end{figure*}
\begin{figure*}
\includegraphics[width=0.4\textwidth]{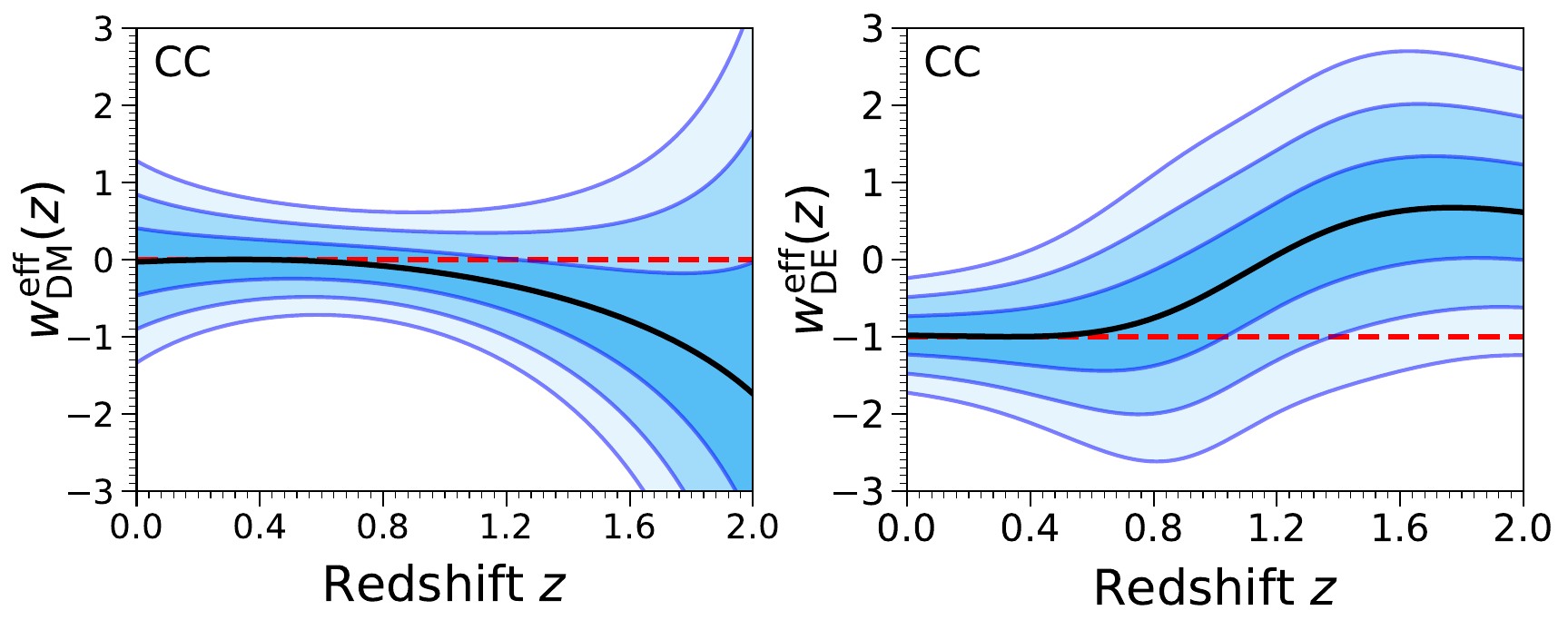}\\
\includegraphics[width=0.4\textwidth]{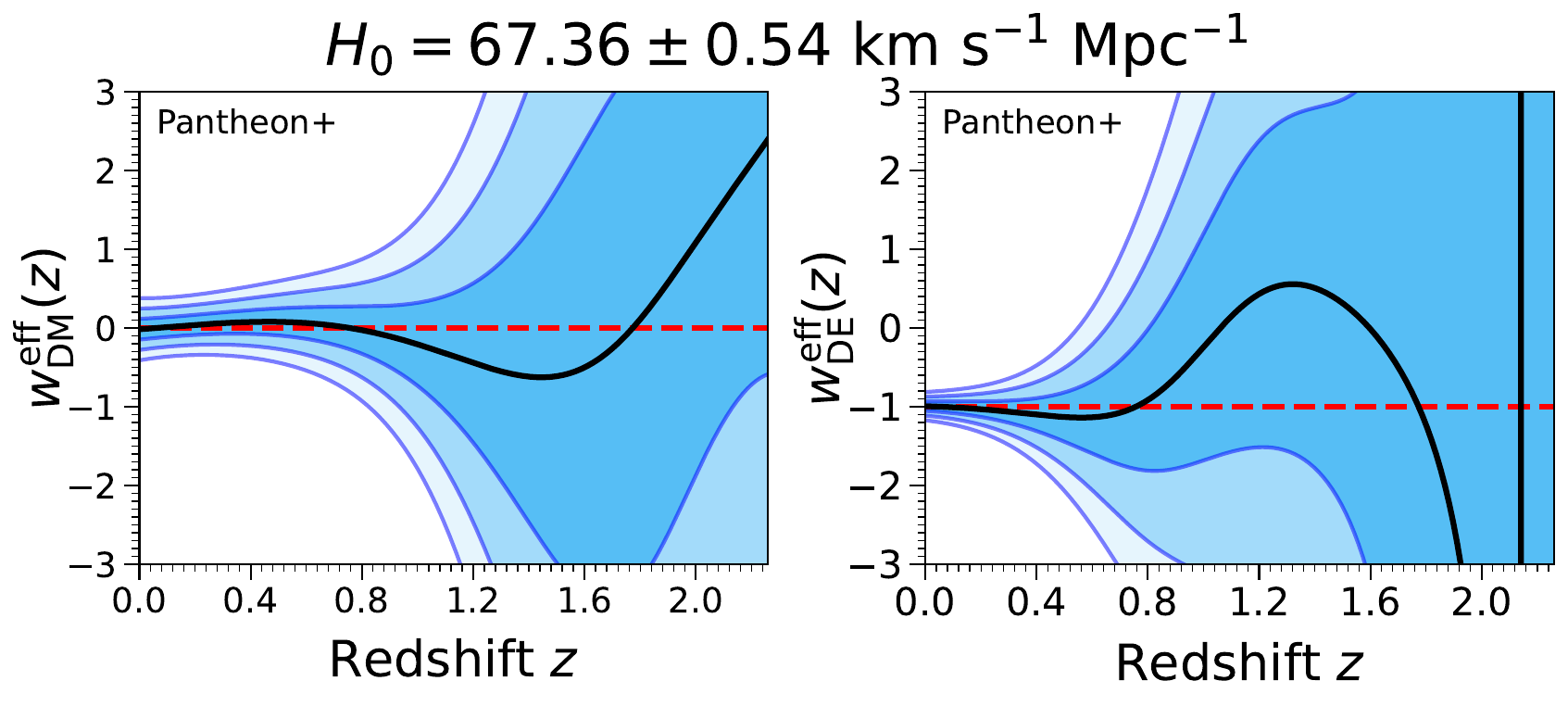}
\includegraphics[width=0.4\textwidth]{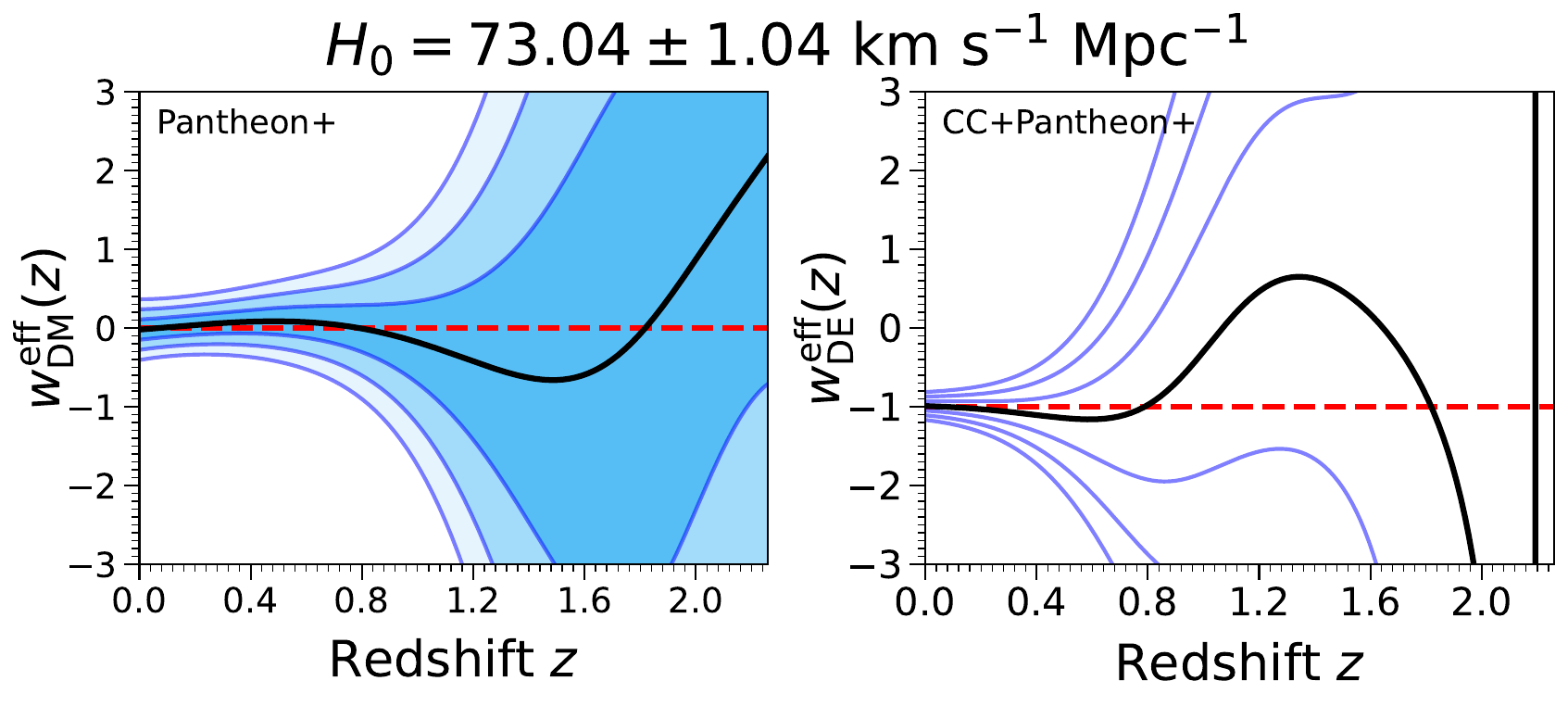}
\includegraphics[width=0.4\textwidth]{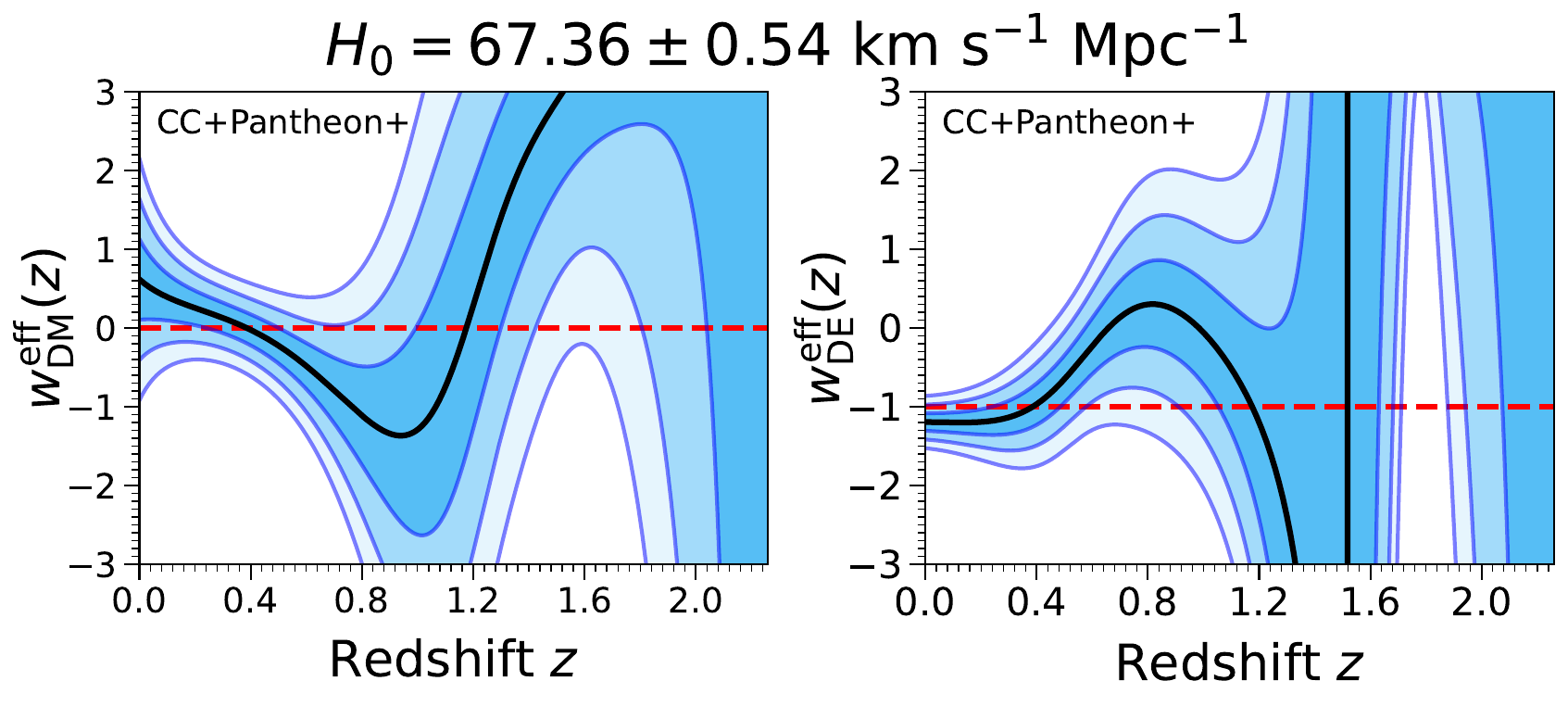}
\includegraphics[width=0.4\textwidth]{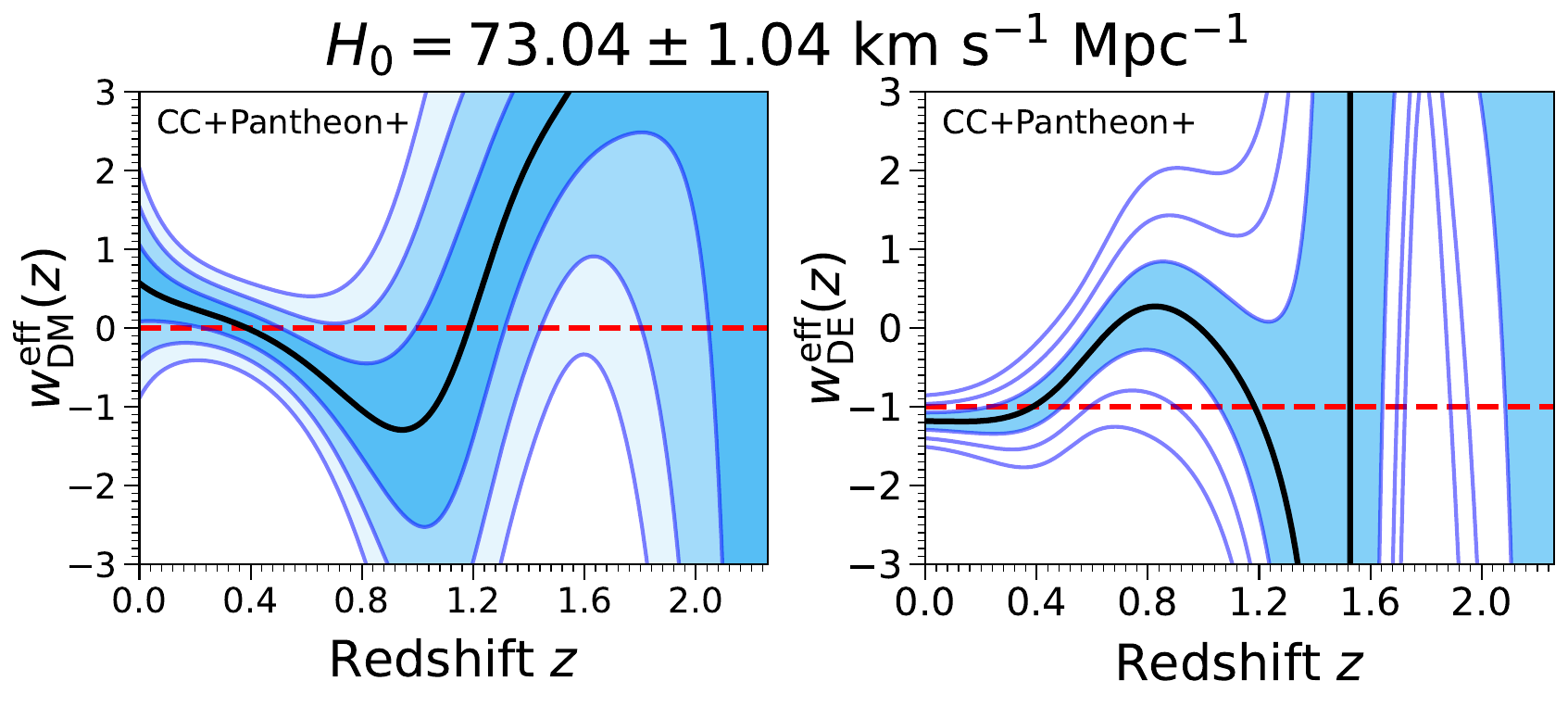}
\caption{Reconstructed effective EoS parameters, $w_{\rm DM}^{\rm eff}$ and $w_{\rm DE}^{\rm eff}$, using the Gaussian process considering 34 CC $H(z)$, Pantheon+, and CC+Pantheon+ data for $w_{\rm DE} =-1$. The Hubble constant $H_{0}=67.36 \pm 0.54$~$\rm km\ s^{-1}\ Mpc^{-1}$ at 68\% CL \protect\cite{Planck:2018vyg} 
and $H_{0}=73.04 \pm 1.04$~$\rm km\ s^{-1}\ Mpc^{-1}$ at 68\% CL 
\protect\cite{Riess:2021jrx} 
are used in the reconstruction of $D(z)$. The horizontal dashed line (red) in the plots for $w_{\rm DM}^{\rm eff}$ ($w_{\rm DE}^{\rm eff}$) corresponds to $w_{\rm DM}^{\rm eff} =0$ ($w_{\rm DE}^{\rm eff} =-1$) and the solid curve (black) stands for the mean curve of the respective reconstructed effective EoS parameter.}
\label{fig:wDMeff-wDEeff-GP}
\end{figure*}
\begin{figure}
    \centering
\includegraphics[width=0.45\textwidth]{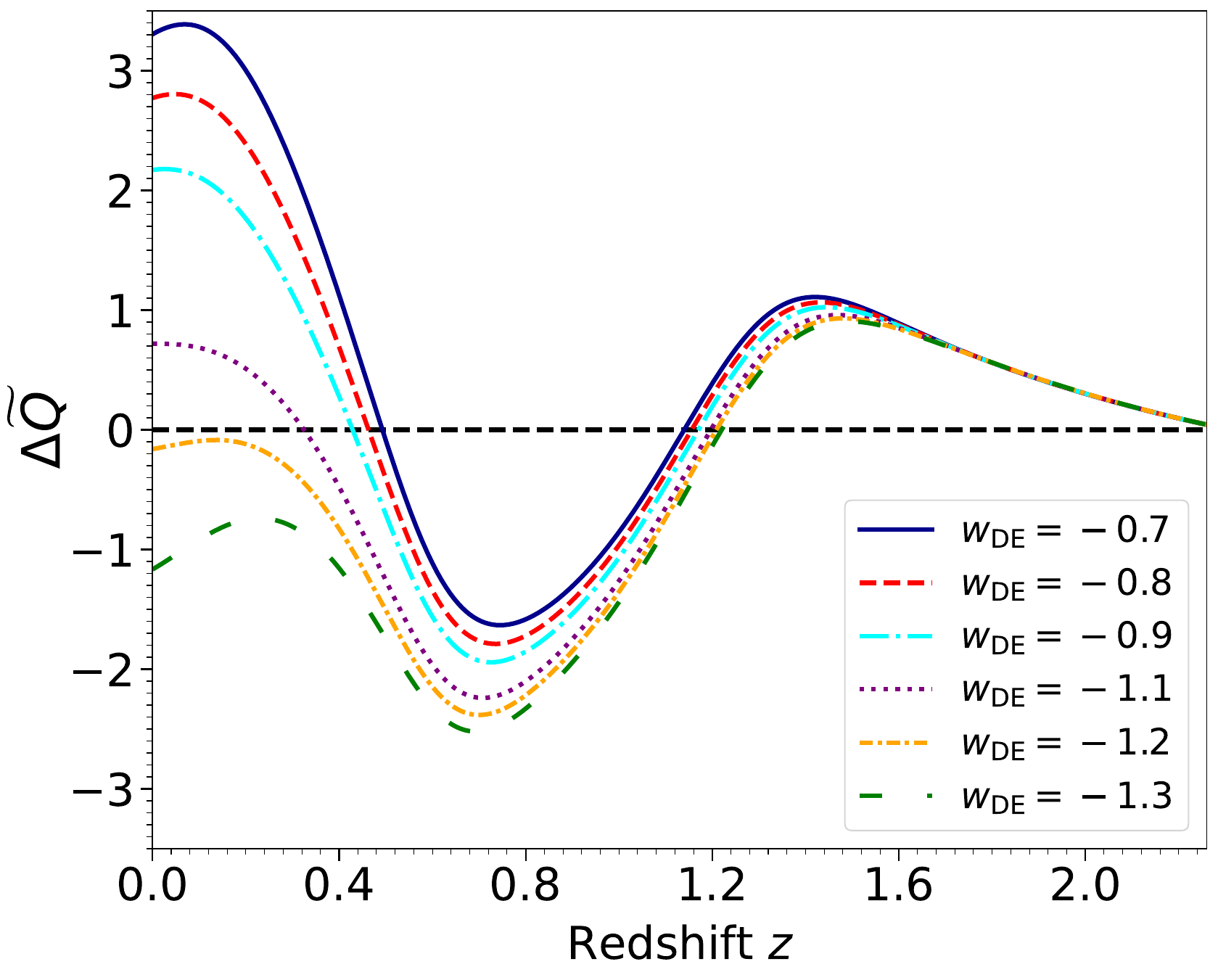}
    \caption{The deviation of $\widetilde{Q}$ from zero, where $\Delta\widetilde{Q} = (\widetilde{Q} (z)-0)/\sigma_{\widetilde{Q}(z)}$ is shown for the GP. The figure corresponds to the  reconstructions using CC+Pantheon+ and for $H_{0}=73.04\pm1.04$ $\rm km\ s^{-1}\ Mpc^{-1}$ ~\protect\cite{Riess:2021jrx}.}
    \label{fig:Q-evolution-diff-w}
\end{figure}
\begin{figure}
\centering
\includegraphics[width=0.45\textwidth]{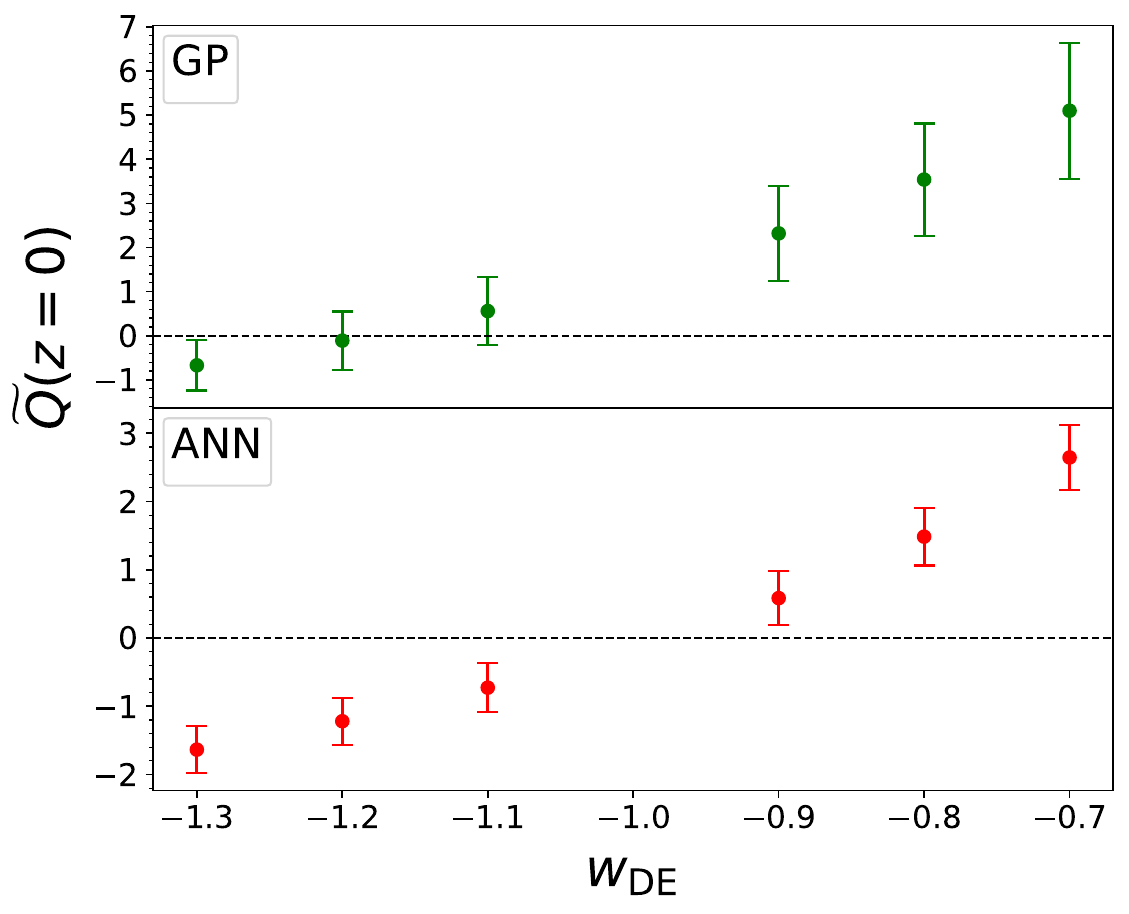}
\caption{Whisker plot showing 68\% CL constraints on $\widetilde{Q} (z =0)$ obtained from GP (upper) and ANN (lower) reconstructions considering different values of $w_{\rm DE}$ and using CC+Pantheon+ with $H_0 = 73.04 \pm 1.04$~$\rm km\ s^{-1}\ Mpc^{-1}$~\protect\cite{Riess:2021jrx}.}
    \label{fig:GP+ANN-Q-at-z=0}
\end{figure}
\begin{figure}
\includegraphics[width=0.45\textwidth]
{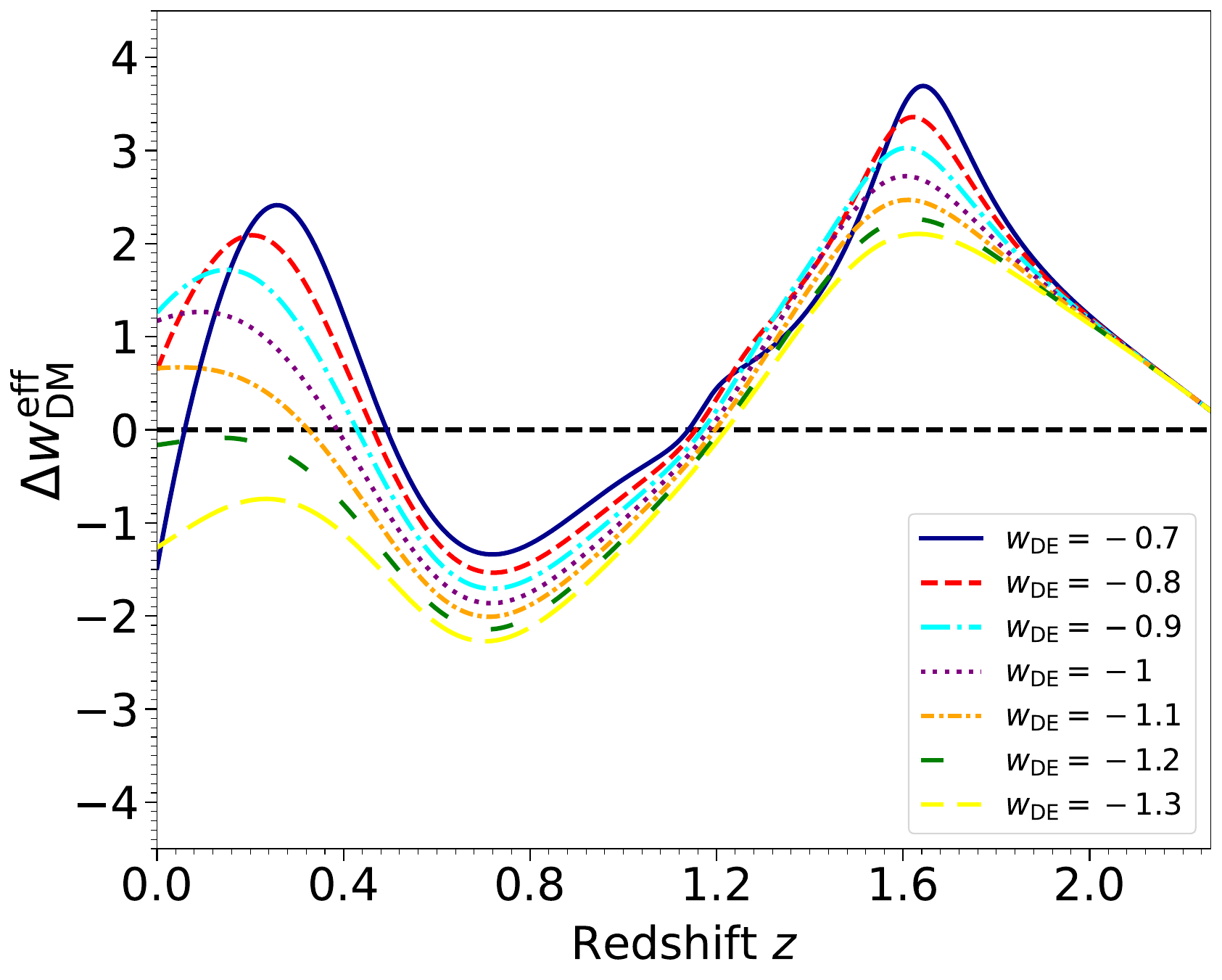}
\caption{The deviation of $w_{\rm DM}^{\rm eff}$ from  $0$  
 where  $\Delta w_{\rm DM}^{\rm eff} = (w_{\rm DM}^{\rm eff} (z) - 0)/\sigma_{w_{\rm DM}^{\rm eff}(z)}$ is shown for CC+Pantheon+ with $H_0 = 73.04 \pm 1.04$~$\rm km\ s^{-1}\ Mpc^{-1}$~ \protect\cite{Riess:2021jrx} considering different values of $w_{\rm DE}$  and using GP.  }
\label{fig:weffDM-weffDE-diffwDE-GP}
\end{figure}

\section{Results}
\label{sec-results}

In this section we present the results on the reconstructed dimensionless interaction function $\widetilde{Q} (z)$  by adopting the Gaussian process and ANN approaches. 
The reconstructions have been done using (i) 34 CC data alone, (ii) 1590 Pantheon+ data alone (excluding the data points affected by peculiar
velocity corrections~\cite{Chan-GyungPark:2024mlx}) distributed in the redshift range $0.01016 \leq z \leq 2.26137$, and finally (iii) using the combined dataset CC+Pantheon+.  It is essential to draw the attention to the readers that while reconstructing $\widetilde{Q} (z)$ using various kernels of the Gaussian process, no significant changes have been observed. The physics of the interaction does not offer any new insights for changing kernels, thus, in this section we present the Gaussian process reconstruction of $\widetilde{Q} (z)$ for the squared exponential kernel. On the other hand, while reconstructing the interaction function using ANN adopting the Softplus and Logsigmoid activation functions, we noticed that the results remain similar irrespective of these activation functions. Thus, considering this issue we present the ANN results for the Softplus activation function. The results are shown in various figures in the main text (Figs. \ref{fig:squared-GP}, \ref{fig:wDMeff-wDEeff-GP}, \ref{fig:Q-evolution-diff-w}, \ref{fig:GP+ANN-Q-at-z=0},  \ref{fig:weffDM-weffDE-diffwDE-GP}, \ref{fig:ANN-CC-PP-CC+PP}, \ref{fig:wDMeff-wDEeff-ANN}, \ref{fig:ANN-CC+PP+vary-w-dev},  \ref{fig:weff_ANN}) and in Appendix-A (Figs. \ref{fig:GP-Q_diff_wde-Appendix-1}, \ref{fig:GP-eff-wdm_wde-Appendix-2}, \ref{fig:ANN-Q_diff_wde-Appendix-1}, \ref{fig:ANN-eff-wdm_wde-Appendix-2}). 
In the following, we present the results of $\widetilde{Q} (z)$ and the effective EoS parameters considering both approaches in presence of the above datasets.

\subsection{Reconstructions using GP} 

\subsubsection{For $w_{\rm DE}=-1$}
\label{subsec-results-GP}

We first start with the reconstructions of $\widetilde{Q} (z)$ using 34 CC data alone. The initial step is to reconstruct the Hubble function $H (z)$ and its first and second derivatives, namely $H' (z)$ and $H''(z)$ (here prime stands for the derivative of $H (z)$ with respect to the redshift $z$) for the 34 CC dataset.  After these reconstructions, we reconstruct the dimensionless variables $E (z) = H(z)/H_0$, $E '(z) = H'(z)/H_0$ and $E ''(z) = H''(z)/H_0$ in which $H_0$ refers to the present value of $H(z)$ (i.e. $H_0 = H (z=0)$). The reconstructions of  $E (z)$, $E'(z)$, and $E''(z)$ are essential because the final form of $\widetilde{Q}(z)$ is expressed in terms of these dimensionless variables (see Eq. (\ref{eqn:Q})). Thus, we can see that the value of $H_0$ which directly enters into the reconstructions through $E (z)$, $E'(z)$, $E''(z)$ seems to affect the reconstructions. It has been  pointed out in Ref. \cite{Wei:2016xti,Wang:2017lri,Wang:2020dbt} that the reconstructions using the Gaussian process might be influenced by the priors of $H_0$. 
Hence, a value of $H_0$ is needed for the reconstructions. Now, from the reconstructed graph of $H (z)$ using CC alone, one can find the value of $H_0$ for the kernel and this can be used for the next step.  Following this methodology, we have reconstructed $\widetilde{Q} (z)$ for the 34 CC data. 
The topmost plot of Fig. \ref{fig:squared-GP} shows the reconstructed interaction function $\widetilde{Q} (z)$ using 34 CC dataset.  In this case, we do not find any strong evidence of an interaction in the dark sector since within 68\% CL, $\widetilde{Q} (z) = 0$ is allowed.

Concerning the reconstructions of $\widetilde{Q} (z)$ using Pantheon+ only, in contrary to the reconstructions using CC alone, here we need to supply an $H_0$ to  reconstruct $D(z)$. 
As the choice of $H_0$ is arbitrary, and also as already argued in \cite{Wei:2016xti,Wang:2017lri,Wang:2020dbt} that the choice of $H_0$ matters in the reconstructions using Gaussian processes, 
therefore, we have taken two distinct 
values of $H_0$, namely, $H_0 = 67.36 \pm 0.54$ $\rm km\ s^{-1}\ Mpc^{-1}$ at 68\% CL \cite{Planck:2018vyg} and $H_0 = 73.04 \pm 1.04$ $\rm km\ s^{-1}\ Mpc^{-1}$ at 68\% CL \cite{Riess:2021jrx}.  Considering these $H_0$ values, we have reconstructed the interaction function $\widetilde{Q} (z)$. In the middle panel of Fig. \ref{fig:squared-GP}, we show the reconstructed graphs of $\widetilde{Q} (z)$ using Pantheon+ alone considering both the values of $H_0$. Focusing on  both the graphs, we do not observe any significant impact caused by the choice of the $H_0$ prior. Moreover, we do not observe any evidence of interaction since within 68\% CL, $\widetilde{Q} (z) = 0$ is recovered.  This is in agreement with \cite{Yang:2015tzc} where the authors performed the reconstruction of $\widetilde{Q} (z)$ considering only Union 2.1 compilation of SNIa~\cite{SupernovaCosmologyProject:2011ycw} and did not get any evidence of interaction.  Although in our case (see the middle panel of Fig. \ref{fig:squared-GP}),  the mean curve has a sign changeable nature, but it is not strong at all since within 68\% CL, we get back the non-interacting scenario.

Finally, we perform the reconstructions of $\widetilde{Q} (z)$, using the combined dataset CC+Pantheon+. 
In  the lower panel of Fig. \ref{fig:squared-GP} we have shown the reconstructed graphs of $\widetilde{Q} (z)$ for both the values of $H_0$, namely,  $H_0 = 67.36 \pm 0.54$ $\rm km\ s^{-1}\ Mpc^{-1}$c at 68\% CL \cite{Planck:2018vyg} (left graph in the lower panel of Fig. \ref{fig:squared-GP}) and $H_0 = 73.04 \pm 1.04$ $\rm km\ s^{-1}\ Mpc^{-1}$ at 68\% CL \cite{Riess:2021jrx} (right graph in the lower panel  of Fig. \ref{fig:squared-GP}), respectively. First of all, we notice that the choice of $H_0$ does not influence the reconstructions, but, interestingly, we find an evidence of interaction at present moment (at more than 68\% CL) and in the intermediate phase at $\sim $ 95\% CL. Similar observation was reported recently by~\cite{Escamilla:2023shf, Mukherjee:2021ggf} in which the authors performed the reconstruction of the interaction function with the use of Pantheon+ sample of SNIa plus other astronomical datasets (baryon acoustic oscillations, CC)\footnote{Note that the datasets used in \cite{Escamilla:2023shf, Mukherjee:2021ggf} are not exactly identical with our datasets. } and found a mild evidence of interaction ($\sim 1\sigma$). However, interestingly,  an evidence of interaction was reported in \cite{Bonilla:2021dql} taking into account of Pantheon sample of SNIa together with baryon acoustic oscillations, CC and  
H0LiCOW ($H_0$ Lenses in COSMOGRAIL's Wellspring) sample \cite{H0LiCOW:2019pvv}.   
Additionally, as depicted in the lower panel of Fig. \ref{fig:squared-GP},
a sign shifting nature of $\widetilde{Q} (z)$ during the evolution of the universe is found. To be more specific, in the high redshift region of our consideration, we first observe a transition of $\widetilde{Q} (z)$ from its earlier positive values (i.e. energy transfer from DM to DE) to negative values (energy transfer from DE to DM). Then $\widetilde{Q} (z)$ again changes its sign and enters into a region with $\widetilde{Q} (z) > 0$ (i.e. energy transfer occurs from DM to DE). At current epoch (i.e. $z =0$), we notice $\widetilde{Q} (z=0) >0$ at more than 68\% CL but less than 95\% CL.

Overall, we notice a mild evidence of interaction  together with its sign changeable. In this context, we would like to remark that he possibility of sign changeable interaction has been proposed by several authors  but assuming different parametric forms of the interaction function~   \cite{Wei:2010fz,Wei:2010cs,Sun:2010vz,Li:2011ga,Guo:2017deu,Arevalo:2019axj,Pan:2019jqh,Arevalo:2022sne}.

We now focus on the behavior of the effective EoS parameters, namely, $w_{\rm DM}^{\rm eff}$ and $w_{\rm DE}^{\rm eff}$ (note that $w_{\rm DE}^{\rm eff}$ does not include $w_{\rm DE}$ explicitly  (see Eqs. (\ref{eff-eos-v1-2}), (\ref{eff-eos-v2-2})) unlike $w_{\rm DM}^{\rm eff}$ in Eqs. (\ref{eff-eos-v1}) or (\ref{eff-eos-v2})). Taking CC, Pantheon+ and CC+Pantheon+,
we have reconstructed these parameters in order to understand how they are influenced when an interaction is considered between DM and DE (see Fig. \ref{fig:wDMeff-wDEeff-GP}). For the reconstructions using CC, we have used the estimated value of $H_0$ which we obtained from the reconstructed graph of $H(z)$ by the square exponential kernel. For the reconstructions using Pantheon+ alone and CC+Pantheon+,  
we have considered same values of $H_0$ which are mentioned earlier, that means,  $H_0 = 67.36 \pm 0.54$ $\rm km\ s^{-1}\ Mpc^{-1}$c at 68\% CL \cite{Planck:2018vyg} and $H_0 = 73.04 \pm 1.04$ $\rm km\ s^{-1}\ Mpc^{-1}$ at 68\% CL \cite{Riess:2021jrx}. Now, according to Fig. \ref{fig:wDMeff-wDEeff-GP}, the reconstructions of the effective EoS parameters of DM and DE for Pantheon+ and CC+Pantheon+ remain unaffected due to the choice of the $H_0$ values.  
Now, concentrating on Fig. \ref{fig:wDMeff-wDEeff-GP}, we notice that for both CC and Pantheon+, $w_{\rm DM}^{\rm eff} = 0$ and $w_{\rm DE}^{\rm eff} = -1$ are found within 68\% CL, and hence,  the results are consistent with the non-interacting $\Lambda$-Cold DM ($\Lambda$CDM) cosmology. 
This is expected because for both CC and Pantheon+, we do not find any evidence of interaction. However, on the other hand, the results of the effective EoS parameters of DM and DE behave in a similar fashion as $\widetilde{Q} (z)$. For CC+Pantheon+, 
a mild evidence of $w_{\rm DM}^{\rm eff} \neq  0$ and $w_{\rm DE}^{\rm eff} \neq  -1$ is observed at slightly more than 68\% CL (at present epoch) and at $\sim$ 95\% CL (in the intermediate redshift regime).  

\subsubsection{For $w_{\rm DE} = \mbox{constant}~ 
 (\neq -1)$}
\label{subsubsec-GP}

The reconstruction of $\widetilde{Q} (z)$, as one can see from Eq. (\ref{eqn:Q}), depends on the EoS of DE. In the earlier section, we reconstructed $\widetilde{Q} (z)$ for $w_{\rm DE} = -1$. While this represents the most simplest interacting scenario, widely known as interacting vacuum scenario (see for instance ~\cite{Martinelli:2019dau,Yang:2021oxc}), however, as the nature of DE is still an yet to be discovered area in modern cosmology, hence, there is no specific reason to stick to $w_{\rm DE} =-1$. In fact,  baryon acoustic oscillations from the Dark Energy Spectroscopic Instrument (DESI) survey~\cite{DESI:2024mwx,DESI:2025zgx,DESI:2025fii,Gu:2025xie} recently indicated that the EoS of DE could be dynamical. That means, reconstruction of the interaction in the dark sector could be performed under the assumption of a dynamical $w_{\rm DE}$. However, the choice of time dependent $w_{\rm DE}$ is not unique. Although the natural parametrization for the dynamical $w_{\rm DE}$ follows the widely used Chevallier-Polarski-Linder (CPL)~\cite{Chevallier:2000qy, Linder:2002et} parametrization given by, $w_{\rm DE} =  w_0 + w_a (1- a/a_0)$ ($w_0$ is the present value of $w_{\rm DE}$ and $w_a = - dw_{\rm DE}/da|_{a =a_0}$ represents the dynamical nature of the DE EoS), however, during the reconstruction, one needs to manually put the values of ($w_0$, $w_a$), and as a result of which, the reconstruction of $\widetilde{Q} (z)$ might be influenced by the parameters.

Thus, considering this fact, in this section we mainly focus on the deviation of $w_{\rm DE}$ from $-1$ without introducing any free parameter, but examine the effects of $w_{\rm DE} \neq -1$ on the reconstruction of the interaction. 
Following this we have reconstructed $\widetilde{Q} (z)$ for a variety of choices of $w_{\rm DE}$ deviating from $-1$, e.g. $w_{\rm DE} = -0.7, -0.8, -0.9, -1.1, -1.2, -1.3$, using CC+Pantheon+\footnote{One can explore the same using CC alone and Pantheon+ alone, however, as CC+Pantheon+ offers the most stringent reconstructions, therefore, we have exercised with this combined analysis only. } and considering the two values of $H_0$. The choices of $H_0$  do not affect the reconstructions. 
In Fig.~\ref{fig:Q-evolution-diff-w} we present the deviation of $\widetilde{Q} (z)$ from zero, characterized by $\Delta\widetilde{Q} = (\widetilde{Q} (z)-0)/\sigma_{\widetilde{Q} (z)}$  for six different values of $w_{\rm DE}$, namely $w_{\rm DE} = -0.7, -0.8, -0.9, -1.1, -1.2, -1.3$ and taking $H_0 = 73.04 \pm 1.04$ $\rm km\ s^{-1}\ Mpc^{-1}$ at 68\% CL~\cite{Riess:2021jrx}\footnote{In Appendix-A we show Fig. \ref{fig:GP-Q_diff_wde-Appendix-1} where we present the reconstruction of $\widetilde{Q} (z)$ for different values of $w_{\rm DE}$ and considering $H_0 = 73.04 \pm 1.04$ $\rm km\ s^{-1}\ Mpc^{-1}$ at 68\% CL~\cite{Riess:2021jrx}. }.
Here, one set of values represents the quintessence nature of $w_{\rm DE}$ (i.e. $w_{\rm DE} > -1$) while the other set of values denotes the phantom nature of $w_{\rm DE}$ (i.e. $w_{\rm DE} < -1$).

From Fig. \ref{fig:Q-evolution-diff-w} it is clearly seen that when $w_{\rm DE}$ starts deviating from $-1$ in the quintessence regime, evidence of interaction gets pronounced and in particularly we notice $\widetilde{Q} (z =0) > 0$ which is an indication of the energy flow from CDM to DE. On the other hand, when $w_{\rm DE}$ assumes its phantom values (i.e. $w_{\rm DE} < -1$), we see that phantom DE may allow both $\widetilde{Q} (z) > 0$ (for $w_{\rm DE} = -1.1$) and 
$\widetilde{Q} (z) < 0$ (for $w_{\rm DE} = -1.2 $ and $-1.3$). This clearly indicates that that the flow of energy-momentum may occur in either direction depending on the strength of the phantom nature of $w_{\rm DE}$. 
For high phantom nature of $w_{\rm DE}$, energy flow occurs from DE to CDM. At this point we recall that considering Union 2.1 sample of SNIa, the authors of \cite{Yang:2015tzc} also found that with the deviation of $w_{\rm DE}$ from $-1$, one can observe an emergence of interaction in the dark sector.

Additionally, in Fig. \ref{fig:GP+ANN-Q-at-z=0} (see the upper graph of this figure; the lower graph corresponds to ANN which we shall discuss in the next section) we show a whisker graph showing the 68\% CL constraints on $\widetilde{Q} (z=0)$ for different values of $w_{\rm DE}$ as used in Fig. \ref{fig:Q-evolution-diff-w}. This figure offers a qualitative nature of the interaction at present moment for the quintessence and phantom DE. One can see that for $w_{\rm DE}$ deviating from $-1$ in the quintessence direction, evidence of interaction is pronounced much than the deviation of $w_{\rm DE}$ from $-1$ in the phantom direction (see again \cite{Yang:2015tzc} reporting similar indications).  

\begin{figure*}
    \centering
    \includegraphics[width=0.35\textwidth]{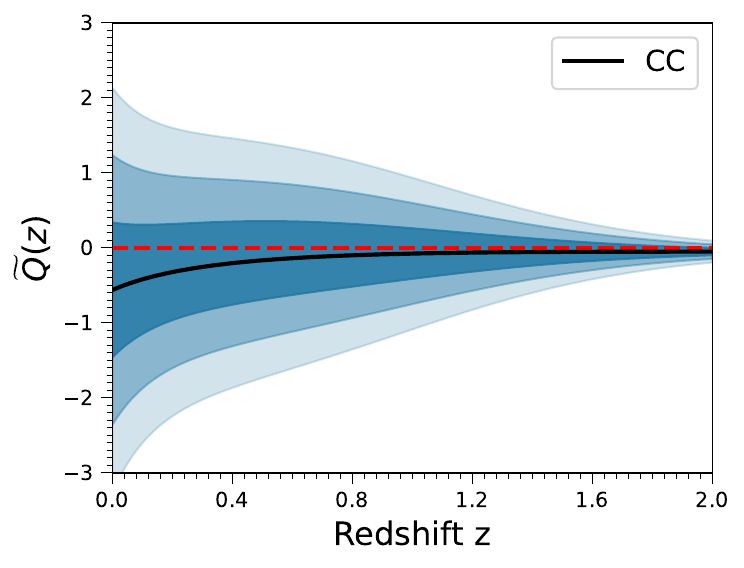} \\
    \includegraphics[width=0.35\textwidth]{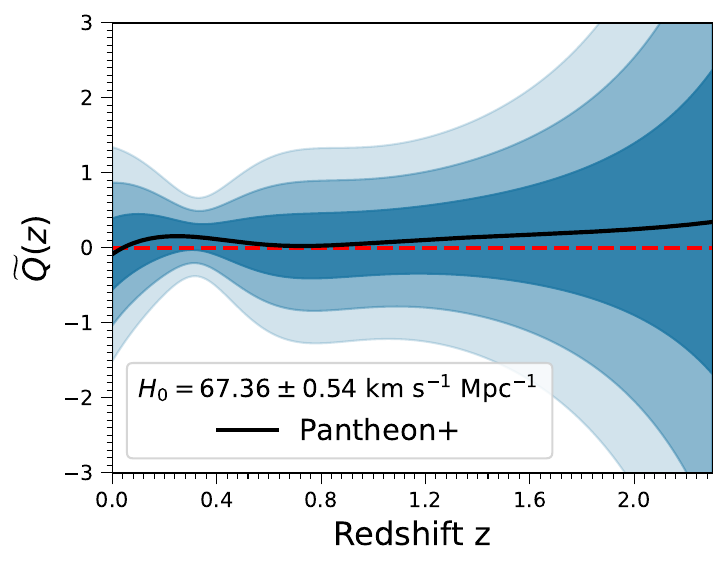}
    \includegraphics[width=0.35\textwidth]{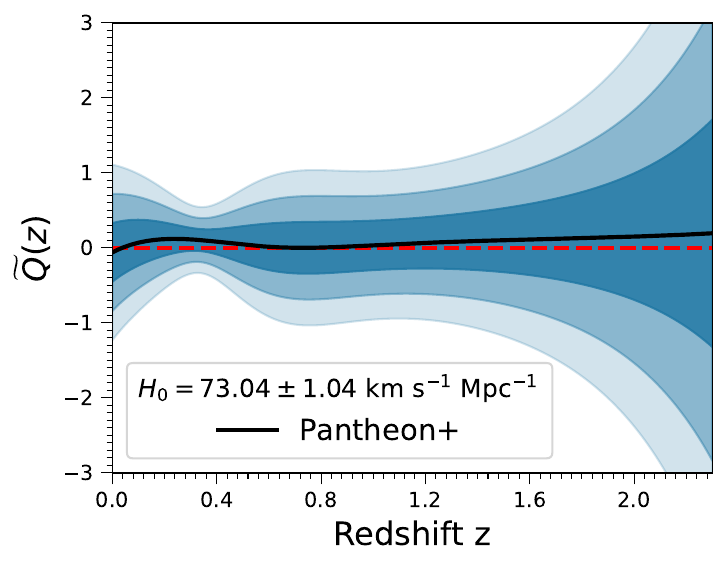}
    \includegraphics[width=0.35\textwidth]{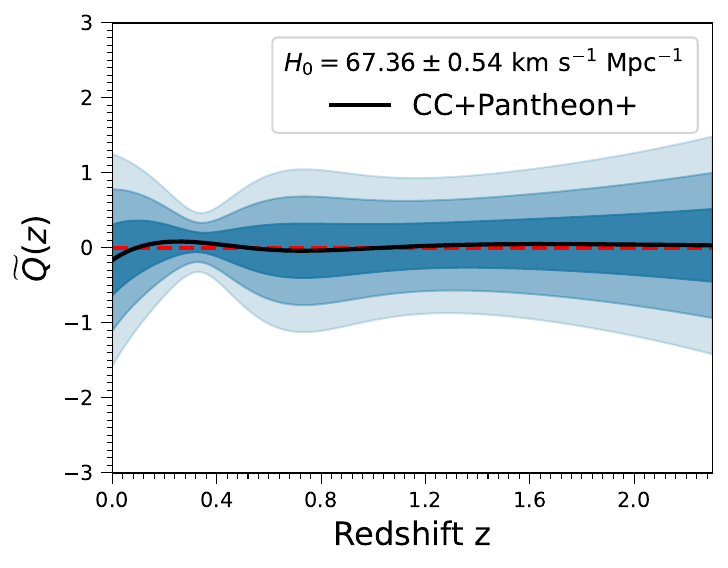}
    \includegraphics[width=0.35\textwidth]{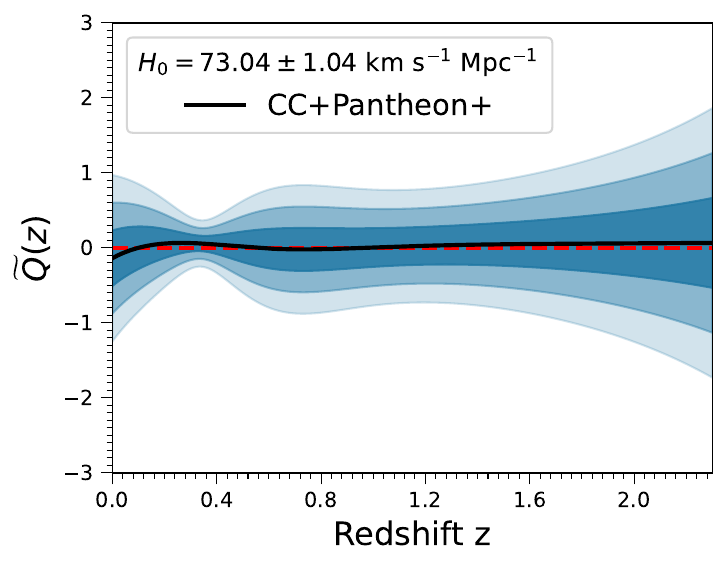}
    \caption{Reconstructed interaction function $\widetilde{Q} (z)$ using the ANN approach considering 34 CC $H(z)$, Pantheon+, and CC+Pantheon+ data  considering $w_{\rm DE} =-1$. The Hubble constant $H_{0}=67.36 \pm 0.54$ $\rm km\ s^{-1}\ Mpc^{-1}$ at 68\% CL~ \protect\cite{Planck:2018vyg} 
    and $H_{0}=73.04 \pm 1.04$ $\rm km\ s^{-1}\ Mpc^{-1}$ at 68\% CL 
   \protect\cite{Riess:2021jrx}
    are used in the reconstruction of $D(z)$. The horizontal dashed line (red) in each plot corresponds to $\widetilde{Q} (z) =0$, i.e. no-interaction between the dark sectors and the solid curve (black) stands for the mean curve for each case.    }
    \label{fig:ANN-CC-PP-CC+PP}
\end{figure*}
\begin{figure*}
    \centering
    \includegraphics[width=0.4\textwidth]{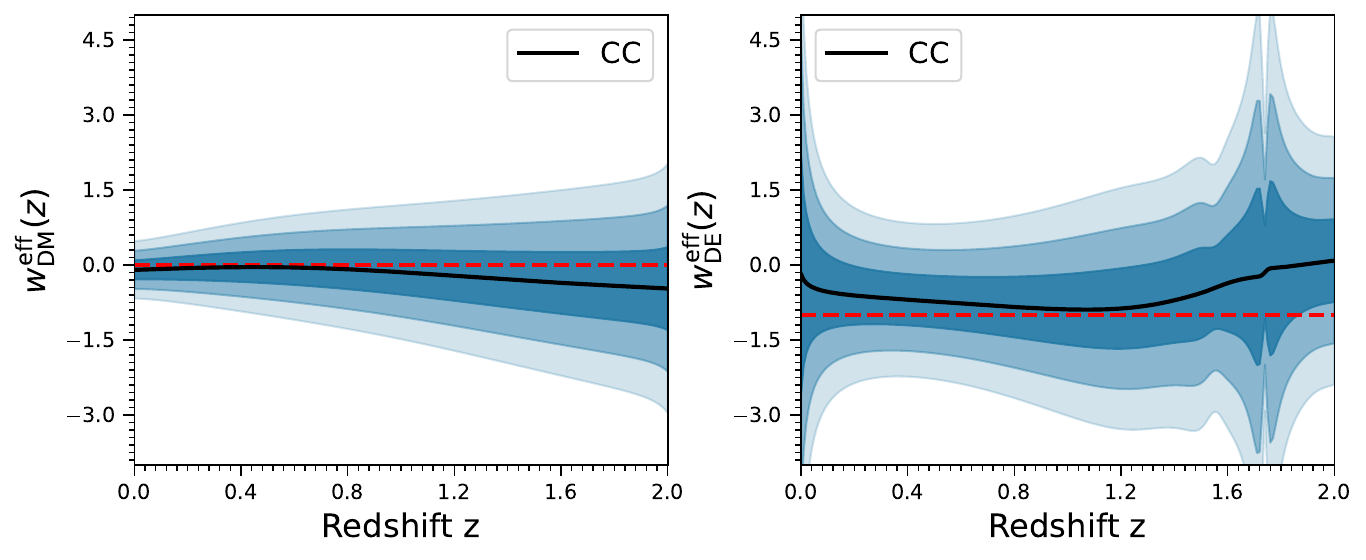}\\
    \includegraphics[width=0.4\textwidth]{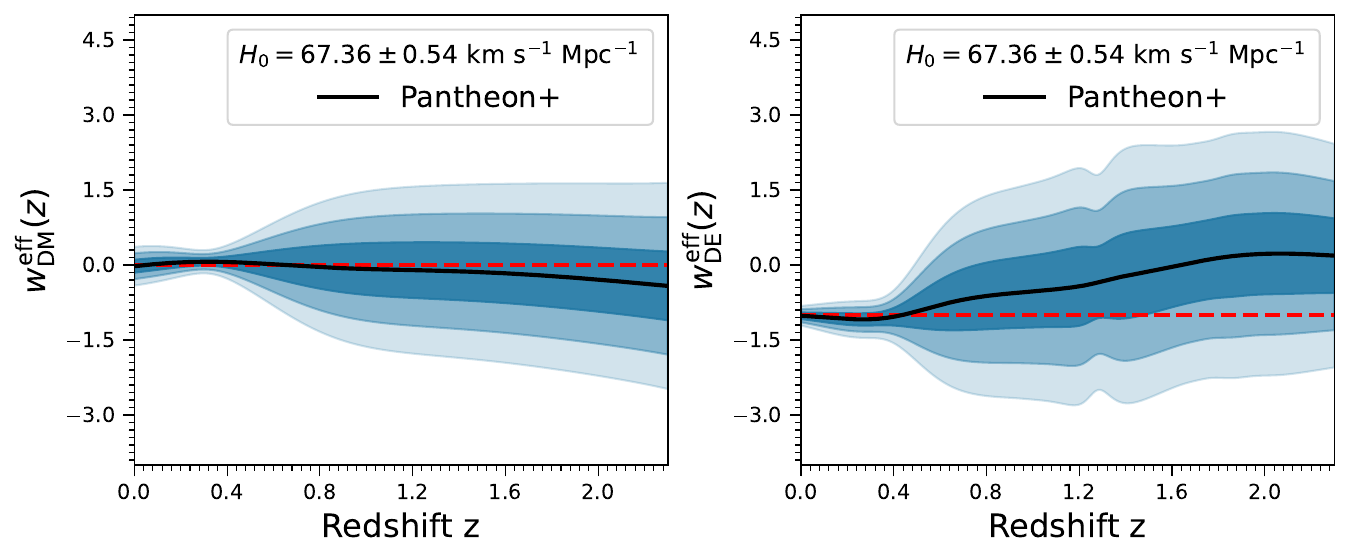}
\includegraphics[width=0.4\textwidth]{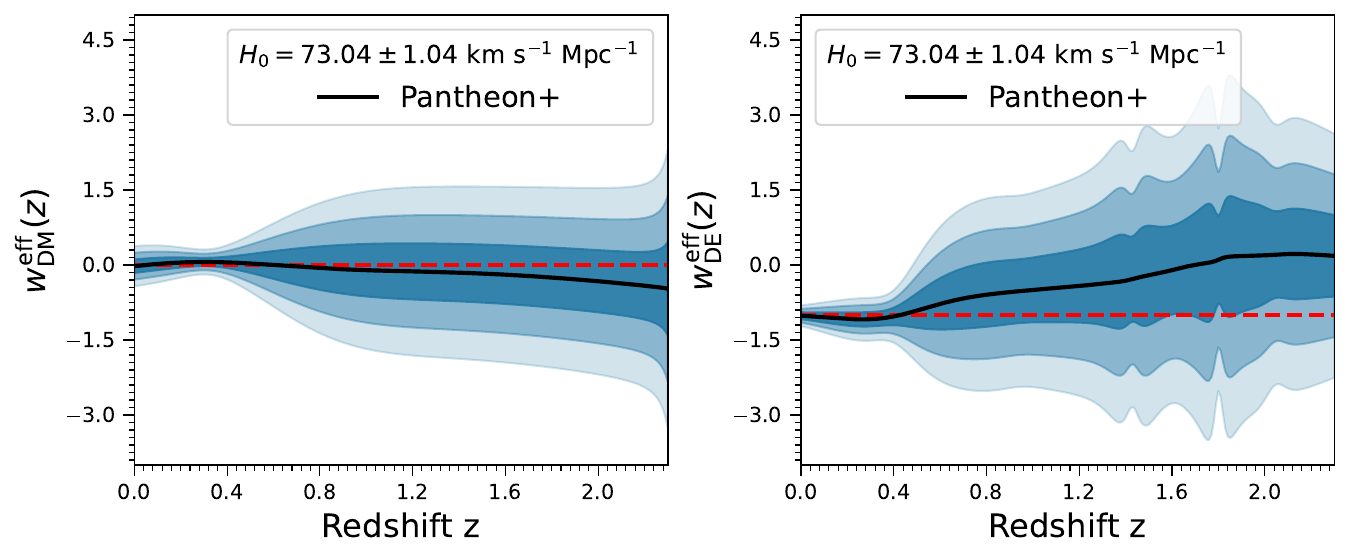}\\
\includegraphics[width=0.4\textwidth]{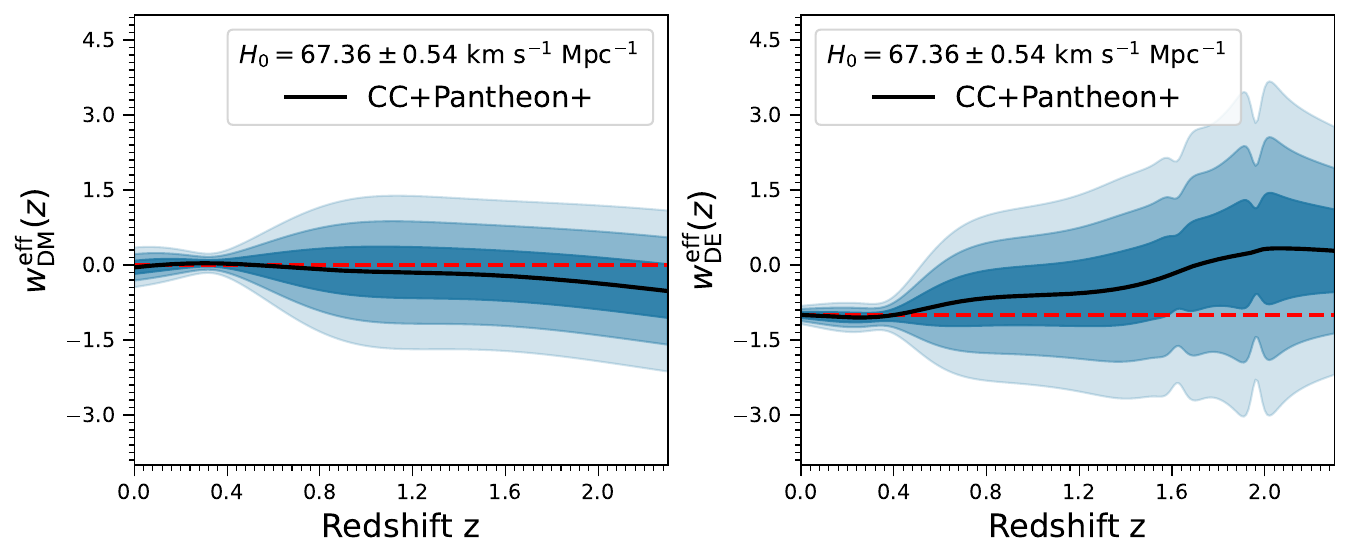}
\includegraphics[width=0.4\textwidth]{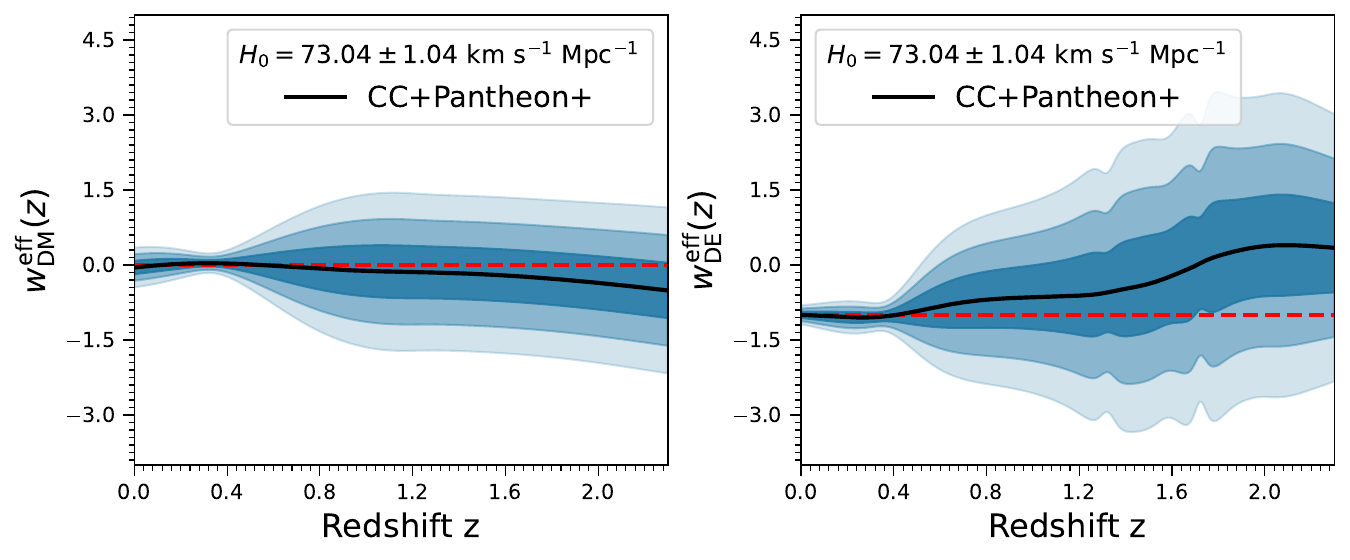}
\caption{Reconstructed effective EoS parameters, $w_{\rm DM}^{\rm eff}$ and $w_{\rm DE}^{\rm eff}$, using ANN considering 34 CC $H(z)$, Pantheon+, and CC+Pantheon+ data. The Hubble constant $H_{0}=67.36 \pm 0.54$~$\rm km\ s^{-1}\ Mpc^{-1}$ at 68\% CL \protect\cite{Planck:2018vyg} 
and $H_{0}=73.04 \pm 1.04$~$\rm km\ s^{-1}\ Mpc^{-1}$ at 68\% CL 
\protect\cite{Riess:2021jrx}  
are used in the reconstruction of $D(z)$. The horizontal dashed line (red) in the plots for $w_{\rm DM}^{\rm eff}$ ($w_{\rm DE}^{\rm eff}$) corresponds to $w_{\rm DM}^{\rm eff} =0$ ($w_{\rm DE}^{\rm eff} =-1$) and the solid curve (black) stands for the mean curve of the respective reconstructed effective EoS parameter.  }
    \label{fig:wDMeff-wDEeff-ANN}
\end{figure*}
\begin{figure}
    \centering
    \includegraphics[width=0.45\textwidth]{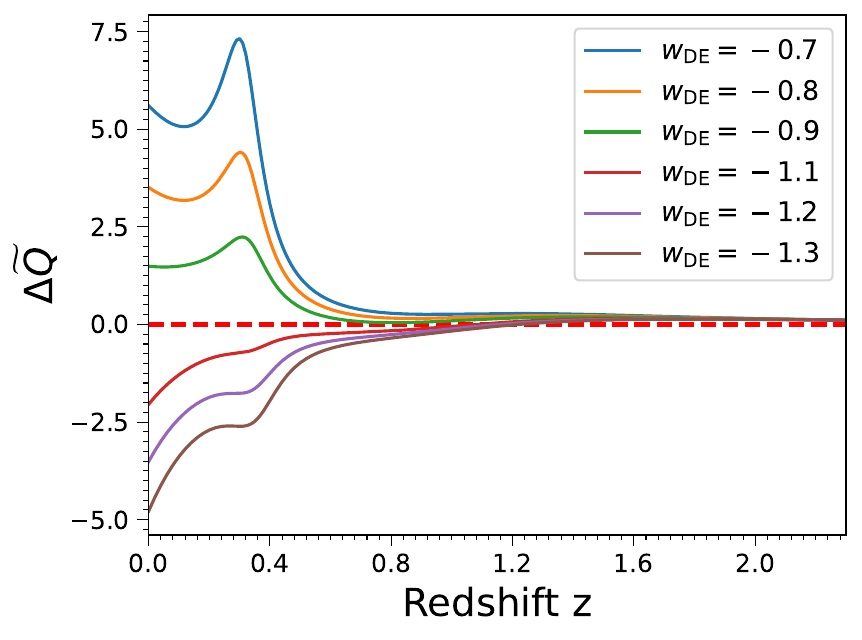}
    \caption{The deviation of $\widetilde{Q}$ from zero, where $\Delta\widetilde{Q} = (\widetilde{Q} (z)-0)/\sigma_{\widetilde{Q}(z)}$ is shown for ANN. The figure corresponds to the  reconstructions using CC+Pantheon+ and for $H_{0}=73.04\pm1.04$ $\rm km\ s^{-1}\ Mpc^{-1}$ \protect\cite{Riess:2021jrx}.    
   }
    \label{fig:ANN-CC+PP+vary-w-dev}
\end{figure}
 \begin{figure}
    \centering
\includegraphics[width=0.45\textwidth]{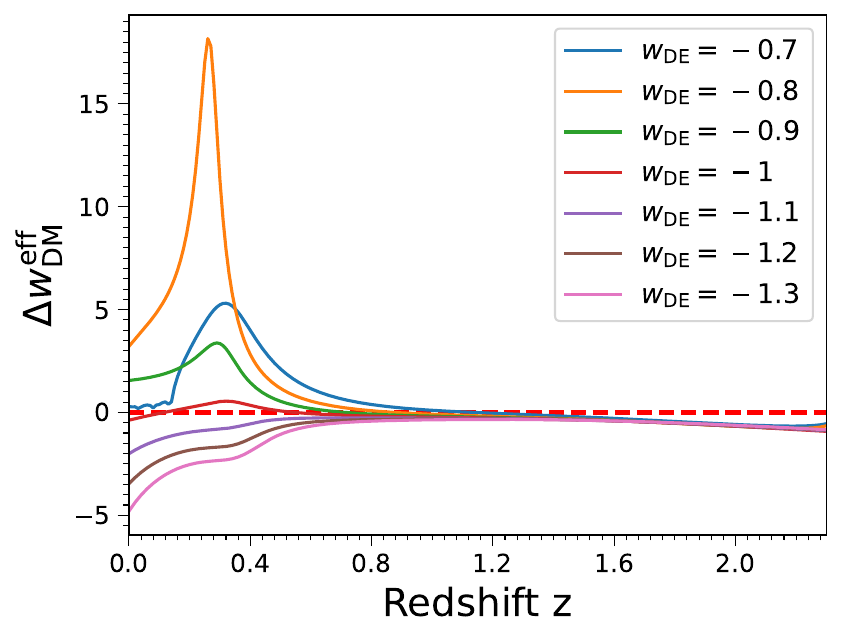}
    \caption{The deviation of  $w_{\rm DM}^{\rm eff}$ from $0$ 
 where  $\Delta w_{\rm DM}^{\rm eff} = (w_{\rm DM}^{\rm eff} (z) - 0)/\sigma_{w_{\rm DM}^{\rm eff}(z)}$ is shown for CC+Pantheon+ with $H_0 = 73.04 \pm 1.04$~$\rm km\ s^{-1}\ Mpc^{-1}$~\protect\cite{Riess:2021jrx} considering different values of $w_{\rm DE}$  and using ANN.   }
    \label{fig:weff_ANN}
\end{figure}
We now focus on the behavior of $w_{\rm DM}^{\rm eff}$ for different values of $w_{\rm DE}$ considering only the CC+Pantheon+ dataset with $H_0 = 73.04 \pm 1.04$ $\rm km\ s^{-1}\ Mpc^{-1}$ at 68\% CL~\cite{Riess:2021jrx}. As already argued, the choice of only one $H_0$ is motivated from the fact that  for different values of $H_0$, the reconstructions do not change. Taking different fixed values of $w_{\rm DE}$, in Fig. \ref{fig:weffDM-weffDE-diffwDE-GP} we present the  deviation of $w_{\rm DM}^{\rm eff}$ from  $0$, defined by 
\begin{eqnarray}
\Delta w_{\rm DM}^{\rm eff} = (w_{\rm DM}^{\rm eff} (z) - 0)/\sigma_{w_{\rm DM}^{\rm eff}(z)}. 
\end{eqnarray}
In Fig. \ref{fig:weffDM-weffDE-diffwDE-GP} we present the evolution of $\Delta w_{\rm DM}^{\rm eff}$ for different values of $w_{\rm DE}$.\footnote{In Appendix-A we show Fig. \ref{fig:GP-eff-wdm_wde-Appendix-2} where we explicitly present the reconstructions of the effective EoS parameter for DM, i.e. $w_{\rm DM}^{\rm eff}$ for different values of $w_{\rm DE}$ and considering CC+Pantheon+ with $H_0 = 73.04 \pm 1.04$ $\rm km\ s^{-1}\ Mpc^{-1}$ at 68\% CL~\cite{Riess:2021jrx}.}
This shows that when $w_{\rm DE}$ deviates from $-1$, effective EoS parameter for DM, $w_{\rm DM}^{\rm eff}$, also deviates from $0$. This is expected since an evidence of interaction has been observed when $w_{\rm DE}$ deviates from $-1$ (see Fig. \ref{fig:Q-evolution-diff-w}).  
However, in order to understand the overall picture clearly, one needs to look at the reconstruction of  $w_{\rm DM}^{\rm eff}$ for different values of $w_{\rm DE}$ shown in Fig. \ref{fig:GP-eff-wdm_wde-Appendix-2}. 
According to the reconstructions of $w_{\rm DM}^{\rm eff}$  (Fig. \ref{fig:GP-eff-wdm_wde-Appendix-2}), we notice that, except in a small part of the high redshift regime (around $z =1.6$) where the evidence of $w_{\rm DM}^{\rm eff} \neq 0$ is noticed for most of the values of $w_{\rm DE}$, in the remaining redshift regime, no such strong evidence of 
$w_{\rm DM}^{\rm eff} \neq 0$ is found; in particularly, in the intermediate redshift regime, an evidence of non-null  $w_{\rm DM}^{\rm eff}$ is found but that persists at $\sim $ 95\% CL ($<$ 95\% CL) for $w_{\rm DE} < -1$ ($w_{\rm DE} > -1$).  The sign changeable behavior  in $w_{\rm DM}^{\rm eff}$ (sort-of oscillatory nature) appears due to the sign shifting behavior of $\widetilde{Q} (z)$ but as we already noticed, this sign-changeable nature is statistically not so strong for all cases (see Fig.~\ref{fig:GP-Q_diff_wde-Appendix-1}) where again we notice that only in the intermediate redshift, evidence of interaction is found at $\sim $ 95\% CL ($< $ 95\% CL) for $w_{\rm DE}< -1$ ($w_{\rm DE}> -1$). Thus, the oscillating nature of $w_{\rm DM}^{\rm eff}$ as exhibited in the right graph of Fig. \ref{fig:weffDM-weffDE-diffwDE-GP} is statistically not so robust according to the present observational datasets.

\subsection{Reconstructions using ANN}

\subsubsection{For $w_{\rm DE}= -1$}
\label{subsec-results-ANN}

In this section we describe the reconstructions of $\widetilde{Q} (z)$  using the ANN approach. We have considered the same datasets, namely, CC alone, Pantheon+ alone and their combined dataset CC+Pantheon+.

We start with the reconstructions of $\widetilde{Q} (z)$ from CC alone shown in the topmost plot of Fig. \ref{fig:ANN-CC-PP-CC+PP}. In this case, we have used $1024$ nodes in the hidden layer and considered Adam as the optimizer~\cite{2014arXiv1412.6980K} and L2 weight decay on the parameters to prevent overfitting which is set to $0.005$. The ANNs are trained with $30,000$ iterations. In this case we do not find any notable evidence of interaction since within 68\% CL,  $\widetilde{Q} (z) =0$ is well recovered.

 When we consider Pantheon+ alone for reconstructing $\widetilde{Q} (z)$, in a similar fashion as in GP, we choose two different values of $H_0$, namely, $H_0 = 67.36 \pm 0.54$ $\rm km\ s^{-1}\ Mpc^{-1}$ at 68\% CL~\cite{Planck:2018vyg} and $H_0 = 73.04 \pm 1.04$ $\rm km\ s^{-1}\ Mpc^{-1}$ at 68\% CL~\cite{Riess:2021jrx}. Let us note that for the cases with Pantheon+ and CC+Pantheon+ (described later in this section) cases, $1024$ nodes are used in the hidden layer and the L2 weight decay follows the Adam optimizer which is set to $0.0005$. The ANNs are trained with 30,000 iterations. 
In the lower panel of Fig. \ref{fig:ANN-CC-PP-CC+PP} we show the reconstructed graphs of $\widetilde{Q} (z)$   
for Pantheon+ only. Our first impression is that the reconstructions do not change for the choice of the $H_0$ priors. In this case we do not find any significant evidence of interaction as $\widetilde{Q} (z) =0$ is well recovered within 68\% CL.

In a similar fashion, we reconstructed the interaction function $\widetilde{Q} (z)$ for CC+Pantheon+ considering again the same $H_0$ values as in the Pantheon+ case. In the lower panel of Fig. \ref{fig:ANN-CC-PP-CC+PP} we show the reconstructed graphs of $\widetilde{Q} (z)$ for both the choices of $H_0$.  Similar to the earlier cases, here too, we do not find any significant evidence of interaction. 

After that we reconstruct $w_{\rm DM}^{\rm eff}$ and $w_{\rm DE}^{\rm eff}$ considering CC, Pantheon+ and CC+Pantheon+ and present them in Fig. \ref{fig:wDMeff-wDEeff-ANN}.\footnote{We again note that $w_{\rm DE}^{\rm eff}$ does not explicitly include $w_{\rm DE}$ and hence the reconstruction of $w_{\rm DE}^{\rm eff}$ only needs the dimensionless variables $E$, $D$ and their derivatives. } During the reconstructions using CC, we have used the estimated value of $H_0$ obtained from the reconstructed graph of $H(z)$ using the SoftPlus activation function. For Pantheon+ and CC+Pantheon+ reconstructions,  
we have considered both the values of $H_0$ used in this article. We noticed that the reconstructions of $w_{\rm DM}^{\rm eff}$ and $w_{\rm DE}^{\rm eff}$ remain same irrespective of the $H_0$ values. 
In this case, for all the datasets, we do not find any evidence of $w_{\rm DM}^{\rm eff} \neq 0$, however, concerning the effective EoS of DE,  a mild deviation from $w_{\rm DE}^{\rm eff} = -1$ is noticed at slightly more than 
68\% CL only in the high redshift regime for Pantheon+ and CC+Pantheon+ datasets, but at present moment, $w_{\rm DE}^{\rm eff} = -1$ is allowed within 68\% CL. That means no deviation from the non-interacting $\Lambda$CDM cosmology is found. This is not surprising because we do not notice any evidence of interaction in this context.

\subsubsection{For $w_{\rm DE}= \mbox{constant}~ 
 (\neq -1)$ }
\label{subsubsec-ANN}

 We repeat the procedure as described in section~\ref{subsubsec-GP} but using the ANN approach. Therefore, considering the same values of $w_{\rm DE}$ as explored in the earlier section~\ref{subsubsec-GP}, we perform the reconstructions of $\widetilde{Q} (z)$ and 
 in Fig. \ref{fig:ANN-CC+PP+vary-w-dev} we show the deviation of $\widetilde{Q} (z)$ from zero considering the combined dataset CC+Pantheon+ and using some typical values of $w_{\rm DE} \neq -1$ and taking  $H_0 = 73.04 \pm 1.04$ $\rm km\ s^{-1}\ Mpc^{-1}$.\footnote{We again note that the choice of $H_0$ does not affect the reconstructions, hence, we considered only one value of $H_0$ for the reconstructions and also we choose the combined dataset CC+Pantheon+ to present the results in order to gain the maximal effect on the reconstructions. In Fig. \ref{fig:ANN-Q_diff_wde-Appendix-1} (see Appendix-A) we present the reconstruction of $\widetilde{Q} (z)$ for the individual values of $w_{\rm DE}$. } 

From Fig. \ref{fig:ANN-CC+PP+vary-w-dev}, we find that  when $w_{\rm DE}$ deviates from $-1$ in either the quintessence or phantom direction, an evidence of interaction is certainly pronounced, similar to what we have noticed in GP reconstructions (see Fig. \ref{fig:Q-evolution-diff-w}), however, here we observe one difference between GP and ANN reconstructions. 
In this case, we notice that for $w_{\rm DE}> -1$ $\widetilde{Q} (z)$ remains positive (i.e. energy flow occurs from CDM to DE)  and for $w_{\rm DE} < -1$, $\widetilde{Q} (z)$ changes its sign from positive (in the high redshift regime) to negative region (in low redshift regime) and thus the direction of energy flow alters accordingly.  While in GP reconstructions (see Fig. \ref{fig:Q-evolution-diff-w}), this is not exactly identical. In both the cases (i.e. quintessence or phantom), evidence of interaction is pronounced positively.   
In the lower portion of Fig. \ref{fig:GP+ANN-Q-at-z=0} we show the whisker graph of $\widetilde{Q} (z =0)$  presenting its 68\% CL constraints. This whisker graph clearly indicates how the interaction at present time is pronounced when  $w_{\rm DE}$ deviates from $-1$.  

Lastly,  in Fig. \ref{fig:weff_ANN} we show the evolution of the effective EoS parameter for DM, in terms of  $\Delta w_{\rm DM}^{\rm eff}$ taking into account of different values of $w_{\rm DE}$ and considering CC+Pantheon+ with $H_0 = 73.04 \pm 1.04$ $\rm km\ s^{-1}\ Mpc^{-1}$~\cite{Riess:2021jrx}.\footnote{We refer to Fig. \ref{fig:ANN-eff-wdm_wde-Appendix-2} in Appendix-A where we explicitly present the evolution of $w_{\rm DM}^{\rm eff}$ for different values of $w_{\rm DE}$.}  In Fig. \ref{fig:weff_ANN}, we show $\Delta w_{\rm DM}^{\rm eff}$ for different values of $w_{\rm DE}$.  We notice that depending on the deviation of $w_{\rm DE}$ from $-1$, $w_{\rm DM}^{\rm eff}$ deviates from $0$. This is expected because when $w_{\rm DE}$ deviates from $-1$, an emergence of interaction is observed (see Fig. \ref{fig:ANN-CC+PP+vary-w-dev}).  

These altogether emphasize that ANN seems to have much constraining nature than the GP reconstructions.

\section{Summary and Conclusions}
\label{sec-summary}

Interaction between the main two dark components of the universe, namely, DM and DE, is the main focus of this work. According to the past records, the cosmological models allowing an interaction between DM and DE have been extensively studied in the literature because of their many appealing properties. In almost all interacting models, a choice of the interaction function is made which is motivated from the phenomenological ground.  However, in the present article we have searched for an interaction between a pressure-less DM and a DE fluid with constant EoS, $w_{\rm DE}$, without assuming any interaction model a priori. 

We make use of two  non-parametric  data driven approaches, namely, GP and ANN which are very well known for the purpose of reconstructing the cosmological variables. We have used mainly the geometrical datasets, namely \textbf{i)} CC alone, \textbf{ii)} Pantheon+ alone and \textbf{iii)} CC+Pantheon+. For the purpose of the reconstructions, 
GP needs a kernel in the background and ANN needs an activation function. Before summarizing the main results, we clarify that out of a variety of kernels (for GP)  and activation functions (for ANN),  we have considered the squared exponential kernel for GP and Softplus activation function for ANN. The reason for such selection is driven by the fact that the reconstructions in GP for the remaining kernels are almost similar (in the sense that the changes in the reconstructions for the other kernels are either indistinguishable or they are very minimal) and the same for ANN as well for the other activation function (LogSigmoid).

On the other hand, another delicate issue, related to the choice of the $H_0$ prior has also been investigated in this work.  
Since the $H_0$ priors have been found to affect the reconstructions \cite{Wei:2016xti,Wang:2017lri,Wang:2020dbt}, therefore,  while reconstructing the interaction function using 
Pantheon+ and CC+Pantheon+, we have taken two different values of $H_0$, namely,   $H_0 = 67.36 \pm 0.54$ $\rm km\ s^{-1}\ Mpc^{-1}$ at 68\% CL from Planck 2018~\cite{Planck:2018vyg} and $H_0 = 73.04 \pm 1.04$ $\rm km\ s^{-1}\ Mpc^{-1}$ at 68\% CL from SH0ES~\cite{Riess:2021jrx}.  However, for the reconstructions using CC alone, we have used the $H_0$ values obtained from the  GP or from ANN.

Now, having these in mind, we reconstructed the interaction function $\widetilde{Q} (z)$.  As the interaction function (see Eq. (\ref{eqn:Q}) or (\ref{geneqn:Q-D_form}))  directly involves $w_{\rm DE}$, therefore, we have given a special attention to the effects of $w_{\rm DE}$ on the reconstruction of the interaction function. For $w_{\rm DE} = -1$, according to our reconstructions from both GP and ANN, we see that ANN does not indicate any interaction for any of the datasets  but from GP reconstructions, evidence of interaction at present epoch (at $<$ 95\% CL) and also in the intermediate redshifts (at $\sim$ 95\% CL) are found, see the plots for CC+Pantheon+ in Figs. \ref{fig:squared-GP}.  
However, an interesting pattern appears in the reconstruction of $\widetilde{Q} (z)$ when the value of $w_{\rm DE}$ deviates from $-1$, see Fig. \ref{fig:Q-evolution-diff-w} (for GP) and Fig. \ref{fig:ANN-CC+PP+vary-w-dev} (for ANN) showing the reconstructions of $\widetilde{Q} (z)$ taking 
several values of $w_{\rm DE}$ other than $-1$.  We have examined this issue considering the quintessence DE ($w_{\rm DE} > -1$) and phantom DE ($w_{\rm DE} < -1$) separately and we found that when $w_{\rm DE}$ deviates from $-1$ in any direction, an evidence of interaction is pronounced. This evidence is much pronounced for ANN reconstructions. We have also examined the interaction at present moment (i.e.  $\widetilde{Q} (z = 0)$) for all possible values of $w_{\rm DE}$ other than $-1$ (see Fig. \ref{fig:GP+ANN-Q-at-z=0}).

We have also examined the effects of interaction on the effective EoS parameters of the dark fluids, namely $w_{\rm DM}^{\rm eff}$ and $w_{\rm DE}^{\rm eff}$. The effective EoS parameter for DM involves $w_{\rm DE}$, however, the effective EoS for DE does not explicitly involve $w_{\rm DE}$ (see Eq. (\ref{eff-eos-v1-2}) or (\ref{eff-eos-v2-2})), therefore, while reconstructing $w_{\rm DM}^{\rm eff}$ we have considered both $w_{\rm DE} =-1$ and $w_{\rm DE} \neq -1$. Focusing on $w_{\rm DE}^{\rm eff}$, we see that for GP, a mild deviation from $-1$ is noticed only for CC+Pantheon+ (Fig. \ref{fig:wDMeff-wDEeff-GP}) but this is not true in the entire redshift regime, while for ANN (Fig. \ref{fig:wDMeff-wDEeff-ANN}), only in the high redshift regime, $w_{\rm DE}^{\rm eff}$ exhibits its deviation from $-1$, but this deviation is statistically very mild (at slightly more than 68\% CL for Pantheon+ and CC+Pantheon+). On the other hand, concentrating on $w_{\rm DM}^{\rm eff}$, we find that for $w_{\rm DE} =-1$, GP shows a very mild deviation (at most 95\% CL) from $w_{\rm DM}^{\rm eff} = 0$ only for CC+Pantheon+ (Fig. \ref{fig:wDMeff-wDEeff-GP}) at certain redshift regimes, while for ANN we do not find any evidence of non-null $w_{\rm DM}^{\rm eff}$. For other constant values of $w_{\rm DE}$, we notice that when $w_{\rm DE}$ deviates from $-1$ either in the quintessence or phantom regime, the effective EoS parameter for DM starts deviating from zero. This is a consequence of the interaction since for different values of $w_{\rm DE}$ other than $-1$, we observe an emergence of interaction in the dark sector.

Summarizing all the results in a nutshell and considering only the present datasets, we find that, 
\begin{itemize}
    \item  if $w_{\rm DE} =-1$, then ANN does not indicate any  evidence of interaction for any of the datasets but concerning GP, only for CC+Pantheon+, an evidence of interaction at present time  ($< 2\sigma$) and  in the intermediate redshift regime ($\sim 2 \sigma$) is found. 

    \item if $w_{\rm DE} \neq -1$, then both GP and ANN predict an evidence of interaction is found when $w_{\rm DE}$ deviates from $-1$; in fact, this evidence is pronounced much for ANN. 

    \item the effective EoS parameters, namely $w_{\rm DM}^{\rm eff}$ and $w_{\rm DE}^{\rm eff}$ are affected by the interaction; we find that with the emergence of interaction, $w_{\rm DM}^{\rm eff}$ and $w_{\rm DE}^{\rm eff}$ deviate from $0$ and $-1$, respectively, but such deviations are not robust according to the current observational datasets.  

    \item ANN seems to offer more stringent constraints than GP, at least in the present interacting context, this is visible. 
    
\end{itemize}
These altogether suggest that $w_{\rm DE}$ seems to play a very striking role in the interaction in the dark sector and this needs special attention. While in this work we consider only CC and Pantheon+, however, based on the results presented in this context,  we believe  that the stage-IV astronomical probes, such as DESI~\cite{DESI:2024mwx,DESI:2025zgx,DESI:2025fii,Gu:2025xie},  Euclid \cite{EUCLID:2011zbd}, the Large Synoptic Survey Telescope (LSST) \cite{LSST:2008ijt}, the Wide-Field InfraRed Survey Telescope (WFIRST)~\cite{Spergel:2013tha}, and the Square Kilometre Array (SKA)~\cite{SKA:2018ckk} can shed more light on the hidden side of the interaction.  
In particularly, the inclusion of baryon acoustic oscillations from DESI from its Data Release 1 and 2~ \cite{DESI:2024mwx,DESI:2025zgx,DESI:2025fii,Gu:2025xie} could be particularly appealing  since DESI indicates the preference for a dynamical DE. Now, an interaction between a pressure-less DM and DE with constant EoS leads to a non-interacting two-fluid system in which the EoS parameters of the dark fluids become dynamical. That means, it may happen that evidence of dynamical EoS parameters of DM and DE is actually a result of interaction between them. So, DESI's inference on the evidence of dynamical DE may probably indicate the presence of an interaction in the dark sector. In a forthcoming article we shall report detailed  reconstructions of the interaction using the baryon acoustic oscillations from DESI~\cite{DESI:2024mwx,DESI:2025zgx}.

\section*{Acknowledgments}
We thank the referee very much for some insightful comments which improved the quality of the manuscript. 
MA acknowledges the Junior Research Fellowship (File No: 08/0155(17246)/2023-EMR-I) from the Council of Scientific and Industrial Research (CSIR), Govt. of India. GJW acknowledges the Africa Europe Cluster of Research Excellence (CoRE-AI) fellowship. YZM acknowledges the support
from National Research Foundation of South Africa with
Grant No. 150580, No. 159044, No. CHN22111069370
and No. ERC23040389081. SP acknowledges the support from the Department of Science and Technology (DST), Govt. of India under the Scheme  ``Fund for Improvement of S\&T Infrastructure (FIST)'' (File No. SR/FST/MS-I/2019/41). We acknowledge computational cluster resources at the Centre for High-Performance Computing, Cape Town, South Africa and the ICT Department of Presidency University, Kolkata.

%-------------------------------------------------------------------------------
\bibliography{references}

\begin{thebibliography}{126}%
\makeatletter
\providecommand \@ifxundefined [1]{%
 \@ifx{#1\undefined}
}%
\providecommand \@ifnum [1]{%
 \ifnum #1\expandafter \@firstoftwo
 \else \expandafter \@secondoftwo
 \fi
}%
\providecommand \@ifx [1]{%
 \ifx #1\expandafter \@firstoftwo
 \else \expandafter \@secondoftwo
 \fi
}%
\providecommand \natexlab [1]{#1}%
\providecommand \enquote  [1]{``#1''}%
\providecommand \bibnamefont  [1]{#1}%
\providecommand \bibfnamefont [1]{#1}%
\providecommand \citenamefont [1]{#1}%
\providecommand \href@noop [0]{\@secondoftwo}%
\providecommand \href [0]{\begingroup \@sanitize@url \@href}%
\providecommand \@href[1]{\@@startlink{#1}\@@href}%
\providecommand \@@href[1]{\endgroup#1\@@endlink}%
\providecommand \@sanitize@url [0]{\catcode `\\12\catcode `\$12\catcode
  `\&12\catcode `\#12\catcode `\^12\catcode `\_12\catcode `\%12\relax}%
\providecommand \@@startlink[1]{}%
\providecommand \@@endlink[0]{}%
\providecommand \url  [0]{\begingroup\@sanitize@url \@url }%
\providecommand \@url [1]{\endgroup\@href {#1}{\urlprefix }}%
\providecommand \urlprefix  [0]{URL }%
\providecommand \Eprint [0]{\href }%
\providecommand \doibase [0]{http://dx.doi.org/}%
\providecommand \selectlanguage [0]{\@gobble}%
\providecommand \bibinfo  [0]{\@secondoftwo}%
\providecommand \bibfield  [0]{\@secondoftwo}%
\providecommand \translation [1]{[#1]}%
\providecommand \BibitemOpen [0]{}%
\providecommand \bibitemStop [0]{}%
\providecommand \bibitemNoStop [0]{.\EOS\space}%
\providecommand \EOS [0]{\spacefactor3000\relax}%
\providecommand \BibitemShut  [1]{\csname bibitem#1\endcsname}%
\let\auto@bib@innerbib\@empty
%</preamble>
\bibitem [{\citenamefont {Copeland}\ \emph {et~al.}(2006)\citenamefont
  {Copeland}, \citenamefont {Sami},\ and\ \citenamefont
  {Tsujikawa}}]{Copeland:2006wr}%
  \BibitemOpen
  \bibfield  {author} {\bibinfo {author} {\bibfnamefont {E.~J.}\ \bibnamefont
  {Copeland}}, \bibinfo {author} {\bibfnamefont {M.}~\bibnamefont {Sami}}, \
  and\ \bibinfo {author} {\bibfnamefont {S.}~\bibnamefont {Tsujikawa}},\ }\href
  {\doibase 10.1142/S021827180600942X} {\bibfield  {journal} {\bibinfo
  {journal} {Int. J. Mod. Phys. D}\ }\textbf {\bibinfo {volume} {15}},\
  \bibinfo {pages} {1753} (\bibinfo {year} {2006})},\ \Eprint
  {http://arxiv.org/abs/hep-th/0603057} {arXiv:hep-th/0603057} \BibitemShut
  {NoStop}%
\bibitem [{\citenamefont {Nojiri}\ and\ \citenamefont
  {Odintsov}(2006)}]{Nojiri:2006ri}%
  \BibitemOpen
  \bibfield  {author} {\bibinfo {author} {\bibfnamefont {S.}~\bibnamefont
  {Nojiri}}\ and\ \bibinfo {author} {\bibfnamefont {S.~D.}\ \bibnamefont
  {Odintsov}},\ }\href {\doibase 10.1142/S0219887807001928} {\bibfield
  {journal} {\bibinfo  {journal} {eConf}\ }\textbf {\bibinfo {volume}
  {C0602061}},\ \bibinfo {pages} {06} (\bibinfo {year} {2006})},\ \Eprint
  {http://arxiv.org/abs/hep-th/0601213} {arXiv:hep-th/0601213} \BibitemShut
  {NoStop}%
\bibitem [{\citenamefont {Sotiriou}\ and\ \citenamefont
  {Faraoni}(2010)}]{Sotiriou:2008rp}%
  \BibitemOpen
  \bibfield  {author} {\bibinfo {author} {\bibfnamefont {T.~P.}\ \bibnamefont
  {Sotiriou}}\ and\ \bibinfo {author} {\bibfnamefont {V.}~\bibnamefont
  {Faraoni}},\ }\href {\doibase 10.1103/RevModPhys.82.451} {\bibfield
  {journal} {\bibinfo  {journal} {Rev. Mod. Phys.}\ }\textbf {\bibinfo {volume}
  {82}},\ \bibinfo {pages} {451} (\bibinfo {year} {2010})},\ \Eprint
  {http://arxiv.org/abs/0805.1726} {arXiv:0805.1726 [gr-qc]} \BibitemShut
  {NoStop}%
\bibitem [{\citenamefont {De~Felice}\ and\ \citenamefont
  {Tsujikawa}(2010)}]{DeFelice:2010aj}%
  \BibitemOpen
  \bibfield  {author} {\bibinfo {author} {\bibfnamefont {A.}~\bibnamefont
  {De~Felice}}\ and\ \bibinfo {author} {\bibfnamefont {S.}~\bibnamefont
  {Tsujikawa}},\ }\href {\doibase 10.12942/lrr-2010-3} {\bibfield  {journal}
  {\bibinfo  {journal} {Living Rev. Rel.}\ }\textbf {\bibinfo {volume} {13}},\
  \bibinfo {pages} {3} (\bibinfo {year} {2010})},\ \Eprint
  {http://arxiv.org/abs/1002.4928} {arXiv:1002.4928 [gr-qc]} \BibitemShut
  {NoStop}%
\bibitem [{\citenamefont {Clifton}\ \emph {et~al.}(2012)\citenamefont
  {Clifton}, \citenamefont {Ferreira}, \citenamefont {Padilla},\ and\
  \citenamefont {Skordis}}]{Clifton:2011jh}%
  \BibitemOpen
  \bibfield  {author} {\bibinfo {author} {\bibfnamefont {T.}~\bibnamefont
  {Clifton}}, \bibinfo {author} {\bibfnamefont {P.~G.}\ \bibnamefont
  {Ferreira}}, \bibinfo {author} {\bibfnamefont {A.}~\bibnamefont {Padilla}}, \
  and\ \bibinfo {author} {\bibfnamefont {C.}~\bibnamefont {Skordis}},\ }\href
  {\doibase 10.1016/j.physrep.2012.01.001} {\bibfield  {journal} {\bibinfo
  {journal} {Phys. Rept.}\ }\textbf {\bibinfo {volume} {513}},\ \bibinfo
  {pages} {1} (\bibinfo {year} {2012})},\ \Eprint
  {http://arxiv.org/abs/1106.2476} {arXiv:1106.2476 [astro-ph.CO]} \BibitemShut
  {NoStop}%
\bibitem [{\citenamefont {Cai}\ \emph {et~al.}(2016{\natexlab{a}})\citenamefont
  {Cai}, \citenamefont {Capozziello}, \citenamefont {De~Laurentis},\ and\
  \citenamefont {Saridakis}}]{Cai:2015emx}%
  \BibitemOpen
  \bibfield  {author} {\bibinfo {author} {\bibfnamefont {Y.-F.}\ \bibnamefont
  {Cai}}, \bibinfo {author} {\bibfnamefont {S.}~\bibnamefont {Capozziello}},
  \bibinfo {author} {\bibfnamefont {M.}~\bibnamefont {De~Laurentis}}, \ and\
  \bibinfo {author} {\bibfnamefont {E.~N.}\ \bibnamefont {Saridakis}},\ }\href
  {\doibase 10.1088/0034-4885/79/10/106901} {\bibfield  {journal} {\bibinfo
  {journal} {Rept. Prog. Phys.}\ }\textbf {\bibinfo {volume} {79}},\ \bibinfo
  {pages} {106901} (\bibinfo {year} {2016}{\natexlab{a}})},\ \Eprint
  {http://arxiv.org/abs/1511.07586} {arXiv:1511.07586 [gr-qc]} \BibitemShut
  {NoStop}%
\bibitem [{\citenamefont {Nojiri}\ \emph {et~al.}(2017)\citenamefont {Nojiri},
  \citenamefont {Odintsov},\ and\ \citenamefont {Oikonomou}}]{Nojiri:2017ncd}%
  \BibitemOpen
  \bibfield  {author} {\bibinfo {author} {\bibfnamefont {S.}~\bibnamefont
  {Nojiri}}, \bibinfo {author} {\bibfnamefont {S.~D.}\ \bibnamefont
  {Odintsov}}, \ and\ \bibinfo {author} {\bibfnamefont {V.~K.}\ \bibnamefont
  {Oikonomou}},\ }\href {\doibase 10.1016/j.physrep.2017.06.001} {\bibfield
  {journal} {\bibinfo  {journal} {Phys. Rept.}\ }\textbf {\bibinfo {volume}
  {692}},\ \bibinfo {pages} {1} (\bibinfo {year} {2017})},\ \Eprint
  {http://arxiv.org/abs/1705.11098} {arXiv:1705.11098 [gr-qc]} \BibitemShut
  {NoStop}%
\bibitem [{\citenamefont {Bahamonde}\ \emph {et~al.}(2023)\citenamefont
  {Bahamonde}, \citenamefont {Dialektopoulos}, \citenamefont
  {Escamilla-Rivera}, \citenamefont {Farrugia}, \citenamefont {Gakis},
  \citenamefont {Hendry}, \citenamefont {Hohmann}, \citenamefont {Levi~Said},
  \citenamefont {Mifsud},\ and\ \citenamefont
  {Di~Valentino}}]{Bahamonde:2021gfp}%
  \BibitemOpen
  \bibfield  {author} {\bibinfo {author} {\bibfnamefont {S.}~\bibnamefont
  {Bahamonde}}, \bibinfo {author} {\bibfnamefont {K.~F.}\ \bibnamefont
  {Dialektopoulos}}, \bibinfo {author} {\bibfnamefont {C.}~\bibnamefont
  {Escamilla-Rivera}}, \bibinfo {author} {\bibfnamefont {G.}~\bibnamefont
  {Farrugia}}, \bibinfo {author} {\bibfnamefont {V.}~\bibnamefont {Gakis}},
  \bibinfo {author} {\bibfnamefont {M.}~\bibnamefont {Hendry}}, \bibinfo
  {author} {\bibfnamefont {M.}~\bibnamefont {Hohmann}}, \bibinfo {author}
  {\bibfnamefont {J.}~\bibnamefont {Levi~Said}}, \bibinfo {author}
  {\bibfnamefont {J.}~\bibnamefont {Mifsud}}, \ and\ \bibinfo {author}
  {\bibfnamefont {E.}~\bibnamefont {Di~Valentino}},\ }\href {\doibase
  10.1088/1361-6633/ac9cef} {\bibfield  {journal} {\bibinfo  {journal} {Rept.
  Prog. Phys.}\ }\textbf {\bibinfo {volume} {86}},\ \bibinfo {pages} {026901}
  (\bibinfo {year} {2023})},\ \Eprint {http://arxiv.org/abs/2106.13793}
  {arXiv:2106.13793 [gr-qc]} \BibitemShut {NoStop}%
\bibitem [{\citenamefont {Amendola}(2000)}]{Amendola:1999er}%
  \BibitemOpen
  \bibfield  {author} {\bibinfo {author} {\bibfnamefont {L.}~\bibnamefont
  {Amendola}},\ }\href {\doibase 10.1103/PhysRevD.62.043511} {\bibfield
  {journal} {\bibinfo  {journal} {Phys. Rev. D}\ }\textbf {\bibinfo {volume}
  {62}},\ \bibinfo {pages} {043511} (\bibinfo {year} {2000})},\ \Eprint
  {http://arxiv.org/abs/astro-ph/9908023} {arXiv:astro-ph/9908023} \BibitemShut
  {NoStop}%
\bibitem [{\citenamefont {Cai}\ and\ \citenamefont {Wang}(2005)}]{Cai:2004dk}%
  \BibitemOpen
  \bibfield  {author} {\bibinfo {author} {\bibfnamefont {R.-G.}\ \bibnamefont
  {Cai}}\ and\ \bibinfo {author} {\bibfnamefont {A.}~\bibnamefont {Wang}},\
  }\href {\doibase 10.1088/1475-7516/2005/03/002} {\bibfield  {journal}
  {\bibinfo  {journal} {JCAP}\ }\textbf {\bibinfo {volume} {03}},\ \bibinfo
  {pages} {002} (\bibinfo {year} {2005})},\ \Eprint
  {http://arxiv.org/abs/hep-th/0411025} {arXiv:hep-th/0411025} \BibitemShut
  {NoStop}%
\bibitem [{\citenamefont {Pavon}\ and\ \citenamefont
  {Zimdahl}(2005)}]{Pavon:2005yx}%
  \BibitemOpen
  \bibfield  {author} {\bibinfo {author} {\bibfnamefont {D.}~\bibnamefont
  {Pavon}}\ and\ \bibinfo {author} {\bibfnamefont {W.}~\bibnamefont
  {Zimdahl}},\ }\href {\doibase 10.1016/j.physletb.2005.08.134} {\bibfield
  {journal} {\bibinfo  {journal} {Phys. Lett. B}\ }\textbf {\bibinfo {volume}
  {628}},\ \bibinfo {pages} {206} (\bibinfo {year} {2005})},\ \Eprint
  {http://arxiv.org/abs/gr-qc/0505020} {arXiv:gr-qc/0505020} \BibitemShut
  {NoStop}%
\bibitem [{\citenamefont {Huey}\ and\ \citenamefont
  {Wandelt}(2006)}]{Huey:2004qv}%
  \BibitemOpen
  \bibfield  {author} {\bibinfo {author} {\bibfnamefont {G.}~\bibnamefont
  {Huey}}\ and\ \bibinfo {author} {\bibfnamefont {B.~D.}\ \bibnamefont
  {Wandelt}},\ }\href {\doibase 10.1103/PhysRevD.74.023519} {\bibfield
  {journal} {\bibinfo  {journal} {Phys. Rev. D}\ }\textbf {\bibinfo {volume}
  {74}},\ \bibinfo {pages} {023519} (\bibinfo {year} {2006})},\ \Eprint
  {http://arxiv.org/abs/astro-ph/0407196} {arXiv:astro-ph/0407196} \BibitemShut
  {NoStop}%
\bibitem [{\citenamefont {del Campo}\ \emph {et~al.}(2008)\citenamefont {del
  Campo}, \citenamefont {Herrera},\ and\ \citenamefont
  {Pavon}}]{delCampo:2008sr}%
  \BibitemOpen
  \bibfield  {author} {\bibinfo {author} {\bibfnamefont {S.}~\bibnamefont {del
  Campo}}, \bibinfo {author} {\bibfnamefont {R.}~\bibnamefont {Herrera}}, \
  and\ \bibinfo {author} {\bibfnamefont {D.}~\bibnamefont {Pavon}},\ }\href
  {\doibase 10.1103/PhysRevD.78.021302} {\bibfield  {journal} {\bibinfo
  {journal} {Phys. Rev. D}\ }\textbf {\bibinfo {volume} {78}},\ \bibinfo
  {pages} {021302} (\bibinfo {year} {2008})},\ \Eprint
  {http://arxiv.org/abs/0806.2116} {arXiv:0806.2116 [astro-ph]} \BibitemShut
  {NoStop}%
\bibitem [{\citenamefont {del Campo}\ \emph {et~al.}(2009)\citenamefont {del
  Campo}, \citenamefont {Herrera},\ and\ \citenamefont
  {Pavon}}]{delCampo:2008jx}%
  \BibitemOpen
  \bibfield  {author} {\bibinfo {author} {\bibfnamefont {S.}~\bibnamefont {del
  Campo}}, \bibinfo {author} {\bibfnamefont {R.}~\bibnamefont {Herrera}}, \
  and\ \bibinfo {author} {\bibfnamefont {D.}~\bibnamefont {Pavon}},\ }\href
  {\doibase 10.1088/1475-7516/2009/01/020} {\bibfield  {journal} {\bibinfo
  {journal} {JCAP}\ }\textbf {\bibinfo {volume} {01}},\ \bibinfo {pages} {020}
  (\bibinfo {year} {2009})},\ \Eprint {http://arxiv.org/abs/0812.2210}
  {arXiv:0812.2210 [gr-qc]} \BibitemShut {NoStop}%
\bibitem [{\citenamefont {Das}\ \emph {et~al.}(2006)\citenamefont {Das},
  \citenamefont {Corasaniti},\ and\ \citenamefont {Khoury}}]{Das:2005yj}%
  \BibitemOpen
  \bibfield  {author} {\bibinfo {author} {\bibfnamefont {S.}~\bibnamefont
  {Das}}, \bibinfo {author} {\bibfnamefont {P.~S.}\ \bibnamefont {Corasaniti}},
  \ and\ \bibinfo {author} {\bibfnamefont {J.}~\bibnamefont {Khoury}},\ }\href
  {\doibase 10.1103/PhysRevD.73.083509} {\bibfield  {journal} {\bibinfo
  {journal} {Phys. Rev. D}\ }\textbf {\bibinfo {volume} {73}},\ \bibinfo
  {pages} {083509} (\bibinfo {year} {2006})},\ \Eprint
  {http://arxiv.org/abs/astro-ph/0510628} {arXiv:astro-ph/0510628} \BibitemShut
  {NoStop}%
\bibitem [{\citenamefont {Wang}\ \emph {et~al.}(2005)\citenamefont {Wang},
  \citenamefont {Gong},\ and\ \citenamefont {Abdalla}}]{Wang:2005jx}%
  \BibitemOpen
  \bibfield  {author} {\bibinfo {author} {\bibfnamefont {B.}~\bibnamefont
  {Wang}}, \bibinfo {author} {\bibfnamefont {Y.-g.}\ \bibnamefont {Gong}}, \
  and\ \bibinfo {author} {\bibfnamefont {E.}~\bibnamefont {Abdalla}},\ }\href
  {\doibase 10.1016/j.physletb.2005.08.008} {\bibfield  {journal} {\bibinfo
  {journal} {Phys. Lett. B}\ }\textbf {\bibinfo {volume} {624}},\ \bibinfo
  {pages} {141} (\bibinfo {year} {2005})},\ \Eprint
  {http://arxiv.org/abs/hep-th/0506069} {arXiv:hep-th/0506069} \BibitemShut
  {NoStop}%
\bibitem [{\citenamefont {Sadjadi}\ and\ \citenamefont
  {Honardoost}(2007)}]{Sadjadi:2006qb}%
  \BibitemOpen
  \bibfield  {author} {\bibinfo {author} {\bibfnamefont {H.~M.}\ \bibnamefont
  {Sadjadi}}\ and\ \bibinfo {author} {\bibfnamefont {M.}~\bibnamefont
  {Honardoost}},\ }\href {\doibase 10.1016/j.physletb.2007.02.016} {\bibfield
  {journal} {\bibinfo  {journal} {Phys. Lett. B}\ }\textbf {\bibinfo {volume}
  {647}},\ \bibinfo {pages} {231} (\bibinfo {year} {2007})},\ \Eprint
  {http://arxiv.org/abs/gr-qc/0609076} {arXiv:gr-qc/0609076} \BibitemShut
  {NoStop}%
\bibitem [{\citenamefont {Kumar}\ and\ \citenamefont
  {Nunes}(2016)}]{Kumar:2016zpg}%
  \BibitemOpen
  \bibfield  {author} {\bibinfo {author} {\bibfnamefont {S.}~\bibnamefont
  {Kumar}}\ and\ \bibinfo {author} {\bibfnamefont {R.~C.}\ \bibnamefont
  {Nunes}},\ }\href {\doibase 10.1103/PhysRevD.94.123511} {\bibfield  {journal}
  {\bibinfo  {journal} {Phys. Rev. D}\ }\textbf {\bibinfo {volume} {94}},\
  \bibinfo {pages} {123511} (\bibinfo {year} {2016})},\ \Eprint
  {http://arxiv.org/abs/1608.02454} {arXiv:1608.02454 [astro-ph.CO]}
  \BibitemShut {NoStop}%
\bibitem [{\citenamefont {Pourtsidou}\ and\ \citenamefont
  {Tram}(2016)}]{Pourtsidou:2016ico}%
  \BibitemOpen
  \bibfield  {author} {\bibinfo {author} {\bibfnamefont {A.}~\bibnamefont
  {Pourtsidou}}\ and\ \bibinfo {author} {\bibfnamefont {T.}~\bibnamefont
  {Tram}},\ }\href {\doibase 10.1103/PhysRevD.94.043518} {\bibfield  {journal}
  {\bibinfo  {journal} {Phys. Rev. D}\ }\textbf {\bibinfo {volume} {94}},\
  \bibinfo {pages} {043518} (\bibinfo {year} {2016})},\ \Eprint
  {http://arxiv.org/abs/1604.04222} {arXiv:1604.04222 [astro-ph.CO]}
  \BibitemShut {NoStop}%
\bibitem [{\citenamefont {An}\ \emph {et~al.}(2018)\citenamefont {An},
  \citenamefont {Feng},\ and\ \citenamefont {Wang}}]{An:2017crg}%
  \BibitemOpen
  \bibfield  {author} {\bibinfo {author} {\bibfnamefont {R.}~\bibnamefont
  {An}}, \bibinfo {author} {\bibfnamefont {C.}~\bibnamefont {Feng}}, \ and\
  \bibinfo {author} {\bibfnamefont {B.}~\bibnamefont {Wang}},\ }\href {\doibase
  10.1088/1475-7516/2018/02/038} {\bibfield  {journal} {\bibinfo  {journal}
  {JCAP}\ }\textbf {\bibinfo {volume} {02}},\ \bibinfo {pages} {038} (\bibinfo
  {year} {2018})},\ \Eprint {http://arxiv.org/abs/1711.06799} {arXiv:1711.06799
  [astro-ph.CO]} \BibitemShut {NoStop}%
\bibitem [{\citenamefont {Di~Valentino}\ \emph {et~al.}(2017)\citenamefont
  {Di~Valentino}, \citenamefont {Melchiorri},\ and\ \citenamefont
  {Mena}}]{DiValentino:2017iww}%
  \BibitemOpen
  \bibfield  {author} {\bibinfo {author} {\bibfnamefont {E.}~\bibnamefont
  {Di~Valentino}}, \bibinfo {author} {\bibfnamefont {A.}~\bibnamefont
  {Melchiorri}}, \ and\ \bibinfo {author} {\bibfnamefont {O.}~\bibnamefont
  {Mena}},\ }\href {\doibase 10.1103/PhysRevD.96.043503} {\bibfield  {journal}
  {\bibinfo  {journal} {Phys. Rev. D}\ }\textbf {\bibinfo {volume} {96}},\
  \bibinfo {pages} {043503} (\bibinfo {year} {2017})},\ \Eprint
  {http://arxiv.org/abs/1704.08342} {arXiv:1704.08342 [astro-ph.CO]}
  \BibitemShut {NoStop}%
\bibitem [{\citenamefont {Yang}\ \emph {et~al.}(2018)\citenamefont {Yang},
  \citenamefont {Pan}, \citenamefont {Di~Valentino}, \citenamefont {Nunes},
  \citenamefont {Vagnozzi},\ and\ \citenamefont {Mota}}]{Yang:2018euj}%
  \BibitemOpen
  \bibfield  {author} {\bibinfo {author} {\bibfnamefont {W.}~\bibnamefont
  {Yang}}, \bibinfo {author} {\bibfnamefont {S.}~\bibnamefont {Pan}}, \bibinfo
  {author} {\bibfnamefont {E.}~\bibnamefont {Di~Valentino}}, \bibinfo {author}
  {\bibfnamefont {R.~C.}\ \bibnamefont {Nunes}}, \bibinfo {author}
  {\bibfnamefont {S.}~\bibnamefont {Vagnozzi}}, \ and\ \bibinfo {author}
  {\bibfnamefont {D.~F.}\ \bibnamefont {Mota}},\ }\href {\doibase
  10.1088/1475-7516/2018/09/019} {\bibfield  {journal} {\bibinfo  {journal}
  {JCAP}\ }\textbf {\bibinfo {volume} {09}},\ \bibinfo {pages} {019} (\bibinfo
  {year} {2018})},\ \Eprint {http://arxiv.org/abs/1805.08252} {arXiv:1805.08252
  [astro-ph.CO]} \BibitemShut {NoStop}%
\bibitem [{\citenamefont {Kumar}\ \emph {et~al.}(2019)\citenamefont {Kumar},
  \citenamefont {Nunes},\ and\ \citenamefont {Yadav}}]{Kumar:2019wfs}%
  \BibitemOpen
  \bibfield  {author} {\bibinfo {author} {\bibfnamefont {S.}~\bibnamefont
  {Kumar}}, \bibinfo {author} {\bibfnamefont {R.~C.}\ \bibnamefont {Nunes}}, \
  and\ \bibinfo {author} {\bibfnamefont {S.~K.}\ \bibnamefont {Yadav}},\ }\href
  {\doibase 10.1140/epjc/s10052-019-7087-7} {\bibfield  {journal} {\bibinfo
  {journal} {Eur. Phys. J. C}\ }\textbf {\bibinfo {volume} {79}},\ \bibinfo
  {pages} {576} (\bibinfo {year} {2019})},\ \Eprint
  {http://arxiv.org/abs/1903.04865} {arXiv:1903.04865 [astro-ph.CO]}
  \BibitemShut {NoStop}%
\bibitem [{\citenamefont {Pan}\ \emph {et~al.}(2019{\natexlab{a}})\citenamefont
  {Pan}, \citenamefont {Yang}, \citenamefont {Di~Valentino}, \citenamefont
  {Saridakis},\ and\ \citenamefont {Chakraborty}}]{Pan:2019gop}%
  \BibitemOpen
  \bibfield  {author} {\bibinfo {author} {\bibfnamefont {S.}~\bibnamefont
  {Pan}}, \bibinfo {author} {\bibfnamefont {W.}~\bibnamefont {Yang}}, \bibinfo
  {author} {\bibfnamefont {E.}~\bibnamefont {Di~Valentino}}, \bibinfo {author}
  {\bibfnamefont {E.~N.}\ \bibnamefont {Saridakis}}, \ and\ \bibinfo {author}
  {\bibfnamefont {S.}~\bibnamefont {Chakraborty}},\ }\href {\doibase
  10.1103/PhysRevD.100.103520} {\bibfield  {journal} {\bibinfo  {journal}
  {Phys. Rev. D}\ }\textbf {\bibinfo {volume} {100}},\ \bibinfo {pages}
  {103520} (\bibinfo {year} {2019}{\natexlab{a}})},\ \Eprint
  {http://arxiv.org/abs/1907.07540} {arXiv:1907.07540 [astro-ph.CO]}
  \BibitemShut {NoStop}%
\bibitem [{\citenamefont {Shah}\ \emph {et~al.}(2024)\citenamefont {Shah},
  \citenamefont {Mukherjee},\ and\ \citenamefont {Pal}}]{Shah:2024rme}%
  \BibitemOpen
  \bibfield  {author} {\bibinfo {author} {\bibfnamefont {R.}~\bibnamefont
  {Shah}}, \bibinfo {author} {\bibfnamefont {P.}~\bibnamefont {Mukherjee}}, \
  and\ \bibinfo {author} {\bibfnamefont {S.}~\bibnamefont {Pal}},\ }\href@noop
  {} {\bibfield  {journal} {\bibinfo  {journal} {2404.06396}\ } (\bibinfo
  {year} {2024})}\BibitemShut {NoStop}%
\bibitem [{\citenamefont {Giar\`e}\ \emph
  {et~al.}(2024{\natexlab{a}})\citenamefont {Giar\`e}, \citenamefont {Sabogal},
  \citenamefont {Nunes},\ and\ \citenamefont {Di~Valentino}}]{Giare:2024smz}%
  \BibitemOpen
  \bibfield  {author} {\bibinfo {author} {\bibfnamefont {W.}~\bibnamefont
  {Giar\`e}}, \bibinfo {author} {\bibfnamefont {M.~A.}\ \bibnamefont
  {Sabogal}}, \bibinfo {author} {\bibfnamefont {R.~C.}\ \bibnamefont {Nunes}},
  \ and\ \bibinfo {author} {\bibfnamefont {E.}~\bibnamefont {Di~Valentino}},\
  }\href {\doibase 10.1103/PhysRevLett.133.251003} {\bibfield  {journal}
  {\bibinfo  {journal} {Phys. Rev. Lett.}\ }\textbf {\bibinfo {volume} {133}},\
  \bibinfo {pages} {251003} (\bibinfo {year} {2024}{\natexlab{a}})},\ \Eprint
  {http://arxiv.org/abs/2404.15232} {arXiv:2404.15232 [astro-ph.CO]}
  \BibitemShut {NoStop}%
\bibitem [{\citenamefont {Di~Valentino}\ \emph {et~al.}(2021)\citenamefont
  {Di~Valentino}, \citenamefont {Mena}, \citenamefont {Pan}, \citenamefont
  {Visinelli}, \citenamefont {Yang}, \citenamefont {Melchiorri}, \citenamefont
  {Mota}, \citenamefont {Riess},\ and\ \citenamefont
  {Silk}}]{DiValentino:2021izs}%
  \BibitemOpen
  \bibfield  {author} {\bibinfo {author} {\bibfnamefont {E.}~\bibnamefont
  {Di~Valentino}}, \bibinfo {author} {\bibfnamefont {O.}~\bibnamefont {Mena}},
  \bibinfo {author} {\bibfnamefont {S.}~\bibnamefont {Pan}}, \bibinfo {author}
  {\bibfnamefont {L.}~\bibnamefont {Visinelli}}, \bibinfo {author}
  {\bibfnamefont {W.}~\bibnamefont {Yang}}, \bibinfo {author} {\bibfnamefont
  {A.}~\bibnamefont {Melchiorri}}, \bibinfo {author} {\bibfnamefont {D.~F.}\
  \bibnamefont {Mota}}, \bibinfo {author} {\bibfnamefont {A.~G.}\ \bibnamefont
  {Riess}}, \ and\ \bibinfo {author} {\bibfnamefont {J.}~\bibnamefont {Silk}},\
  }\href {\doibase 10.1088/1361-6382/ac086d} {\bibfield  {journal} {\bibinfo
  {journal} {Class. Quant. Grav.}\ }\textbf {\bibinfo {volume} {38}},\ \bibinfo
  {pages} {153001} (\bibinfo {year} {2021})},\ \Eprint
  {http://arxiv.org/abs/2103.01183} {arXiv:2103.01183 [astro-ph.CO]}
  \BibitemShut {NoStop}%
\bibitem [{\citenamefont {Seikel}\ \emph
  {et~al.}(2012{\natexlab{a}})\citenamefont {Seikel}, \citenamefont
  {Clarkson},\ and\ \citenamefont {Smith}}]{Seikel:2012uu}%
  \BibitemOpen
  \bibfield  {author} {\bibinfo {author} {\bibfnamefont {M.}~\bibnamefont
  {Seikel}}, \bibinfo {author} {\bibfnamefont {C.}~\bibnamefont {Clarkson}}, \
  and\ \bibinfo {author} {\bibfnamefont {M.}~\bibnamefont {Smith}},\ }\href
  {\doibase 10.1088/1475-7516/2012/06/036} {\bibfield  {journal} {\bibinfo
  {journal} {JCAP}\ }\textbf {\bibinfo {volume} {06}},\ \bibinfo {pages} {036}
  (\bibinfo {year} {2012}{\natexlab{a}})},\ \Eprint
  {http://arxiv.org/abs/1204.2832} {arXiv:1204.2832 [astro-ph.CO]} \BibitemShut
  {NoStop}%
\bibitem [{\citenamefont {Seikel}\ \emph
  {et~al.}(2012{\natexlab{b}})\citenamefont {Seikel}, \citenamefont {Yahya},
  \citenamefont {Maartens},\ and\ \citenamefont {Clarkson}}]{Seikel:2012cs}%
  \BibitemOpen
  \bibfield  {author} {\bibinfo {author} {\bibfnamefont {M.}~\bibnamefont
  {Seikel}}, \bibinfo {author} {\bibfnamefont {S.}~\bibnamefont {Yahya}},
  \bibinfo {author} {\bibfnamefont {R.}~\bibnamefont {Maartens}}, \ and\
  \bibinfo {author} {\bibfnamefont {C.}~\bibnamefont {Clarkson}},\ }\href
  {\doibase 10.1103/PhysRevD.86.083001} {\bibfield  {journal} {\bibinfo
  {journal} {Phys. Rev. D}\ }\textbf {\bibinfo {volume} {86}},\ \bibinfo
  {pages} {083001} (\bibinfo {year} {2012}{\natexlab{b}})},\ \Eprint
  {http://arxiv.org/abs/1205.3431} {arXiv:1205.3431 [astro-ph.CO]} \BibitemShut
  {NoStop}%
\bibitem [{\citenamefont {Yahya}\ \emph {et~al.}(2014)\citenamefont {Yahya},
  \citenamefont {Seikel}, \citenamefont {Clarkson}, \citenamefont {Maartens},\
  and\ \citenamefont {Smith}}]{Yahya:2013xma}%
  \BibitemOpen
  \bibfield  {author} {\bibinfo {author} {\bibfnamefont {S.}~\bibnamefont
  {Yahya}}, \bibinfo {author} {\bibfnamefont {M.}~\bibnamefont {Seikel}},
  \bibinfo {author} {\bibfnamefont {C.}~\bibnamefont {Clarkson}}, \bibinfo
  {author} {\bibfnamefont {R.}~\bibnamefont {Maartens}}, \ and\ \bibinfo
  {author} {\bibfnamefont {M.}~\bibnamefont {Smith}},\ }\href {\doibase
  10.1103/PhysRevD.89.023503} {\bibfield  {journal} {\bibinfo  {journal} {Phys.
  Rev. D}\ }\textbf {\bibinfo {volume} {89}},\ \bibinfo {pages} {023503}
  (\bibinfo {year} {2014})},\ \Eprint {http://arxiv.org/abs/1308.4099}
  {arXiv:1308.4099 [astro-ph.CO]} \BibitemShut {NoStop}%
\bibitem [{\citenamefont {Li}\ \emph {et~al.}(2016)\citenamefont {Li},
  \citenamefont {Gonzalez}, \citenamefont {Yu}, \citenamefont {Zhu},\ and\
  \citenamefont {Alcaniz}}]{Li:2015nta}%
  \BibitemOpen
  \bibfield  {author} {\bibinfo {author} {\bibfnamefont {Z.}~\bibnamefont
  {Li}}, \bibinfo {author} {\bibfnamefont {J.~E.}\ \bibnamefont {Gonzalez}},
  \bibinfo {author} {\bibfnamefont {H.}~\bibnamefont {Yu}}, \bibinfo {author}
  {\bibfnamefont {Z.-H.}\ \bibnamefont {Zhu}}, \ and\ \bibinfo {author}
  {\bibfnamefont {J.~S.}\ \bibnamefont {Alcaniz}},\ }\href {\doibase
  10.1103/PhysRevD.93.043014} {\bibfield  {journal} {\bibinfo  {journal} {Phys.
  Rev. D}\ }\textbf {\bibinfo {volume} {93}},\ \bibinfo {pages} {043014}
  (\bibinfo {year} {2016})},\ \Eprint {http://arxiv.org/abs/1504.03269}
  {arXiv:1504.03269 [astro-ph.CO]} \BibitemShut {NoStop}%
\bibitem [{\citenamefont {Cai}\ \emph {et~al.}(2016{\natexlab{b}})\citenamefont
  {Cai}, \citenamefont {Guo},\ and\ \citenamefont {Yang}}]{Cai:2015pia}%
  \BibitemOpen
  \bibfield  {author} {\bibinfo {author} {\bibfnamefont {R.-G.}\ \bibnamefont
  {Cai}}, \bibinfo {author} {\bibfnamefont {Z.-K.}\ \bibnamefont {Guo}}, \ and\
  \bibinfo {author} {\bibfnamefont {T.}~\bibnamefont {Yang}},\ }\href {\doibase
  10.1103/PhysRevD.93.043517} {\bibfield  {journal} {\bibinfo  {journal} {Phys.
  Rev. D}\ }\textbf {\bibinfo {volume} {93}},\ \bibinfo {pages} {043517}
  (\bibinfo {year} {2016}{\natexlab{b}})},\ \Eprint
  {http://arxiv.org/abs/1509.06283} {arXiv:1509.06283 [astro-ph.CO]}
  \BibitemShut {NoStop}%
\bibitem [{\citenamefont {Cai}\ \emph {et~al.}(2016{\natexlab{c}})\citenamefont
  {Cai}, \citenamefont {Guo},\ and\ \citenamefont {Yang}}]{Cai:2016vmn}%
  \BibitemOpen
  \bibfield  {author} {\bibinfo {author} {\bibfnamefont {R.-G.}\ \bibnamefont
  {Cai}}, \bibinfo {author} {\bibfnamefont {Z.-K.}\ \bibnamefont {Guo}}, \ and\
  \bibinfo {author} {\bibfnamefont {T.}~\bibnamefont {Yang}},\ }\href {\doibase
  10.1088/1475-7516/2016/08/016} {\bibfield  {journal} {\bibinfo  {journal}
  {JCAP}\ }\textbf {\bibinfo {volume} {08}},\ \bibinfo {pages} {016} (\bibinfo
  {year} {2016}{\natexlab{c}})},\ \Eprint {http://arxiv.org/abs/1601.05497}
  {arXiv:1601.05497 [astro-ph.CO]} \BibitemShut {NoStop}%
\bibitem [{\citenamefont {Wei}\ and\ \citenamefont {Wu}(2017)}]{Wei:2016xti}%
  \BibitemOpen
  \bibfield  {author} {\bibinfo {author} {\bibfnamefont {J.-J.}\ \bibnamefont
  {Wei}}\ and\ \bibinfo {author} {\bibfnamefont {X.-F.}\ \bibnamefont {Wu}},\
  }\href {\doibase 10.3847/1538-4357/aa674b} {\bibfield  {journal} {\bibinfo
  {journal} {Astrophys. J.}\ }\textbf {\bibinfo {volume} {838}},\ \bibinfo
  {pages} {160} (\bibinfo {year} {2017})},\ \Eprint
  {http://arxiv.org/abs/1611.00904} {arXiv:1611.00904 [astro-ph.CO]}
  \BibitemShut {NoStop}%
\bibitem [{\citenamefont {Yu}\ \emph {et~al.}(2018)\citenamefont {Yu},
  \citenamefont {Ratra},\ and\ \citenamefont {Wang}}]{Yu:2017iju}%
  \BibitemOpen
  \bibfield  {author} {\bibinfo {author} {\bibfnamefont {H.}~\bibnamefont
  {Yu}}, \bibinfo {author} {\bibfnamefont {B.}~\bibnamefont {Ratra}}, \ and\
  \bibinfo {author} {\bibfnamefont {F.-Y.}\ \bibnamefont {Wang}},\ }\href
  {\doibase 10.3847/1538-4357/aab0a2} {\bibfield  {journal} {\bibinfo
  {journal} {Astrophys. J.}\ }\textbf {\bibinfo {volume} {856}},\ \bibinfo
  {pages} {3} (\bibinfo {year} {2018})},\ \Eprint
  {http://arxiv.org/abs/1711.03437} {arXiv:1711.03437 [astro-ph.CO]}
  \BibitemShut {NoStop}%
\bibitem [{\citenamefont {Wang}\ \emph {et~al.}(2017)\citenamefont {Wang},
  \citenamefont {Wei}, \citenamefont {Li}, \citenamefont {Xia},\ and\
  \citenamefont {Zhu}}]{Wang:2017lri}%
  \BibitemOpen
  \bibfield  {author} {\bibinfo {author} {\bibfnamefont {G.-J.}\ \bibnamefont
  {Wang}}, \bibinfo {author} {\bibfnamefont {J.-J.}\ \bibnamefont {Wei}},
  \bibinfo {author} {\bibfnamefont {Z.-X.}\ \bibnamefont {Li}}, \bibinfo
  {author} {\bibfnamefont {J.-Q.}\ \bibnamefont {Xia}}, \ and\ \bibinfo
  {author} {\bibfnamefont {Z.-H.}\ \bibnamefont {Zhu}},\ }\href {\doibase
  10.3847/1538-4357/aa8725} {\bibfield  {journal} {\bibinfo  {journal}
  {Astrophys. J.}\ }\textbf {\bibinfo {volume} {847}},\ \bibinfo {pages} {45}
  (\bibinfo {year} {2017})},\ \Eprint {http://arxiv.org/abs/1709.07258}
  {arXiv:1709.07258 [astro-ph.CO]} \BibitemShut {NoStop}%
\bibitem [{\citenamefont {Mukherjee}\ and\ \citenamefont
  {Banerjee}(2022)}]{Mukherjee:2022ujw}%
  \BibitemOpen
  \bibfield  {author} {\bibinfo {author} {\bibfnamefont {P.}~\bibnamefont
  {Mukherjee}}\ and\ \bibinfo {author} {\bibfnamefont {N.}~\bibnamefont
  {Banerjee}},\ }\href {\doibase 10.1103/PhysRevD.105.063516} {\bibfield
  {journal} {\bibinfo  {journal} {Phys. Rev. D}\ }\textbf {\bibinfo {volume}
  {105}},\ \bibinfo {pages} {063516} (\bibinfo {year} {2022})},\ \Eprint
  {http://arxiv.org/abs/2202.07886} {arXiv:2202.07886 [astro-ph.CO]}
  \BibitemShut {NoStop}%
\bibitem [{\citenamefont {Busti}\ \emph {et~al.}(2014)\citenamefont {Busti},
  \citenamefont {Clarkson},\ and\ \citenamefont {Seikel}}]{Busti:2014dua}%
  \BibitemOpen
  \bibfield  {author} {\bibinfo {author} {\bibfnamefont {V.~C.}\ \bibnamefont
  {Busti}}, \bibinfo {author} {\bibfnamefont {C.}~\bibnamefont {Clarkson}}, \
  and\ \bibinfo {author} {\bibfnamefont {M.}~\bibnamefont {Seikel}},\ }\href
  {\doibase 10.1093/mnrasl/slu035} {\bibfield  {journal} {\bibinfo  {journal}
  {Mon. Not. Roy. Astron. Soc.}\ }\textbf {\bibinfo {volume} {441}},\ \bibinfo
  {pages} {11} (\bibinfo {year} {2014})},\ \Eprint
  {http://arxiv.org/abs/1402.5429} {arXiv:1402.5429 [astro-ph.CO]} \BibitemShut
  {NoStop}%
\bibitem [{\citenamefont {G\'omez-Valent}\ and\ \citenamefont
  {Amendola}(2018)}]{Gomez-Valent:2018hwc}%
  \BibitemOpen
  \bibfield  {author} {\bibinfo {author} {\bibfnamefont {A.}~\bibnamefont
  {G\'omez-Valent}}\ and\ \bibinfo {author} {\bibfnamefont {L.}~\bibnamefont
  {Amendola}},\ }\href {\doibase 10.1088/1475-7516/2018/04/051} {\bibfield
  {journal} {\bibinfo  {journal} {JCAP}\ }\textbf {\bibinfo {volume} {04}},\
  \bibinfo {pages} {051} (\bibinfo {year} {2018})},\ \Eprint
  {http://arxiv.org/abs/1802.01505} {arXiv:1802.01505 [astro-ph.CO]}
  \BibitemShut {NoStop}%
\bibitem [{\citenamefont {Shafieloo}\ \emph {et~al.}(2013)\citenamefont
  {Shafieloo}, \citenamefont {Kim},\ and\ \citenamefont
  {Linder}}]{Shafieloo:2012ms}%
  \BibitemOpen
  \bibfield  {author} {\bibinfo {author} {\bibfnamefont {A.}~\bibnamefont
  {Shafieloo}}, \bibinfo {author} {\bibfnamefont {A.~G.}\ \bibnamefont {Kim}},
  \ and\ \bibinfo {author} {\bibfnamefont {E.~V.}\ \bibnamefont {Linder}},\
  }\href {\doibase 10.1103/PhysRevD.87.023520} {\bibfield  {journal} {\bibinfo
  {journal} {Phys. Rev. D}\ }\textbf {\bibinfo {volume} {87}},\ \bibinfo
  {pages} {023520} (\bibinfo {year} {2013})},\ \Eprint
  {http://arxiv.org/abs/1211.6128} {arXiv:1211.6128 [astro-ph.CO]} \BibitemShut
  {NoStop}%
\bibitem [{\citenamefont {Gonzalez}(2017)}]{Gonzalez:2017fra}%
  \BibitemOpen
  \bibfield  {author} {\bibinfo {author} {\bibfnamefont {J.~E.}\ \bibnamefont
  {Gonzalez}},\ }\href {\doibase 10.1103/PhysRevD.96.123501} {\bibfield
  {journal} {\bibinfo  {journal} {Phys. Rev. D}\ }\textbf {\bibinfo {volume}
  {96}},\ \bibinfo {pages} {123501} (\bibinfo {year} {2017})},\ \Eprint
  {http://arxiv.org/abs/1710.07656} {arXiv:1710.07656 [astro-ph.CO]}
  \BibitemShut {NoStop}%
\bibitem [{\citenamefont {Santos-da Costa}\ \emph {et~al.}(2015)\citenamefont
  {Santos-da Costa}, \citenamefont {Busti},\ and\ \citenamefont
  {Holanda}}]{Santos-da-Costa:2015kmv}%
  \BibitemOpen
  \bibfield  {author} {\bibinfo {author} {\bibfnamefont {S.}~\bibnamefont
  {Santos-da Costa}}, \bibinfo {author} {\bibfnamefont {V.~C.}\ \bibnamefont
  {Busti}}, \ and\ \bibinfo {author} {\bibfnamefont {R.~F.~L.}\ \bibnamefont
  {Holanda}},\ }\href {\doibase 10.1088/1475-7516/2015/10/061} {\bibfield
  {journal} {\bibinfo  {journal} {JCAP}\ }\textbf {\bibinfo {volume} {10}},\
  \bibinfo {pages} {061} (\bibinfo {year} {2015})},\ \Eprint
  {http://arxiv.org/abs/1506.00145} {arXiv:1506.00145 [astro-ph.CO]}
  \BibitemShut {NoStop}%
\bibitem [{\citenamefont {Li}\ and\ \citenamefont {Lin}(2018)}]{Li:2017zrx}%
  \BibitemOpen
  \bibfield  {author} {\bibinfo {author} {\bibfnamefont {X.}~\bibnamefont
  {Li}}\ and\ \bibinfo {author} {\bibfnamefont {H.~N.}\ \bibnamefont {Lin}},\
  }\href {\doibase 10.1093/mnras/stx2810} {\bibfield  {journal} {\bibinfo
  {journal} {Mon. Not. Roy. Astron. Soc.}\ }\textbf {\bibinfo {volume} {474}},\
  \bibinfo {pages} {313} (\bibinfo {year} {2018})},\ \Eprint
  {http://arxiv.org/abs/1710.11361} {arXiv:1710.11361 [astro-ph.CO]}
  \BibitemShut {NoStop}%
\bibitem [{\citenamefont {Mukherjee}\ and\ \citenamefont
  {Mukherjee}(2021)}]{Mukherjee:2021kcu}%
  \BibitemOpen
  \bibfield  {author} {\bibinfo {author} {\bibfnamefont {P.}~\bibnamefont
  {Mukherjee}}\ and\ \bibinfo {author} {\bibfnamefont {A.}~\bibnamefont
  {Mukherjee}},\ }\href {\doibase 10.1093/mnras/stab1054} {\bibfield  {journal}
  {\bibinfo  {journal} {Mon. Not. Roy. Astron. Soc.}\ }\textbf {\bibinfo
  {volume} {504}},\ \bibinfo {pages} {3938} (\bibinfo {year} {2021})},\ \Eprint
  {http://arxiv.org/abs/2104.06066} {arXiv:2104.06066 [astro-ph.CO]}
  \BibitemShut {NoStop}%
\bibitem [{\citenamefont {Yang}\ \emph {et~al.}(2024)\citenamefont {Yang},
  \citenamefont {Ren}, \citenamefont {Wang}, \citenamefont {Lu}, \citenamefont
  {Zhang}, \citenamefont {Cai},\ and\ \citenamefont
  {Saridakis}}]{Yang:2024kdo}%
  \BibitemOpen
  \bibfield  {author} {\bibinfo {author} {\bibfnamefont {Y.}~\bibnamefont
  {Yang}}, \bibinfo {author} {\bibfnamefont {X.}~\bibnamefont {Ren}}, \bibinfo
  {author} {\bibfnamefont {Q.}~\bibnamefont {Wang}}, \bibinfo {author}
  {\bibfnamefont {Z.}~\bibnamefont {Lu}}, \bibinfo {author} {\bibfnamefont
  {D.}~\bibnamefont {Zhang}}, \bibinfo {author} {\bibfnamefont {Y.-F.}\
  \bibnamefont {Cai}}, \ and\ \bibinfo {author} {\bibfnamefont {E.~N.}\
  \bibnamefont {Saridakis}},\ }\href {\doibase 10.1016/j.scib.2024.07.029}
  {\bibfield  {journal} {\bibinfo  {journal} {Sci. Bull.}\ }\textbf {\bibinfo
  {volume} {69}},\ \bibinfo {pages} {2698} (\bibinfo {year} {2024})},\ \Eprint
  {http://arxiv.org/abs/2404.19437} {arXiv:2404.19437 [astro-ph.CO]}
  \BibitemShut {NoStop}%
\bibitem [{\citenamefont {Yang}\ \emph
  {et~al.}(2025{\natexlab{a}})\citenamefont {Yang}, \citenamefont {Wang},
  \citenamefont {Li}, \citenamefont {Yuan}, \citenamefont {Ren}, \citenamefont
  {Saridakis},\ and\ \citenamefont {Cai}}]{Yang:2025kgc}%
  \BibitemOpen
  \bibfield  {author} {\bibinfo {author} {\bibfnamefont {Y.}~\bibnamefont
  {Yang}}, \bibinfo {author} {\bibfnamefont {Q.}~\bibnamefont {Wang}}, \bibinfo
  {author} {\bibfnamefont {C.}~\bibnamefont {Li}}, \bibinfo {author}
  {\bibfnamefont {P.}~\bibnamefont {Yuan}}, \bibinfo {author} {\bibfnamefont
  {X.}~\bibnamefont {Ren}}, \bibinfo {author} {\bibfnamefont {E.~N.}\
  \bibnamefont {Saridakis}}, \ and\ \bibinfo {author} {\bibfnamefont {Y.-F.}\
  \bibnamefont {Cai}},\ }\href@noop {} {\enquote {\bibinfo {title}
  {{Gaussian-process reconstructions and model building of quintom dark energy
  from latest cosmological observations}},}\ } (\bibinfo {year}
  {2025}{\natexlab{a}}),\ \Eprint {http://arxiv.org/abs/2501.18336}
  {arXiv:2501.18336 [astro-ph.CO]} \BibitemShut {NoStop}%
\bibitem [{\citenamefont {Yang}\ \emph
  {et~al.}(2025{\natexlab{b}})\citenamefont {Yang}, \citenamefont {Wang},
  \citenamefont {Ren}, \citenamefont {Saridakis},\ and\ \citenamefont
  {Cai}}]{Yang:2025mws}%
  \BibitemOpen
  \bibfield  {author} {\bibinfo {author} {\bibfnamefont {Y.}~\bibnamefont
  {Yang}}, \bibinfo {author} {\bibfnamefont {Q.}~\bibnamefont {Wang}}, \bibinfo
  {author} {\bibfnamefont {X.}~\bibnamefont {Ren}}, \bibinfo {author}
  {\bibfnamefont {E.~N.}\ \bibnamefont {Saridakis}}, \ and\ \bibinfo {author}
  {\bibfnamefont {Y.-F.}\ \bibnamefont {Cai}},\ }\href@noop {} {\enquote
  {\bibinfo {title} {{Modified gravity realizations of quintom dark energy
  after DESI DR2}},}\ } (\bibinfo {year} {2025}{\natexlab{b}}),\ \Eprint
  {http://arxiv.org/abs/2504.06784} {arXiv:2504.06784 [astro-ph.CO]}
  \BibitemShut {NoStop}%
\bibitem [{\citenamefont {Yang}\ \emph {et~al.}(2015)\citenamefont {Yang},
  \citenamefont {Guo},\ and\ \citenamefont {Cai}}]{Yang:2015tzc}%
  \BibitemOpen
  \bibfield  {author} {\bibinfo {author} {\bibfnamefont {T.}~\bibnamefont
  {Yang}}, \bibinfo {author} {\bibfnamefont {Z.-K.}\ \bibnamefont {Guo}}, \
  and\ \bibinfo {author} {\bibfnamefont {R.-G.}\ \bibnamefont {Cai}},\ }\href
  {\doibase 10.1103/PhysRevD.91.123533} {\bibfield  {journal} {\bibinfo
  {journal} {Phys. Rev. D}\ }\textbf {\bibinfo {volume} {91}},\ \bibinfo
  {pages} {123533} (\bibinfo {year} {2015})},\ \Eprint
  {http://arxiv.org/abs/1505.04443} {arXiv:1505.04443 [astro-ph.CO]}
  \BibitemShut {NoStop}%
\bibitem [{\citenamefont {Cai}\ \emph {et~al.}(2017)\citenamefont {Cai},
  \citenamefont {Tamanini},\ and\ \citenamefont {Yang}}]{Cai:2017yww}%
  \BibitemOpen
  \bibfield  {author} {\bibinfo {author} {\bibfnamefont {R.-G.}\ \bibnamefont
  {Cai}}, \bibinfo {author} {\bibfnamefont {N.}~\bibnamefont {Tamanini}}, \
  and\ \bibinfo {author} {\bibfnamefont {T.}~\bibnamefont {Yang}},\ }\href
  {\doibase 10.1088/1475-7516/2017/05/031} {\bibfield  {journal} {\bibinfo
  {journal} {JCAP}\ }\textbf {\bibinfo {volume} {05}},\ \bibinfo {pages} {031}
  (\bibinfo {year} {2017})},\ \Eprint {http://arxiv.org/abs/1703.07323}
  {arXiv:1703.07323 [astro-ph.CO]} \BibitemShut {NoStop}%
\bibitem [{\citenamefont {Aljaf}\ \emph {et~al.}(2021)\citenamefont {Aljaf},
  \citenamefont {Gregoris},\ and\ \citenamefont {Khurshudyan}}]{Aljaf:2020eqh}%
  \BibitemOpen
  \bibfield  {author} {\bibinfo {author} {\bibfnamefont {M.}~\bibnamefont
  {Aljaf}}, \bibinfo {author} {\bibfnamefont {D.}~\bibnamefont {Gregoris}}, \
  and\ \bibinfo {author} {\bibfnamefont {M.}~\bibnamefont {Khurshudyan}},\
  }\href {\doibase 10.1140/epjc/s10052-021-09306-2} {\bibfield  {journal}
  {\bibinfo  {journal} {Eur. Phys. J. C}\ }\textbf {\bibinfo {volume} {81}},\
  \bibinfo {pages} {544} (\bibinfo {year} {2021})},\ \Eprint
  {http://arxiv.org/abs/2005.01891} {arXiv:2005.01891 [astro-ph.CO]}
  \BibitemShut {NoStop}%
\bibitem [{\citenamefont {Mukherjee}\ and\ \citenamefont
  {Banerjee}(2021)}]{Mukherjee:2021ggf}%
  \BibitemOpen
  \bibfield  {author} {\bibinfo {author} {\bibfnamefont {P.}~\bibnamefont
  {Mukherjee}}\ and\ \bibinfo {author} {\bibfnamefont {N.}~\bibnamefont
  {Banerjee}},\ }\href {\doibase 10.1103/PhysRevD.103.123530} {\bibfield
  {journal} {\bibinfo  {journal} {Phys. Rev. D}\ }\textbf {\bibinfo {volume}
  {103}},\ \bibinfo {pages} {123530} (\bibinfo {year} {2021})},\ \Eprint
  {http://arxiv.org/abs/2105.09995} {arXiv:2105.09995 [astro-ph.CO]}
  \BibitemShut {NoStop}%
\bibitem [{\citenamefont {Bonilla}\ \emph {et~al.}(2022)\citenamefont
  {Bonilla}, \citenamefont {Kumar}, \citenamefont {Nunes},\ and\ \citenamefont
  {Pan}}]{Bonilla:2021dql}%
  \BibitemOpen
  \bibfield  {author} {\bibinfo {author} {\bibfnamefont {A.}~\bibnamefont
  {Bonilla}}, \bibinfo {author} {\bibfnamefont {S.}~\bibnamefont {Kumar}},
  \bibinfo {author} {\bibfnamefont {R.~C.}\ \bibnamefont {Nunes}}, \ and\
  \bibinfo {author} {\bibfnamefont {S.}~\bibnamefont {Pan}},\ }\href {\doibase
  10.1093/mnras/stac687} {\bibfield  {journal} {\bibinfo  {journal} {Mon. Not.
  Roy. Astron. Soc.}\ }\textbf {\bibinfo {volume} {512}},\ \bibinfo {pages}
  {4231} (\bibinfo {year} {2022})},\ \Eprint {http://arxiv.org/abs/2102.06149}
  {arXiv:2102.06149 [astro-ph.CO]} \BibitemShut {NoStop}%
\bibitem [{\citenamefont {Escamilla}\ \emph {et~al.}(2023)\citenamefont
  {Escamilla}, \citenamefont {Akarsu}, \citenamefont {Di~Valentino},\ and\
  \citenamefont {Vazquez}}]{Escamilla:2023shf}%
  \BibitemOpen
  \bibfield  {author} {\bibinfo {author} {\bibfnamefont {L.~A.}\ \bibnamefont
  {Escamilla}}, \bibinfo {author} {\bibfnamefont {O.}~\bibnamefont {Akarsu}},
  \bibinfo {author} {\bibfnamefont {E.}~\bibnamefont {Di~Valentino}}, \ and\
  \bibinfo {author} {\bibfnamefont {J.~A.}\ \bibnamefont {Vazquez}},\ }\href
  {\doibase 10.1088/1475-7516/2023/11/051} {\bibfield  {journal} {\bibinfo
  {journal} {JCAP}\ }\textbf {\bibinfo {volume} {11}},\ \bibinfo {pages} {051}
  (\bibinfo {year} {2023})},\ \Eprint {http://arxiv.org/abs/2305.16290}
  {arXiv:2305.16290 [astro-ph.CO]} \BibitemShut {NoStop}%
\bibitem [{\citenamefont {Wang}\ \emph
  {et~al.}(2020{\natexlab{a}})\citenamefont {Wang}, \citenamefont {Ma},
  \citenamefont {Li},\ and\ \citenamefont {Xia}}]{Wang:2019vxv}%
  \BibitemOpen
  \bibfield  {author} {\bibinfo {author} {\bibfnamefont {G.-J.}\ \bibnamefont
  {Wang}}, \bibinfo {author} {\bibfnamefont {X.-J.}\ \bibnamefont {Ma}},
  \bibinfo {author} {\bibfnamefont {S.-Y.}\ \bibnamefont {Li}}, \ and\ \bibinfo
  {author} {\bibfnamefont {J.-Q.}\ \bibnamefont {Xia}},\ }\href {\doibase
  10.3847/1538-4365/ab620b} {\bibfield  {journal} {\bibinfo  {journal}
  {Astrophys. J. Suppl.}\ }\textbf {\bibinfo {volume} {246}},\ \bibinfo {pages}
  {13} (\bibinfo {year} {2020}{\natexlab{a}})},\ \Eprint
  {http://arxiv.org/abs/1910.03636} {arXiv:1910.03636 [astro-ph.CO]}
  \BibitemShut {NoStop}%
\bibitem [{\citenamefont {Wang}\ \emph {et~al.}(2021)\citenamefont {Wang},
  \citenamefont {Ma},\ and\ \citenamefont {Xia}}]{Wang:2020dbt}%
  \BibitemOpen
  \bibfield  {author} {\bibinfo {author} {\bibfnamefont {G.-J.}\ \bibnamefont
  {Wang}}, \bibinfo {author} {\bibfnamefont {X.-J.}\ \bibnamefont {Ma}}, \ and\
  \bibinfo {author} {\bibfnamefont {J.-Q.}\ \bibnamefont {Xia}},\ }\href
  {\doibase 10.1093/mnras/staa4044} {\bibfield  {journal} {\bibinfo  {journal}
  {Mon. Not. Roy. Astron. Soc.}\ }\textbf {\bibinfo {volume} {501}},\ \bibinfo
  {pages} {5714} (\bibinfo {year} {2021})},\ \Eprint
  {http://arxiv.org/abs/2004.13913} {arXiv:2004.13913 [astro-ph.CO]}
  \BibitemShut {NoStop}%
\bibitem [{\citenamefont {Wang}\ \emph
  {et~al.}(2020{\natexlab{b}})\citenamefont {Wang}, \citenamefont {Li},\ and\
  \citenamefont {Xia}}]{Wang:2020hmn}%
  \BibitemOpen
  \bibfield  {author} {\bibinfo {author} {\bibfnamefont {G.-J.}\ \bibnamefont
  {Wang}}, \bibinfo {author} {\bibfnamefont {S.-Y.}\ \bibnamefont {Li}}, \ and\
  \bibinfo {author} {\bibfnamefont {J.-Q.}\ \bibnamefont {Xia}},\ }\href
  {\doibase 10.3847/1538-4365/aba190} {\bibfield  {journal} {\bibinfo
  {journal} {Astrophys. J. Suppl.}\ }\textbf {\bibinfo {volume} {249}},\
  \bibinfo {pages} {25} (\bibinfo {year} {2020}{\natexlab{b}})},\ \Eprint
  {http://arxiv.org/abs/2005.07089} {arXiv:2005.07089 [astro-ph.CO]}
  \BibitemShut {NoStop}%
\bibitem [{\citenamefont {Dialektopoulos}\ \emph {et~al.}(2022)\citenamefont
  {Dialektopoulos}, \citenamefont {Said}, \citenamefont {Mifsud}, \citenamefont
  {Sultana},\ and\ \citenamefont {Adami}}]{Dialektopoulos:2021wde}%
  \BibitemOpen
  \bibfield  {author} {\bibinfo {author} {\bibfnamefont {K.}~\bibnamefont
  {Dialektopoulos}}, \bibinfo {author} {\bibfnamefont {J.~L.}\ \bibnamefont
  {Said}}, \bibinfo {author} {\bibfnamefont {J.}~\bibnamefont {Mifsud}},
  \bibinfo {author} {\bibfnamefont {J.}~\bibnamefont {Sultana}}, \ and\
  \bibinfo {author} {\bibfnamefont {K.~Z.}\ \bibnamefont {Adami}},\ }\href
  {\doibase 10.1088/1475-7516/2022/02/023} {\bibfield  {journal} {\bibinfo
  {journal} {JCAP}\ }\textbf {\bibinfo {volume} {02}},\ \bibinfo {pages} {023}
  (\bibinfo {year} {2022})},\ \Eprint {http://arxiv.org/abs/2111.11462}
  {arXiv:2111.11462 [astro-ph.CO]} \BibitemShut {NoStop}%
\bibitem [{\citenamefont {Wang}\ \emph {et~al.}(2022)\citenamefont {Wang},
  \citenamefont {Cheng}, \citenamefont {Ma},\ and\ \citenamefont
  {Xia}}]{Wang:2022qta}%
  \BibitemOpen
  \bibfield  {author} {\bibinfo {author} {\bibfnamefont {G.-J.}\ \bibnamefont
  {Wang}}, \bibinfo {author} {\bibfnamefont {C.}~\bibnamefont {Cheng}},
  \bibinfo {author} {\bibfnamefont {Y.-Z.}\ \bibnamefont {Ma}}, \ and\ \bibinfo
  {author} {\bibfnamefont {J.-Q.}\ \bibnamefont {Xia}},\ }\href {\doibase
  10.3847/1538-4365/ac7da1} {\bibfield  {journal} {\bibinfo  {journal}
  {Astrophys. J. Supp.}\ }\textbf {\bibinfo {volume} {262}},\ \bibinfo {pages}
  {24} (\bibinfo {year} {2022})},\ \Eprint {http://arxiv.org/abs/2207.00185}
  {arXiv:2207.00185 [astro-ph.CO]} \BibitemShut {NoStop}%
\bibitem [{\citenamefont {Wang}\ \emph {et~al.}(2023)\citenamefont {Wang},
  \citenamefont {Cheng}, \citenamefont {Ma}, \citenamefont {Xia}, \citenamefont
  {Abebe},\ and\ \citenamefont {Beesham}}]{Wang:2023vej}%
  \BibitemOpen
  \bibfield  {author} {\bibinfo {author} {\bibfnamefont {G.-J.}\ \bibnamefont
  {Wang}}, \bibinfo {author} {\bibfnamefont {C.}~\bibnamefont {Cheng}},
  \bibinfo {author} {\bibfnamefont {Y.-Z.}\ \bibnamefont {Ma}}, \bibinfo
  {author} {\bibfnamefont {J.-Q.}\ \bibnamefont {Xia}}, \bibinfo {author}
  {\bibfnamefont {A.}~\bibnamefont {Abebe}}, \ and\ \bibinfo {author}
  {\bibfnamefont {A.}~\bibnamefont {Beesham}},\ }\href {\doibase
  10.3847/1538-4365/ace113} {\bibfield  {journal} {\bibinfo  {journal}
  {Astrophys. J. Suppl.}\ }\textbf {\bibinfo {volume} {268}},\ \bibinfo {pages}
  {7} (\bibinfo {year} {2023})},\ \Eprint {http://arxiv.org/abs/2306.11102}
  {arXiv:2306.11102 [astro-ph.CO]} \BibitemShut {NoStop}%
\bibitem [{\citenamefont {Qi}\ \emph {et~al.}(2023)\citenamefont {Qi},
  \citenamefont {Meng}, \citenamefont {Zhang},\ and\ \citenamefont
  {Zhang}}]{Qi:2023oxv}%
  \BibitemOpen
  \bibfield  {author} {\bibinfo {author} {\bibfnamefont {J.-Z.}\ \bibnamefont
  {Qi}}, \bibinfo {author} {\bibfnamefont {P.}~\bibnamefont {Meng}}, \bibinfo
  {author} {\bibfnamefont {J.-F.}\ \bibnamefont {Zhang}}, \ and\ \bibinfo
  {author} {\bibfnamefont {X.}~\bibnamefont {Zhang}},\ }\href {\doibase
  10.1103/PhysRevD.108.063522} {\bibfield  {journal} {\bibinfo  {journal}
  {Phys. Rev. D}\ }\textbf {\bibinfo {volume} {108}},\ \bibinfo {pages}
  {063522} (\bibinfo {year} {2023})},\ \Eprint
  {http://arxiv.org/abs/2302.08889} {arXiv:2302.08889 [astro-ph.CO]}
  \BibitemShut {NoStop}%
\bibitem [{\citenamefont {Giar\`e}\ \emph
  {et~al.}(2024{\natexlab{b}})\citenamefont {Giar\`e}, \citenamefont {Betts},
  \citenamefont {van~de Bruck},\ and\ \citenamefont
  {Di~Valentino}}]{Giare:2024syw}%
  \BibitemOpen
  \bibfield  {author} {\bibinfo {author} {\bibfnamefont {W.}~\bibnamefont
  {Giar\`e}}, \bibinfo {author} {\bibfnamefont {J.}~\bibnamefont {Betts}},
  \bibinfo {author} {\bibfnamefont {C.}~\bibnamefont {van~de Bruck}}, \ and\
  \bibinfo {author} {\bibfnamefont {E.}~\bibnamefont {Di~Valentino}},\
  }\href@noop {} {\enquote {\bibinfo {title} {{A model-independent test of
  pre-recombination New Physics: Machine Learning based estimate of the Sound
  Horizon from Gravitational Wave Standard Sirens and the Baryon Acoustic
  Oscillation Angular Scale}},}\ } (\bibinfo {year} {2024}{\natexlab{b}}),\
  \Eprint {http://arxiv.org/abs/2406.07493} {arXiv:2406.07493 [astro-ph.CO]}
  \BibitemShut {NoStop}%
\bibitem [{\citenamefont {Mukherjee}\ \emph {et~al.}(2022)\citenamefont
  {Mukherjee}, \citenamefont {Levi~Said},\ and\ \citenamefont
  {Mifsud}}]{Mukherjee:2022yyq}%
  \BibitemOpen
  \bibfield  {author} {\bibinfo {author} {\bibfnamefont {P.}~\bibnamefont
  {Mukherjee}}, \bibinfo {author} {\bibfnamefont {J.}~\bibnamefont
  {Levi~Said}}, \ and\ \bibinfo {author} {\bibfnamefont {J.}~\bibnamefont
  {Mifsud}},\ }\href {\doibase 10.1088/1475-7516/2022/12/029} {\bibfield
  {journal} {\bibinfo  {journal} {JCAP}\ }\textbf {\bibinfo {volume} {12}},\
  \bibinfo {pages} {029} (\bibinfo {year} {2022})},\ \Eprint
  {http://arxiv.org/abs/2209.01113} {arXiv:2209.01113 [astro-ph.CO]}
  \BibitemShut {NoStop}%
\bibitem [{\citenamefont {Dialektopoulos}\ \emph {et~al.}(2024)\citenamefont
  {Dialektopoulos}, \citenamefont {Mukherjee}, \citenamefont {Levi~Said},\ and\
  \citenamefont {Mifsud}}]{Dialektopoulos:2023jam}%
  \BibitemOpen
  \bibfield  {author} {\bibinfo {author} {\bibfnamefont {K.~F.}\ \bibnamefont
  {Dialektopoulos}}, \bibinfo {author} {\bibfnamefont {P.}~\bibnamefont
  {Mukherjee}}, \bibinfo {author} {\bibfnamefont {J.}~\bibnamefont
  {Levi~Said}}, \ and\ \bibinfo {author} {\bibfnamefont {J.}~\bibnamefont
  {Mifsud}},\ }\href {\doibase 10.1016/j.dark.2023.101383} {\bibfield
  {journal} {\bibinfo  {journal} {Phys. Dark Univ.}\ }\textbf {\bibinfo
  {volume} {43}},\ \bibinfo {pages} {101383} (\bibinfo {year} {2024})},\
  \Eprint {http://arxiv.org/abs/2305.15500} {arXiv:2305.15500 [gr-qc]}
  \BibitemShut {NoStop}%
\bibitem [{\citenamefont {Kim}\ \emph {et~al.}(2015)\citenamefont {Kim},
  \citenamefont {Harry}, \citenamefont {Hodge}, \citenamefont {Kim},
  \citenamefont {Lee}, \citenamefont {Lee}, \citenamefont {Oh}, \citenamefont
  {Oh},\ and\ \citenamefont {Son}}]{Kim:2014nba}%
  \BibitemOpen
  \bibfield  {author} {\bibinfo {author} {\bibfnamefont {K.}~\bibnamefont
  {Kim}}, \bibinfo {author} {\bibfnamefont {I.~W.}\ \bibnamefont {Harry}},
  \bibinfo {author} {\bibfnamefont {K.~A.}\ \bibnamefont {Hodge}}, \bibinfo
  {author} {\bibfnamefont {Y.-M.}\ \bibnamefont {Kim}}, \bibinfo {author}
  {\bibfnamefont {C.-H.}\ \bibnamefont {Lee}}, \bibinfo {author} {\bibfnamefont
  {H.~K.}\ \bibnamefont {Lee}}, \bibinfo {author} {\bibfnamefont {J.~J.}\
  \bibnamefont {Oh}}, \bibinfo {author} {\bibfnamefont {S.~H.}\ \bibnamefont
  {Oh}}, \ and\ \bibinfo {author} {\bibfnamefont {E.~J.}\ \bibnamefont {Son}},\
  }\href {\doibase 10.1088/0264-9381/32/24/245002} {\bibfield  {journal}
  {\bibinfo  {journal} {Class. Quant. Grav.}\ }\textbf {\bibinfo {volume}
  {32}},\ \bibinfo {pages} {245002} (\bibinfo {year} {2015})},\ \Eprint
  {http://arxiv.org/abs/1410.6878} {arXiv:1410.6878 [astro-ph.IM]} \BibitemShut
  {NoStop}%
\bibitem [{\citenamefont {Cheng}\ \emph {et~al.}(2018)\citenamefont {Cheng},
  \citenamefont {Feng}, \citenamefont {Zhai},\ and\ \citenamefont
  {Li}}]{Cheng:2018nhz}%
  \BibitemOpen
  \bibfield  {author} {\bibinfo {author} {\bibfnamefont {Q.-B.}\ \bibnamefont
  {Cheng}}, \bibinfo {author} {\bibfnamefont {C.-J.}\ \bibnamefont {Feng}},
  \bibinfo {author} {\bibfnamefont {X.-H.}\ \bibnamefont {Zhai}}, \ and\
  \bibinfo {author} {\bibfnamefont {X.-Z.}\ \bibnamefont {Li}},\ }\href
  {\doibase 10.1103/PhysRevD.97.123530} {\bibfield  {journal} {\bibinfo
  {journal} {Phys. Rev. D}\ }\textbf {\bibinfo {volume} {97}},\ \bibinfo
  {pages} {123530} (\bibinfo {year} {2018})},\ \Eprint
  {http://arxiv.org/abs/1801.01723} {arXiv:1801.01723 [astro-ph.CO]}
  \BibitemShut {NoStop}%
\bibitem [{\citenamefont {Cheng}\ \emph {et~al.}(2021)\citenamefont {Cheng},
  \citenamefont {Feng}, \citenamefont {Zhai},\ and\ \citenamefont
  {Li}}]{Cheng:2020gec}%
  \BibitemOpen
  \bibfield  {author} {\bibinfo {author} {\bibfnamefont {Q.-B.}\ \bibnamefont
  {Cheng}}, \bibinfo {author} {\bibfnamefont {C.-J.}\ \bibnamefont {Feng}},
  \bibinfo {author} {\bibfnamefont {X.-H.}\ \bibnamefont {Zhai}}, \ and\
  \bibinfo {author} {\bibfnamefont {X.-Z.}\ \bibnamefont {Li}},\ }\href
  {\doibase 10.1142/S0217732321501492} {\bibfield  {journal} {\bibinfo
  {journal} {Mod. Phys. Lett. A}\ }\textbf {\bibinfo {volume} {36}},\ \bibinfo
  {pages} {2150149} (\bibinfo {year} {2021})},\ \Eprint
  {http://arxiv.org/abs/2004.04382} {arXiv:2004.04382 [astro-ph.CO]}
  \BibitemShut {NoStop}%
\bibitem [{\citenamefont {Choudhury}\ \emph {et~al.}(2020)\citenamefont
  {Choudhury}, \citenamefont {Datta},\ and\ \citenamefont
  {Chakraborty}}]{Choudhury:2019vat}%
  \BibitemOpen
  \bibfield  {author} {\bibinfo {author} {\bibfnamefont {M.}~\bibnamefont
  {Choudhury}}, \bibinfo {author} {\bibfnamefont {A.}~\bibnamefont {Datta}}, \
  and\ \bibinfo {author} {\bibfnamefont {A.}~\bibnamefont {Chakraborty}},\
  }\href {\doibase 10.1093/mnras/stz3107} {\bibfield  {journal} {\bibinfo
  {journal} {Mon. Not. Roy. Astron. Soc.}\ }\textbf {\bibinfo {volume} {491}},\
  \bibinfo {pages} {4031} (\bibinfo {year} {2020})},\ \Eprint
  {http://arxiv.org/abs/1911.02580} {arXiv:1911.02580 [astro-ph.CO]}
  \BibitemShut {NoStop}%
\bibitem [{\citenamefont {Choudhury}\ \emph {et~al.}(2021)\citenamefont
  {Choudhury}, \citenamefont {Chatterjee}, \citenamefont {Datta},\ and\
  \citenamefont {Choudhury}}]{Choudhury:2020azd}%
  \BibitemOpen
  \bibfield  {author} {\bibinfo {author} {\bibfnamefont {M.}~\bibnamefont
  {Choudhury}}, \bibinfo {author} {\bibfnamefont {A.}~\bibnamefont
  {Chatterjee}}, \bibinfo {author} {\bibfnamefont {A.}~\bibnamefont {Datta}}, \
  and\ \bibinfo {author} {\bibfnamefont {T.~R.}\ \bibnamefont {Choudhury}},\
  }\href {\doibase 10.1093/mnras/stab180} {\bibfield  {journal} {\bibinfo
  {journal} {Mon. Not. Roy. Astron. Soc.}\ }\textbf {\bibinfo {volume} {502}},\
  \bibinfo {pages} {2815} (\bibinfo {year} {2021})},\ \Eprint
  {http://arxiv.org/abs/2012.00028} {arXiv:2012.00028 [astro-ph.CO]}
  \BibitemShut {NoStop}%
\bibitem [{\citenamefont {Zhang}\ \emph {et~al.}(2022)\citenamefont {Zhang},
  \citenamefont {Jiao}, \citenamefont {Zhang}, \citenamefont {Zhang},\ and\
  \citenamefont {Yu}}]{Zhang:2022caa}%
  \BibitemOpen
  \bibfield  {author} {\bibinfo {author} {\bibfnamefont {J.-C.}\ \bibnamefont
  {Zhang}}, \bibinfo {author} {\bibfnamefont {K.}~\bibnamefont {Jiao}},
  \bibinfo {author} {\bibfnamefont {T.}~\bibnamefont {Zhang}}, \bibinfo
  {author} {\bibfnamefont {T.-J.}\ \bibnamefont {Zhang}}, \ and\ \bibinfo
  {author} {\bibfnamefont {B.}~\bibnamefont {Yu}},\ }\href {\doibase
  10.3847/1538-4357/ac85aa} {\bibfield  {journal} {\bibinfo  {journal}
  {Astrophys. J.}\ }\textbf {\bibinfo {volume} {936}},\ \bibinfo {pages} {21}
  (\bibinfo {year} {2022})},\ \Eprint {http://arxiv.org/abs/2208.03960}
  {arXiv:2208.03960 [astro-ph.CO]} \BibitemShut {NoStop}%
\bibitem [{\citenamefont {Pal}\ \emph {et~al.}(2023)\citenamefont {Pal},
  \citenamefont {Chanda},\ and\ \citenamefont {Saha}}]{Pal:2022hpi}%
  \BibitemOpen
  \bibfield  {author} {\bibinfo {author} {\bibfnamefont {S.}~\bibnamefont
  {Pal}}, \bibinfo {author} {\bibfnamefont {P.}~\bibnamefont {Chanda}}, \ and\
  \bibinfo {author} {\bibfnamefont {R.}~\bibnamefont {Saha}},\ }\href {\doibase
  10.3847/1538-4357/acb4ee} {\bibfield  {journal} {\bibinfo  {journal}
  {Astrophys. J.}\ }\textbf {\bibinfo {volume} {945}},\ \bibinfo {pages} {77}
  (\bibinfo {year} {2023})},\ \Eprint {http://arxiv.org/abs/2203.14060}
  {arXiv:2203.14060 [astro-ph.CO]} \BibitemShut {NoStop}%
\bibitem [{\citenamefont {Sikder}\ \emph {et~al.}(2023)\citenamefont {Sikder},
  \citenamefont {Barkana}, \citenamefont {Reis},\ and\ \citenamefont
  {Fialkov}}]{Sikder:2022hzk}%
  \BibitemOpen
  \bibfield  {author} {\bibinfo {author} {\bibfnamefont {S.}~\bibnamefont
  {Sikder}}, \bibinfo {author} {\bibfnamefont {R.}~\bibnamefont {Barkana}},
  \bibinfo {author} {\bibfnamefont {I.}~\bibnamefont {Reis}}, \ and\ \bibinfo
  {author} {\bibfnamefont {A.}~\bibnamefont {Fialkov}},\ }\href {\doibase
  10.1093/mnras/stad3699} {\bibfield  {journal} {\bibinfo  {journal} {Mon. Not.
  Roy. Astron. Soc.}\ }\textbf {\bibinfo {volume} {527}},\ \bibinfo {pages}
  {9977} (\bibinfo {year} {2023})},\ \Eprint {http://arxiv.org/abs/2201.08205}
  {arXiv:2201.08205 [astro-ph.CO]} \BibitemShut {NoStop}%
\bibitem [{\citenamefont {G\'omez-Vargas}\ \emph {et~al.}(2023)\citenamefont
  {G\'omez-Vargas}, \citenamefont {Esquivel}, \citenamefont
  {Garc\'\i{}a-Salcedo},\ and\ \citenamefont
  {V\'azquez}}]{Gomez-Vargas:2021zyl}%
  \BibitemOpen
  \bibfield  {author} {\bibinfo {author} {\bibfnamefont {I.}~\bibnamefont
  {G\'omez-Vargas}}, \bibinfo {author} {\bibfnamefont {R.~M.}\ \bibnamefont
  {Esquivel}}, \bibinfo {author} {\bibfnamefont {R.}~\bibnamefont
  {Garc\'\i{}a-Salcedo}}, \ and\ \bibinfo {author} {\bibfnamefont {J.~A.}\
  \bibnamefont {V\'azquez}},\ }\href {\doibase 10.1140/epjc/s10052-023-11435-9}
  {\bibfield  {journal} {\bibinfo  {journal} {Eur. Phys. J. C}\ }\textbf
  {\bibinfo {volume} {83}},\ \bibinfo {pages} {304} (\bibinfo {year} {2023})},\
  \Eprint {http://arxiv.org/abs/2104.00595} {arXiv:2104.00595 [astro-ph.CO]}
  \BibitemShut {NoStop}%
\bibitem [{\citenamefont {Ran}\ and\ \citenamefont {Wei}(2024)}]{Ran:2023jmh}%
  \BibitemOpen
  \bibfield  {author} {\bibinfo {author} {\bibfnamefont {J.-Y.}\ \bibnamefont
  {Ran}}\ and\ \bibinfo {author} {\bibfnamefont {J.-J.}\ \bibnamefont {Wei}},\
  }\href {\doibase 10.1103/PhysRevD.109.043001} {\bibfield  {journal} {\bibinfo
   {journal} {Phys. Rev. D}\ }\textbf {\bibinfo {volume} {109}},\ \bibinfo
  {pages} {043001} (\bibinfo {year} {2024})},\ \Eprint
  {http://arxiv.org/abs/2309.11810} {arXiv:2309.11810 [astro-ph.CO]}
  \BibitemShut {NoStop}%
\bibitem [{\citenamefont {Liu}\ \emph {et~al.}(2024)\citenamefont {Liu},
  \citenamefont {Cao}, \citenamefont {Biesiada}, \citenamefont {Zhang},\ and\
  \citenamefont {Wang}}]{Liu:2024gne}%
  \BibitemOpen
  \bibfield  {author} {\bibinfo {author} {\bibfnamefont {T.}~\bibnamefont
  {Liu}}, \bibinfo {author} {\bibfnamefont {S.}~\bibnamefont {Cao}}, \bibinfo
  {author} {\bibfnamefont {M.}~\bibnamefont {Biesiada}}, \bibinfo {author}
  {\bibfnamefont {Y.}~\bibnamefont {Zhang}}, \ and\ \bibinfo {author}
  {\bibfnamefont {J.}~\bibnamefont {Wang}},\ }\href {\doibase
  10.3847/2041-8213/ad3553} {\bibfield  {journal} {\bibinfo  {journal}
  {Astrophys. J. Lett.}\ }\textbf {\bibinfo {volume} {965}},\ \bibinfo {pages}
  {L11} (\bibinfo {year} {2024})},\ \Eprint {http://arxiv.org/abs/2404.07419}
  {arXiv:2404.07419 [astro-ph.CO]} \BibitemShut {NoStop}%
\bibitem [{\citenamefont {Fortunato}\ \emph {et~al.}(2024)\citenamefont
  {Fortunato}, \citenamefont {Bacon}, \citenamefont {Hip\'olito-Ricaldi},\ and\
  \citenamefont {Wands}}]{Fortunato:2024hfm}%
  \BibitemOpen
  \bibfield  {author} {\bibinfo {author} {\bibfnamefont {J.~A.~S.}\
  \bibnamefont {Fortunato}}, \bibinfo {author} {\bibfnamefont {D.~J.}\
  \bibnamefont {Bacon}}, \bibinfo {author} {\bibfnamefont {W.~S.}\ \bibnamefont
  {Hip\'olito-Ricaldi}}, \ and\ \bibinfo {author} {\bibfnamefont
  {D.}~\bibnamefont {Wands}},\ }\href@noop {} {\bibfield  {journal} {\bibinfo
  {journal} {2407.03532}\ } (\bibinfo {year} {2024})}\BibitemShut {NoStop}%
\bibitem [{\citenamefont {Qi}\ \emph {et~al.}(2024)\citenamefont {Qi},
  \citenamefont {Jiang}, \citenamefont {Hou},\ and\ \citenamefont
  {Zhang}}]{Qi:2024acx}%
  \BibitemOpen
  \bibfield  {author} {\bibinfo {author} {\bibfnamefont {J.-Z.}\ \bibnamefont
  {Qi}}, \bibinfo {author} {\bibfnamefont {Y.-F.}\ \bibnamefont {Jiang}},
  \bibinfo {author} {\bibfnamefont {W.-T.}\ \bibnamefont {Hou}}, \ and\
  \bibinfo {author} {\bibfnamefont {X.}~\bibnamefont {Zhang}},\ }\href@noop {}
  {\bibfield  {journal} {\bibinfo  {journal} {2407.07336}\ } (\bibinfo {year}
  {2024})}\BibitemShut {NoStop}%
\bibitem [{\citenamefont {Hogg}(1999)}]{Hogg:1999ad}%
  \BibitemOpen
  \bibfield  {author} {\bibinfo {author} {\bibfnamefont {D.~W.}\ \bibnamefont
  {Hogg}},\ }\href@noop {} {\bibfield  {journal} {\bibinfo  {journal}
  {astro-ph/9905116}\ } (\bibinfo {year} {1999})}\BibitemShut {NoStop}%
\bibitem [{\citenamefont {Rasmussen}\ and\ \citenamefont
  {Williams}(2006)}]{books/lib/RasmussenW06}%
  \BibitemOpen
  \bibfield  {author} {\bibinfo {author} {\bibfnamefont {C.~E.}\ \bibnamefont
  {Rasmussen}}\ and\ \bibinfo {author} {\bibfnamefont {C.~K.~I.}\ \bibnamefont
  {Williams}},\ }\href@noop {} {\emph {\bibinfo {title} {Gaussian processes for
  machine learning.}}},\ Adaptive computation and machine learning\ (\bibinfo
  {publisher} {MIT Press},\ \bibinfo {year} {2006})\ pp.\ \bibinfo {pages}
  {I--XVIII, 1--248}\BibitemShut {NoStop}%
\bibitem [{\citenamefont {{Seikel}}\ \emph {et~al.}(2012)\citenamefont
  {{Seikel}}, \citenamefont {{Clarkson}},\ and\ \citenamefont
  {{Smith}}}]{2012JCAP...06..036S}%
  \BibitemOpen
  \bibfield  {author} {\bibinfo {author} {\bibfnamefont {M.}~\bibnamefont
  {{Seikel}}}, \bibinfo {author} {\bibfnamefont {C.}~\bibnamefont
  {{Clarkson}}}, \ and\ \bibinfo {author} {\bibfnamefont {M.}~\bibnamefont
  {{Smith}}},\ }\href {\doibase 10.1088/1475-7516/2012/06/036} {\bibfield
  {journal} {\bibinfo  {journal} {jcap}\ }\textbf {\bibinfo {volume} {2012}},\
  \bibinfo {eid} {036} (\bibinfo {year} {2012})},\ \Eprint
  {http://arxiv.org/abs/1204.2832} {arXiv:1204.2832 [astro-ph.CO]} \BibitemShut
  {NoStop}%
\bibitem [{\citenamefont {Williams}(1997)}]{e6e4c2b74de0412ab4feae906164e191}%
  \BibitemOpen
  \bibfield  {author} {\bibinfo {author} {\bibfnamefont {C.}~\bibnamefont
  {Williams}},\ }{\selectlanguage {English}\enquote {\bibinfo {title}
  {Prediction with gaussian processes: From linear regression to linear
  prediction and beyond},}\ }in\ \href {\doibase 10.1007/978-94-011-5014-9_23}
  {{\selectlanguage {English}\emph {\bibinfo {booktitle} {Learning in Graphical
  Models}}}},\ \bibinfo {series and number} {NATO ASI Series D: Behavioural and
  Social Sciences}\ (\bibinfo  {publisher} {Springer Netherlands},\ \bibinfo
  {year} {1997})\ pp.\ \bibinfo {pages} {599--621}\BibitemShut {NoStop}%
\bibitem [{\citenamefont {MacKay}(2003)}]{MacKay2003}%
  \BibitemOpen
  \bibfield  {author} {\bibinfo {author} {\bibfnamefont {D.~J.~C.}\
  \bibnamefont {MacKay}},\ }\href@noop {} {\emph {\bibinfo {title} {Information
  Theory, Inference, and Learning Algorithms}}}\ (\bibinfo  {publisher}
  {Copyright Cambridge University Press},\ \bibinfo {year} {2003})\BibitemShut
  {NoStop}%
\bibitem [{\citenamefont {{Girshick}}(2015)}]{2015arXiv150408083G}%
  \BibitemOpen
  \bibfield  {author} {\bibinfo {author} {\bibfnamefont {R.}~\bibnamefont
  {{Girshick}}},\ }\href {\doibase 10.48550/arXiv.1504.08083} {\bibfield
  {journal} {\bibinfo  {journal} {arXiv e-prints}\ ,\ \bibinfo {eid}
  {arXiv:1504.08083}} (\bibinfo {year} {2015})},\ \Eprint
  {http://arxiv.org/abs/1504.08083} {arXiv:1504.08083 [cs.CV]} \BibitemShut
  {NoStop}%
\bibitem [{\citenamefont {{Kingma}}\ and\ \citenamefont
  {{Ba}}(2015)}]{2014arXiv1412.6980K}%
  \BibitemOpen
  \bibfield  {author} {\bibinfo {author} {\bibfnamefont {D.~P.}\ \bibnamefont
  {{Kingma}}}\ and\ \bibinfo {author} {\bibfnamefont {J.}~\bibnamefont
  {{Ba}}},\ }\bibfield  {booktitle} {\emph {\bibinfo {booktitle} {3rd
  International Conference on Learning Representations, {ICLR} 2015, San Diego,
  CA, USA, May 7-9, 2015, Conference Track Proceedings}},\ }\href
  {http://arxiv.org/abs/1412.6980} {\  (\bibinfo {year} {2015})},\ \Eprint
  {http://arxiv.org/abs/1412.6980} {arXiv:1412.6980 [cs.LG]} \BibitemShut
  {NoStop}%
\bibitem [{\citenamefont {Loshchilov}\ and\ \citenamefont
  {Hutter}(2017)}]{Loshchilov:2017bsp}%
  \BibitemOpen
  \bibfield  {author} {\bibinfo {author} {\bibfnamefont {I.}~\bibnamefont
  {Loshchilov}}\ and\ \bibinfo {author} {\bibfnamefont {F.}~\bibnamefont
  {Hutter}}\ }(\bibinfo {year} {2017})\ \Eprint
  {http://arxiv.org/abs/1711.05101} {arXiv:1711.05101 [cs.LG]} \BibitemShut
  {NoStop}%
\bibitem [{\citenamefont {Zheng}\ \emph {et~al.}(2015)\citenamefont {Zheng},
  \citenamefont {Yang}, \citenamefont {Liu}, \citenamefont {Liang},\ and\
  \citenamefont {Li}}]{7280459}%
  \BibitemOpen
  \bibfield  {author} {\bibinfo {author} {\bibfnamefont {H.}~\bibnamefont
  {Zheng}}, \bibinfo {author} {\bibfnamefont {Z.}~\bibnamefont {Yang}},
  \bibinfo {author} {\bibfnamefont {W.}~\bibnamefont {Liu}}, \bibinfo {author}
  {\bibfnamefont {J.}~\bibnamefont {Liang}}, \ and\ \bibinfo {author}
  {\bibfnamefont {Y.}~\bibnamefont {Li}},\ }in\ \href {\doibase
  10.1109/IJCNN.2015.7280459} {\emph {\bibinfo {booktitle} {2015 International
  Joint Conference on Neural Networks (IJCNN)}}}\ (\bibinfo {year} {2015})\
  pp.\ \bibinfo {pages} {1--4}\BibitemShut {NoStop}%
\bibitem [{\citenamefont {Vithanage}\ \emph {et~al.}(2014)\citenamefont
  {Vithanage}, \citenamefont {Udugama},\ and\ \citenamefont
  {Ratnayake}}]{inproceedings}%
  \BibitemOpen
  \bibfield  {author} {\bibinfo {author} {\bibfnamefont {N.}~\bibnamefont
  {Vithanage}}, \bibinfo {author} {\bibfnamefont {L.}~\bibnamefont {Udugama}},
  \ and\ \bibinfo {author} {\bibfnamefont {U.}~\bibnamefont {Ratnayake}},\ }in\
  \href@noop {} {\emph {\bibinfo {booktitle} {International Conference on
  Communication and Computing (ICC2014)}}},\ \bibinfo {editor} {edited by\
  \bibinfo {editor} {\bibfnamefont {K.~R.}\ \bibnamefont {Venugopal}}\ and\
  \bibinfo {editor} {\bibfnamefont {A.~C.}\ \bibnamefont {Ramachandra}}}\
  (\bibinfo {year} {2014})\ pp.\ \bibinfo {pages} {269--274}\BibitemShut
  {NoStop}%
\bibitem [{\citenamefont {Ratsimbazafy}\ \emph {et~al.}(2017)\citenamefont
  {Ratsimbazafy}, \citenamefont {Loubser}, \citenamefont {Crawford},
  \citenamefont {Cress}, \citenamefont {Bassett}, \citenamefont {Nichol},\ and\
  \citenamefont {V\"ais\"anen}}]{Ratsimbazafy:2017vga}%
  \BibitemOpen
  \bibfield  {author} {\bibinfo {author} {\bibfnamefont {A.~L.}\ \bibnamefont
  {Ratsimbazafy}}, \bibinfo {author} {\bibfnamefont {S.~I.}\ \bibnamefont
  {Loubser}}, \bibinfo {author} {\bibfnamefont {S.~M.}\ \bibnamefont
  {Crawford}}, \bibinfo {author} {\bibfnamefont {C.~M.}\ \bibnamefont {Cress}},
  \bibinfo {author} {\bibfnamefont {B.~A.}\ \bibnamefont {Bassett}}, \bibinfo
  {author} {\bibfnamefont {R.~C.}\ \bibnamefont {Nichol}}, \ and\ \bibinfo
  {author} {\bibfnamefont {P.}~\bibnamefont {V\"ais\"anen}},\ }\href {\doibase
  10.1093/mnras/stx301} {\bibfield  {journal} {\bibinfo  {journal} {Mon. Not.
  Roy. Astron. Soc.}\ }\textbf {\bibinfo {volume} {467}},\ \bibinfo {pages}
  {3239} (\bibinfo {year} {2017})},\ \Eprint {http://arxiv.org/abs/1702.00418}
  {arXiv:1702.00418 [astro-ph.CO]} \BibitemShut {NoStop}%
\bibitem [{\citenamefont {Jiao}\ \emph {et~al.}(2023)\citenamefont {Jiao},
  \citenamefont {Borghi}, \citenamefont {Moresco},\ and\ \citenamefont
  {Zhang}}]{Jiao:2022aep}%
  \BibitemOpen
  \bibfield  {author} {\bibinfo {author} {\bibfnamefont {K.}~\bibnamefont
  {Jiao}}, \bibinfo {author} {\bibfnamefont {N.}~\bibnamefont {Borghi}},
  \bibinfo {author} {\bibfnamefont {M.}~\bibnamefont {Moresco}}, \ and\
  \bibinfo {author} {\bibfnamefont {T.-J.}\ \bibnamefont {Zhang}},\ }\href
  {\doibase 10.3847/1538-4365/acbc77} {\bibfield  {journal} {\bibinfo
  {journal} {Astrophys. J. Suppl.}\ }\textbf {\bibinfo {volume} {265}},\
  \bibinfo {pages} {48} (\bibinfo {year} {2023})},\ \Eprint
  {http://arxiv.org/abs/2205.05701} {arXiv:2205.05701 [astro-ph.CO]}
  \BibitemShut {NoStop}%
\bibitem [{\citenamefont {Jimenez}\ \emph {et~al.}(2003)\citenamefont
  {Jimenez}, \citenamefont {Verde}, \citenamefont {Treu},\ and\ \citenamefont
  {Stern}}]{Jimenez:2003iv}%
  \BibitemOpen
  \bibfield  {author} {\bibinfo {author} {\bibfnamefont {R.}~\bibnamefont
  {Jimenez}}, \bibinfo {author} {\bibfnamefont {L.}~\bibnamefont {Verde}},
  \bibinfo {author} {\bibfnamefont {T.}~\bibnamefont {Treu}}, \ and\ \bibinfo
  {author} {\bibfnamefont {D.}~\bibnamefont {Stern}},\ }\href {\doibase
  10.1086/376595} {\bibfield  {journal} {\bibinfo  {journal} {Astrophys. J.}\
  }\textbf {\bibinfo {volume} {593}},\ \bibinfo {pages} {622} (\bibinfo {year}
  {2003})},\ \Eprint {http://arxiv.org/abs/astro-ph/0302560}
  {arXiv:astro-ph/0302560} \BibitemShut {NoStop}%
\bibitem [{\citenamefont {Simon}\ \emph {et~al.}(2005)\citenamefont {Simon},
  \citenamefont {Verde},\ and\ \citenamefont {Jimenez}}]{Simon:2004tf}%
  \BibitemOpen
  \bibfield  {author} {\bibinfo {author} {\bibfnamefont {J.}~\bibnamefont
  {Simon}}, \bibinfo {author} {\bibfnamefont {L.}~\bibnamefont {Verde}}, \ and\
  \bibinfo {author} {\bibfnamefont {R.}~\bibnamefont {Jimenez}},\ }\href
  {\doibase 10.1103/PhysRevD.71.123001} {\bibfield  {journal} {\bibinfo
  {journal} {Phys. Rev. D}\ }\textbf {\bibinfo {volume} {71}},\ \bibinfo
  {pages} {123001} (\bibinfo {year} {2005})},\ \Eprint
  {http://arxiv.org/abs/astro-ph/0412269} {arXiv:astro-ph/0412269} \BibitemShut
  {NoStop}%
\bibitem [{\citenamefont {Stern}\ \emph {et~al.}(2010)\citenamefont {Stern},
  \citenamefont {Jimenez}, \citenamefont {Verde}, \citenamefont
  {Kamionkowski},\ and\ \citenamefont {Stanford}}]{Stern:2009ep}%
  \BibitemOpen
  \bibfield  {author} {\bibinfo {author} {\bibfnamefont {D.}~\bibnamefont
  {Stern}}, \bibinfo {author} {\bibfnamefont {R.}~\bibnamefont {Jimenez}},
  \bibinfo {author} {\bibfnamefont {L.}~\bibnamefont {Verde}}, \bibinfo
  {author} {\bibfnamefont {M.}~\bibnamefont {Kamionkowski}}, \ and\ \bibinfo
  {author} {\bibfnamefont {S.~A.}\ \bibnamefont {Stanford}},\ }\href {\doibase
  10.1088/1475-7516/2010/02/008} {\bibfield  {journal} {\bibinfo  {journal}
  {JCAP}\ }\textbf {\bibinfo {volume} {02}},\ \bibinfo {pages} {008} (\bibinfo
  {year} {2010})},\ \Eprint {http://arxiv.org/abs/0907.3149} {arXiv:0907.3149
  [astro-ph.CO]} \BibitemShut {NoStop}%
\bibitem [{\citenamefont {Moresco}\ \emph {et~al.}(2012)\citenamefont
  {Moresco}, \citenamefont {Verde}, \citenamefont {Pozzetti}, \citenamefont
  {Jimenez},\ and\ \citenamefont {Cimatti}}]{Moresco:2012by}%
  \BibitemOpen
  \bibfield  {author} {\bibinfo {author} {\bibfnamefont {M.}~\bibnamefont
  {Moresco}}, \bibinfo {author} {\bibfnamefont {L.}~\bibnamefont {Verde}},
  \bibinfo {author} {\bibfnamefont {L.}~\bibnamefont {Pozzetti}}, \bibinfo
  {author} {\bibfnamefont {R.}~\bibnamefont {Jimenez}}, \ and\ \bibinfo
  {author} {\bibfnamefont {A.}~\bibnamefont {Cimatti}},\ }\href {\doibase
  10.1088/1475-7516/2012/07/053} {\bibfield  {journal} {\bibinfo  {journal}
  {JCAP}\ }\textbf {\bibinfo {volume} {07}},\ \bibinfo {pages} {053} (\bibinfo
  {year} {2012})},\ \Eprint {http://arxiv.org/abs/1201.6658} {arXiv:1201.6658
  [astro-ph.CO]} \BibitemShut {NoStop}%
\bibitem [{\citenamefont {Zhang}\ \emph {et~al.}(2014)\citenamefont {Zhang},
  \citenamefont {Zhang}, \citenamefont {Yuan}, \citenamefont {Zhang},\ and\
  \citenamefont {Sun}}]{Zhang:2012mp}%
  \BibitemOpen
  \bibfield  {author} {\bibinfo {author} {\bibfnamefont {C.}~\bibnamefont
  {Zhang}}, \bibinfo {author} {\bibfnamefont {H.}~\bibnamefont {Zhang}},
  \bibinfo {author} {\bibfnamefont {S.}~\bibnamefont {Yuan}}, \bibinfo {author}
  {\bibfnamefont {T.-J.}\ \bibnamefont {Zhang}}, \ and\ \bibinfo {author}
  {\bibfnamefont {Y.-C.}\ \bibnamefont {Sun}},\ }\href {\doibase
  10.1088/1674-4527/14/10/002} {\bibfield  {journal} {\bibinfo  {journal} {Res.
  Astron. Astrophys.}\ }\textbf {\bibinfo {volume} {14}},\ \bibinfo {pages}
  {1221} (\bibinfo {year} {2014})},\ \Eprint {http://arxiv.org/abs/1207.4541}
  {arXiv:1207.4541 [astro-ph.CO]} \BibitemShut {NoStop}%
\bibitem [{\citenamefont {Moresco}(2015)}]{Moresco:2015cya}%
  \BibitemOpen
  \bibfield  {author} {\bibinfo {author} {\bibfnamefont {M.}~\bibnamefont
  {Moresco}},\ }\href {\doibase 10.1093/mnrasl/slv037} {\bibfield  {journal}
  {\bibinfo  {journal} {Mon. Not. Roy. Astron. Soc.}\ }\textbf {\bibinfo
  {volume} {450}},\ \bibinfo {pages} {L16} (\bibinfo {year} {2015})},\ \Eprint
  {http://arxiv.org/abs/1503.01116} {arXiv:1503.01116 [astro-ph.CO]}
  \BibitemShut {NoStop}%
\bibitem [{\citenamefont {Moresco}\ \emph {et~al.}(2016)\citenamefont
  {Moresco}, \citenamefont {Pozzetti}, \citenamefont {Cimatti}, \citenamefont
  {Jimenez}, \citenamefont {Maraston}, \citenamefont {Verde}, \citenamefont
  {Thomas}, \citenamefont {Citro}, \citenamefont {Tojeiro},\ and\ \citenamefont
  {Wilkinson}}]{Moresco:2016mzx}%
  \BibitemOpen
  \bibfield  {author} {\bibinfo {author} {\bibfnamefont {M.}~\bibnamefont
  {Moresco}}, \bibinfo {author} {\bibfnamefont {L.}~\bibnamefont {Pozzetti}},
  \bibinfo {author} {\bibfnamefont {A.}~\bibnamefont {Cimatti}}, \bibinfo
  {author} {\bibfnamefont {R.}~\bibnamefont {Jimenez}}, \bibinfo {author}
  {\bibfnamefont {C.}~\bibnamefont {Maraston}}, \bibinfo {author}
  {\bibfnamefont {L.}~\bibnamefont {Verde}}, \bibinfo {author} {\bibfnamefont
  {D.}~\bibnamefont {Thomas}}, \bibinfo {author} {\bibfnamefont
  {A.}~\bibnamefont {Citro}}, \bibinfo {author} {\bibfnamefont
  {R.}~\bibnamefont {Tojeiro}}, \ and\ \bibinfo {author} {\bibfnamefont
  {D.}~\bibnamefont {Wilkinson}},\ }\href {\doibase
  10.1088/1475-7516/2016/05/014} {\bibfield  {journal} {\bibinfo  {journal}
  {JCAP}\ }\textbf {\bibinfo {volume} {05}},\ \bibinfo {pages} {014} (\bibinfo
  {year} {2016})},\ \Eprint {http://arxiv.org/abs/1601.01701} {arXiv:1601.01701
  [astro-ph.CO]} \BibitemShut {NoStop}%
\bibitem [{\citenamefont {Borghi}\ \emph {et~al.}(2022)\citenamefont {Borghi},
  \citenamefont {Moresco},\ and\ \citenamefont {Cimatti}}]{Borghi:2021rft}%
  \BibitemOpen
  \bibfield  {author} {\bibinfo {author} {\bibfnamefont {N.}~\bibnamefont
  {Borghi}}, \bibinfo {author} {\bibfnamefont {M.}~\bibnamefont {Moresco}}, \
  and\ \bibinfo {author} {\bibfnamefont {A.}~\bibnamefont {Cimatti}},\ }\href
  {\doibase 10.3847/2041-8213/ac3fb2} {\bibfield  {journal} {\bibinfo
  {journal} {Astrophys. J. Lett.}\ }\textbf {\bibinfo {volume} {928}},\
  \bibinfo {pages} {L4} (\bibinfo {year} {2022})},\ \Eprint
  {http://arxiv.org/abs/2110.04304} {arXiv:2110.04304 [astro-ph.CO]}
  \BibitemShut {NoStop}%
\bibitem [{\citenamefont {Tomasetti}\ \emph {et~al.}(2023)\citenamefont
  {Tomasetti}, \citenamefont {Moresco}, \citenamefont {Borghi}, \citenamefont
  {Jiao}, \citenamefont {Cimatti}, \citenamefont {Pozzetti}, \citenamefont
  {Carnall}, \citenamefont {McLure},\ and\ \citenamefont
  {Pentericci}}]{Tomasetti:2023kek}%
  \BibitemOpen
  \bibfield  {author} {\bibinfo {author} {\bibfnamefont {E.}~\bibnamefont
  {Tomasetti}}, \bibinfo {author} {\bibfnamefont {M.}~\bibnamefont {Moresco}},
  \bibinfo {author} {\bibfnamefont {N.}~\bibnamefont {Borghi}}, \bibinfo
  {author} {\bibfnamefont {K.}~\bibnamefont {Jiao}}, \bibinfo {author}
  {\bibfnamefont {A.}~\bibnamefont {Cimatti}}, \bibinfo {author} {\bibfnamefont
  {L.}~\bibnamefont {Pozzetti}}, \bibinfo {author} {\bibfnamefont {A.~C.}\
  \bibnamefont {Carnall}}, \bibinfo {author} {\bibfnamefont {R.~J.}\
  \bibnamefont {McLure}}, \ and\ \bibinfo {author} {\bibfnamefont
  {L.}~\bibnamefont {Pentericci}},\ }\href {\doibase
  10.1051/0004-6361/202346992} {\bibfield  {journal} {\bibinfo  {journal}
  {Astron. Astrophys.}\ }\textbf {\bibinfo {volume} {679}},\ \bibinfo {pages}
  {A96} (\bibinfo {year} {2023})},\ \Eprint {http://arxiv.org/abs/2305.16387}
  {arXiv:2305.16387 [astro-ph.CO]} \BibitemShut {NoStop}%
\bibitem [{\citenamefont {Brout}\ \emph {et~al.}(2022)\citenamefont {Brout}
  \emph {et~al.}}]{Brout:2022vxf}%
  \BibitemOpen
  \bibfield  {author} {\bibinfo {author} {\bibfnamefont {D.}~\bibnamefont
  {Brout}} \emph {et~al.},\ }\href {\doibase 10.3847/1538-4357/ac8e04}
  {\bibfield  {journal} {\bibinfo  {journal} {Astrophys. J.}\ }\textbf
  {\bibinfo {volume} {938}},\ \bibinfo {pages} {110} (\bibinfo {year}
  {2022})},\ \Eprint {http://arxiv.org/abs/2202.04077} {arXiv:2202.04077
  [astro-ph.CO]} \BibitemShut {NoStop}%
\bibitem [{\citenamefont {Park}\ \emph {et~al.}(2024)\citenamefont {Park},
  \citenamefont {de~Cruz~P\'erez},\ and\ \citenamefont
  {Ratra}}]{Chan-GyungPark:2024mlx}%
  \BibitemOpen
  \bibfield  {author} {\bibinfo {author} {\bibfnamefont {C.-G.}\ \bibnamefont
  {Park}}, \bibinfo {author} {\bibfnamefont {J.}~\bibnamefont
  {de~Cruz~P\'erez}}, \ and\ \bibinfo {author} {\bibfnamefont {B.}~\bibnamefont
  {Ratra}},\ }\href {\doibase 10.1103/PhysRevD.110.123533} {\bibfield
  {journal} {\bibinfo  {journal} {Phys. Rev. D}\ }\textbf {\bibinfo {volume}
  {110}},\ \bibinfo {pages} {123533} (\bibinfo {year} {2024})},\ \Eprint
  {http://arxiv.org/abs/2405.00502} {arXiv:2405.00502 [astro-ph.CO]}
  \BibitemShut {NoStop}%
\bibitem [{\citenamefont {Perivolaropoulos}\ and\ \citenamefont
  {Skara}(2022)}]{Perivolaropoulos:2021jda}%
  \BibitemOpen
  \bibfield  {author} {\bibinfo {author} {\bibfnamefont {L.}~\bibnamefont
  {Perivolaropoulos}}\ and\ \bibinfo {author} {\bibfnamefont {F.}~\bibnamefont
  {Skara}},\ }\href {\doibase 10.1016/j.newar.2022.101659} {\bibfield
  {journal} {\bibinfo  {journal} {New Astron. Rev.}\ }\textbf {\bibinfo
  {volume} {95}},\ \bibinfo {pages} {101659} (\bibinfo {year} {2022})},\
  \Eprint {http://arxiv.org/abs/2105.05208} {arXiv:2105.05208 [astro-ph.CO]}
  \BibitemShut {NoStop}%
\bibitem [{\citenamefont {Sch\"oneberg}\ \emph {et~al.}(2022)\citenamefont
  {Sch\"oneberg}, \citenamefont {Franco~Abell\'an}, \citenamefont
  {P\'erez~S\'anchez}, \citenamefont {Witte}, \citenamefont {Poulin},\ and\
  \citenamefont {Lesgourgues}}]{Schoneberg:2021qvd}%
  \BibitemOpen
  \bibfield  {author} {\bibinfo {author} {\bibfnamefont {N.}~\bibnamefont
  {Sch\"oneberg}}, \bibinfo {author} {\bibfnamefont {G.}~\bibnamefont
  {Franco~Abell\'an}}, \bibinfo {author} {\bibfnamefont {A.}~\bibnamefont
  {P\'erez~S\'anchez}}, \bibinfo {author} {\bibfnamefont {S.~J.}\ \bibnamefont
  {Witte}}, \bibinfo {author} {\bibfnamefont {V.}~\bibnamefont {Poulin}}, \
  and\ \bibinfo {author} {\bibfnamefont {J.}~\bibnamefont {Lesgourgues}},\
  }\href {\doibase 10.1016/j.physrep.2022.07.001} {\bibfield  {journal}
  {\bibinfo  {journal} {Phys. Rept.}\ }\textbf {\bibinfo {volume} {984}},\
  \bibinfo {pages} {1} (\bibinfo {year} {2022})},\ \Eprint
  {http://arxiv.org/abs/2107.10291} {arXiv:2107.10291 [astro-ph.CO]}
  \BibitemShut {NoStop}%
\bibitem [{\citenamefont {Abdalla}\ \emph {et~al.}(2022)\citenamefont {Abdalla}
  \emph {et~al.}}]{Abdalla:2022yfr}%
  \BibitemOpen
  \bibfield  {author} {\bibinfo {author} {\bibfnamefont {E.}~\bibnamefont
  {Abdalla}} \emph {et~al.},\ }\href {\doibase 10.1016/j.jheap.2022.04.002}
  {\bibfield  {journal} {\bibinfo  {journal} {JHEAp}\ }\textbf {\bibinfo
  {volume} {34}},\ \bibinfo {pages} {49} (\bibinfo {year} {2022})},\ \Eprint
  {http://arxiv.org/abs/2203.06142} {arXiv:2203.06142 [astro-ph.CO]}
  \BibitemShut {NoStop}%
\bibitem [{\citenamefont {Aghanim}\ \emph {et~al.}(2020)\citenamefont {Aghanim}
  \emph {et~al.}}]{Planck:2018vyg}%
  \BibitemOpen
  \bibfield  {author} {\bibinfo {author} {\bibfnamefont {N.}~\bibnamefont
  {Aghanim}} \emph {et~al.} (\bibinfo {collaboration} {Planck}),\ }\href
  {\doibase 10.1051/0004-6361/201833910} {\bibfield  {journal} {\bibinfo
  {journal} {Astron. Astrophys.}\ }\textbf {\bibinfo {volume} {641}},\ \bibinfo
  {pages} {A6} (\bibinfo {year} {2020})},\ \bibinfo {note} {[Erratum:
  Astron.Astrophys. 652, C4 (2021)]},\ \Eprint
  {http://arxiv.org/abs/1807.06209} {arXiv:1807.06209 [astro-ph.CO]}
  \BibitemShut {NoStop}%
\bibitem [{\citenamefont {Riess}\ \emph {et~al.}(2022)\citenamefont {Riess}
  \emph {et~al.}}]{Riess:2021jrx}%
  \BibitemOpen
  \bibfield  {author} {\bibinfo {author} {\bibfnamefont {A.~G.}\ \bibnamefont
  {Riess}} \emph {et~al.},\ }\href {\doibase 10.3847/2041-8213/ac5c5b}
  {\bibfield  {journal} {\bibinfo  {journal} {Astrophys. J. Lett.}\ }\textbf
  {\bibinfo {volume} {934}},\ \bibinfo {pages} {L7} (\bibinfo {year} {2022})},\
  \Eprint {http://arxiv.org/abs/2112.04510} {arXiv:2112.04510 [astro-ph.CO]}
  \BibitemShut {NoStop}%
\bibitem [{\citenamefont {Suzuki}\ \emph {et~al.}(2012)\citenamefont {Suzuki}
  \emph {et~al.}}]{SupernovaCosmologyProject:2011ycw}%
  \BibitemOpen
  \bibfield  {author} {\bibinfo {author} {\bibfnamefont {N.}~\bibnamefont
  {Suzuki}} \emph {et~al.} (\bibinfo {collaboration} {Supernova Cosmology
  Project}),\ }\href {\doibase 10.1088/0004-637X/746/1/85} {\bibfield
  {journal} {\bibinfo  {journal} {Astrophys. J.}\ }\textbf {\bibinfo {volume}
  {746}},\ \bibinfo {pages} {85} (\bibinfo {year} {2012})},\ \Eprint
  {http://arxiv.org/abs/1105.3470} {arXiv:1105.3470 [astro-ph.CO]} \BibitemShut
  {NoStop}%
\bibitem [{\citenamefont {Wong}\ \emph {et~al.}(2020)\citenamefont {Wong} \emph
  {et~al.}}]{H0LiCOW:2019pvv}%
  \BibitemOpen
  \bibfield  {author} {\bibinfo {author} {\bibfnamefont {K.~C.}\ \bibnamefont
  {Wong}} \emph {et~al.} (\bibinfo {collaboration} {H0LiCOW}),\ }\href
  {\doibase 10.1093/mnras/stz3094} {\bibfield  {journal} {\bibinfo  {journal}
  {Mon. Not. Roy. Astron. Soc.}\ }\textbf {\bibinfo {volume} {498}},\ \bibinfo
  {pages} {1420} (\bibinfo {year} {2020})},\ \Eprint
  {http://arxiv.org/abs/1907.04869} {arXiv:1907.04869 [astro-ph.CO]}
  \BibitemShut {NoStop}%
\bibitem [{\citenamefont {Wei}(2011{\natexlab{a}})}]{Wei:2010fz}%
  \BibitemOpen
  \bibfield  {author} {\bibinfo {author} {\bibfnamefont {H.}~\bibnamefont
  {Wei}},\ }\href {\doibase 10.1016/j.nuclphysb.2010.12.010} {\bibfield
  {journal} {\bibinfo  {journal} {Nucl. Phys. B}\ }\textbf {\bibinfo {volume}
  {845}},\ \bibinfo {pages} {381} (\bibinfo {year} {2011}{\natexlab{a}})},\
  \Eprint {http://arxiv.org/abs/1008.4968} {arXiv:1008.4968 [gr-qc]}
  \BibitemShut {NoStop}%
\bibitem [{\citenamefont {Wei}(2011{\natexlab{b}})}]{Wei:2010cs}%
  \BibitemOpen
  \bibfield  {author} {\bibinfo {author} {\bibfnamefont {H.}~\bibnamefont
  {Wei}},\ }\href {\doibase 10.1088/0253-6102/56/5/29} {\bibfield  {journal}
  {\bibinfo  {journal} {Commun. Theor. Phys.}\ }\textbf {\bibinfo {volume}
  {56}},\ \bibinfo {pages} {972} (\bibinfo {year} {2011}{\natexlab{b}})},\
  \Eprint {http://arxiv.org/abs/1010.1074} {arXiv:1010.1074 [gr-qc]}
  \BibitemShut {NoStop}%
\bibitem [{\citenamefont {Sun}\ and\ \citenamefont {Yue}(2012)}]{Sun:2010vz}%
  \BibitemOpen
  \bibfield  {author} {\bibinfo {author} {\bibfnamefont {C.-Y.}\ \bibnamefont
  {Sun}}\ and\ \bibinfo {author} {\bibfnamefont {R.-H.}\ \bibnamefont {Yue}},\
  }\href {\doibase 10.1103/PhysRevD.85.043010} {\bibfield  {journal} {\bibinfo
  {journal} {Phys. Rev. D}\ }\textbf {\bibinfo {volume} {85}},\ \bibinfo
  {pages} {043010} (\bibinfo {year} {2012})},\ \Eprint
  {http://arxiv.org/abs/1009.1214} {arXiv:1009.1214 [gr-qc]} \BibitemShut
  {NoStop}%
\bibitem [{\citenamefont {Li}\ and\ \citenamefont {Zhang}(2011)}]{Li:2011ga}%
  \BibitemOpen
  \bibfield  {author} {\bibinfo {author} {\bibfnamefont {Y.-H.}\ \bibnamefont
  {Li}}\ and\ \bibinfo {author} {\bibfnamefont {X.}~\bibnamefont {Zhang}},\
  }\href {\doibase 10.1140/epjc/s10052-011-1700-8} {\bibfield  {journal}
  {\bibinfo  {journal} {Eur. Phys. J. C}\ }\textbf {\bibinfo {volume} {71}},\
  \bibinfo {pages} {1700} (\bibinfo {year} {2011})},\ \Eprint
  {http://arxiv.org/abs/1103.3185} {arXiv:1103.3185 [astro-ph.CO]} \BibitemShut
  {NoStop}%
\bibitem [{\citenamefont {Guo}\ \emph {et~al.}(2018)\citenamefont {Guo},
  \citenamefont {Zhang}, \citenamefont {Li}, \citenamefont {He},\ and\
  \citenamefont {Zhang}}]{Guo:2017deu}%
  \BibitemOpen
  \bibfield  {author} {\bibinfo {author} {\bibfnamefont {J.-J.}\ \bibnamefont
  {Guo}}, \bibinfo {author} {\bibfnamefont {J.-F.}\ \bibnamefont {Zhang}},
  \bibinfo {author} {\bibfnamefont {Y.-H.}\ \bibnamefont {Li}}, \bibinfo
  {author} {\bibfnamefont {D.-Z.}\ \bibnamefont {He}}, \ and\ \bibinfo {author}
  {\bibfnamefont {X.}~\bibnamefont {Zhang}},\ }\href {\doibase
  10.1007/s11433-017-9131-9} {\bibfield  {journal} {\bibinfo  {journal} {Sci.
  China Phys. Mech. Astron.}\ }\textbf {\bibinfo {volume} {61}},\ \bibinfo
  {pages} {030011} (\bibinfo {year} {2018})},\ \Eprint
  {http://arxiv.org/abs/1710.03068} {arXiv:1710.03068 [astro-ph.CO]}
  \BibitemShut {NoStop}%
\bibitem [{\citenamefont {Arevalo}\ \emph {et~al.}(2019)\citenamefont
  {Arevalo}, \citenamefont {Cid}, \citenamefont {Chimento},\ and\ \citenamefont
  {Mella}}]{Arevalo:2019axj}%
  \BibitemOpen
  \bibfield  {author} {\bibinfo {author} {\bibfnamefont {F.}~\bibnamefont
  {Arevalo}}, \bibinfo {author} {\bibfnamefont {A.}~\bibnamefont {Cid}},
  \bibinfo {author} {\bibfnamefont {L.~P.}\ \bibnamefont {Chimento}}, \ and\
  \bibinfo {author} {\bibfnamefont {P.}~\bibnamefont {Mella}},\ }\href
  {\doibase 10.1140/epjc/s10052-019-6872-7} {\bibfield  {journal} {\bibinfo
  {journal} {Eur. Phys. J. C}\ }\textbf {\bibinfo {volume} {79}},\ \bibinfo
  {pages} {355} (\bibinfo {year} {2019})},\ \Eprint
  {http://arxiv.org/abs/1901.04300} {arXiv:1901.04300 [gr-qc]} \BibitemShut
  {NoStop}%
\bibitem [{\citenamefont {Pan}\ \emph {et~al.}(2019{\natexlab{b}})\citenamefont
  {Pan}, \citenamefont {Yang}, \citenamefont {Singha},\ and\ \citenamefont
  {Saridakis}}]{Pan:2019jqh}%
  \BibitemOpen
  \bibfield  {author} {\bibinfo {author} {\bibfnamefont {S.}~\bibnamefont
  {Pan}}, \bibinfo {author} {\bibfnamefont {W.}~\bibnamefont {Yang}}, \bibinfo
  {author} {\bibfnamefont {C.}~\bibnamefont {Singha}}, \ and\ \bibinfo {author}
  {\bibfnamefont {E.~N.}\ \bibnamefont {Saridakis}},\ }\href {\doibase
  10.1103/PhysRevD.100.083539} {\bibfield  {journal} {\bibinfo  {journal}
  {Phys. Rev. D}\ }\textbf {\bibinfo {volume} {100}},\ \bibinfo {pages}
  {083539} (\bibinfo {year} {2019}{\natexlab{b}})},\ \Eprint
  {http://arxiv.org/abs/1903.10969} {arXiv:1903.10969 [astro-ph.CO]}
  \BibitemShut {NoStop}%
\bibitem [{\citenamefont {Arevalo}\ and\ \citenamefont
  {Cid}(2022)}]{Arevalo:2022sne}%
  \BibitemOpen
  \bibfield  {author} {\bibinfo {author} {\bibfnamefont {F.}~\bibnamefont
  {Arevalo}}\ and\ \bibinfo {author} {\bibfnamefont {A.}~\bibnamefont {Cid}},\
  }\href {\doibase 10.1140/epjc/s10052-022-10898-6} {\bibfield  {journal}
  {\bibinfo  {journal} {Eur. Phys. J. C}\ }\textbf {\bibinfo {volume} {82}},\
  \bibinfo {pages} {946} (\bibinfo {year} {2022})},\ \Eprint
  {http://arxiv.org/abs/2202.05130} {arXiv:2202.05130 [astro-ph.CO]}
  \BibitemShut {NoStop}%
\bibitem [{\citenamefont {Martinelli}\ \emph {et~al.}(2019)\citenamefont
  {Martinelli}, \citenamefont {Hogg}, \citenamefont {Peirone}, \citenamefont
  {Bruni},\ and\ \citenamefont {Wands}}]{Martinelli:2019dau}%
  \BibitemOpen
  \bibfield  {author} {\bibinfo {author} {\bibfnamefont {M.}~\bibnamefont
  {Martinelli}}, \bibinfo {author} {\bibfnamefont {N.~B.}\ \bibnamefont
  {Hogg}}, \bibinfo {author} {\bibfnamefont {S.}~\bibnamefont {Peirone}},
  \bibinfo {author} {\bibfnamefont {M.}~\bibnamefont {Bruni}}, \ and\ \bibinfo
  {author} {\bibfnamefont {D.}~\bibnamefont {Wands}},\ }\href {\doibase
  10.1093/mnras/stz1915} {\bibfield  {journal} {\bibinfo  {journal} {Mon. Not.
  Roy. Astron. Soc.}\ }\textbf {\bibinfo {volume} {488}},\ \bibinfo {pages}
  {3423} (\bibinfo {year} {2019})},\ \Eprint {http://arxiv.org/abs/1902.10694}
  {arXiv:1902.10694 [astro-ph.CO]} \BibitemShut {NoStop}%
\bibitem [{\citenamefont {Yang}\ \emph {et~al.}(2021)\citenamefont {Yang},
  \citenamefont {Pan}, \citenamefont {Arest\'e~Sal\'o},\ and\ \citenamefont
  {de~Haro}}]{Yang:2021oxc}%
  \BibitemOpen
  \bibfield  {author} {\bibinfo {author} {\bibfnamefont {W.}~\bibnamefont
  {Yang}}, \bibinfo {author} {\bibfnamefont {S.}~\bibnamefont {Pan}}, \bibinfo
  {author} {\bibfnamefont {L.}~\bibnamefont {Arest\'e~Sal\'o}}, \ and\ \bibinfo
  {author} {\bibfnamefont {J.}~\bibnamefont {de~Haro}},\ }\href {\doibase
  10.1103/PhysRevD.103.083520} {\bibfield  {journal} {\bibinfo  {journal}
  {Phys. Rev. D}\ }\textbf {\bibinfo {volume} {103}},\ \bibinfo {pages}
  {083520} (\bibinfo {year} {2021})},\ \Eprint
  {http://arxiv.org/abs/2104.04505} {arXiv:2104.04505 [astro-ph.CO]}
  \BibitemShut {NoStop}%
\bibitem [{\citenamefont {Adame}\ \emph {et~al.}(2024)\citenamefont {Adame}
  \emph {et~al.}}]{DESI:2024mwx}%
  \BibitemOpen
  \bibfield  {author} {\bibinfo {author} {\bibfnamefont {A.~G.}\ \bibnamefont
  {Adame}} \emph {et~al.} (\bibinfo {collaboration} {DESI}),\ }\href@noop {}
  {\bibfield  {journal} {\bibinfo  {journal} {2404.03002}\ } (\bibinfo {year}
  {2024})}\BibitemShut {NoStop}%
\bibitem [{\citenamefont {Abdul~Karim}\ \emph {et~al.}(2025)\citenamefont
  {Abdul~Karim} \emph {et~al.}}]{DESI:2025zgx}%
  \BibitemOpen
  \bibfield  {author} {\bibinfo {author} {\bibfnamefont {M.}~\bibnamefont
  {Abdul~Karim}} \emph {et~al.} (\bibinfo {collaboration} {DESI}),\ }\href@noop
  {} {\enquote {\bibinfo {title} {{DESI DR2 Results II: Measurements of Baryon
  Acoustic Oscillations and Cosmological Constraints}},}\ } (\bibinfo {year}
  {2025}),\ \Eprint {http://arxiv.org/abs/2503.14738} {arXiv:2503.14738
  [astro-ph.CO]} \BibitemShut {NoStop}%
\bibitem [{\citenamefont {Lodha}\ \emph {et~al.}(2025)\citenamefont {Lodha}
  \emph {et~al.}}]{DESI:2025fii}%
  \BibitemOpen
  \bibfield  {author} {\bibinfo {author} {\bibfnamefont {K.}~\bibnamefont
  {Lodha}} \emph {et~al.} (\bibinfo {collaboration} {DESI}),\ }\href@noop {}
  {\enquote {\bibinfo {title} {{Extended Dark Energy analysis using DESI DR2
  BAO measurements}},}\ } (\bibinfo {year} {2025}),\ \Eprint
  {http://arxiv.org/abs/2503.14743} {arXiv:2503.14743 [astro-ph.CO]}
  \BibitemShut {NoStop}%
\bibitem [{\citenamefont {Gu}\ \emph {et~al.}(2025)\citenamefont {Gu} \emph
  {et~al.}}]{Gu:2025xie}%
  \BibitemOpen
  \bibfield  {author} {\bibinfo {author} {\bibfnamefont {G.}~\bibnamefont {Gu}}
  \emph {et~al.},\ }\href@noop {} {\enquote {\bibinfo {title} {{Dynamical Dark
  Energy in light of the DESI DR2 Baryonic Acoustic Oscillations
  Measurements}},}\ } (\bibinfo {year} {2025}),\ \Eprint
  {http://arxiv.org/abs/2504.06118} {arXiv:2504.06118 [astro-ph.CO]}
  \BibitemShut {NoStop}%
\bibitem [{\citenamefont {Chevallier}\ and\ \citenamefont
  {Polarski}(2001)}]{Chevallier:2000qy}%
  \BibitemOpen
  \bibfield  {author} {\bibinfo {author} {\bibfnamefont {M.}~\bibnamefont
  {Chevallier}}\ and\ \bibinfo {author} {\bibfnamefont {D.}~\bibnamefont
  {Polarski}},\ }\href {\doibase 10.1142/S0218271801000822} {\bibfield
  {journal} {\bibinfo  {journal} {Int. J. Mod. Phys. D}\ }\textbf {\bibinfo
  {volume} {10}},\ \bibinfo {pages} {213} (\bibinfo {year} {2001})},\ \Eprint
  {http://arxiv.org/abs/gr-qc/0009008} {arXiv:gr-qc/0009008} \BibitemShut
  {NoStop}%
\bibitem [{\citenamefont {Linder}(2003)}]{Linder:2002et}%
  \BibitemOpen
  \bibfield  {author} {\bibinfo {author} {\bibfnamefont {E.~V.}\ \bibnamefont
  {Linder}},\ }\href {\doibase 10.1103/PhysRevLett.90.091301} {\bibfield
  {journal} {\bibinfo  {journal} {Phys. Rev. Lett.}\ }\textbf {\bibinfo
  {volume} {90}},\ \bibinfo {pages} {091301} (\bibinfo {year} {2003})},\
  \Eprint {http://arxiv.org/abs/astro-ph/0208512} {arXiv:astro-ph/0208512}
  \BibitemShut {NoStop}%
\bibitem [{\citenamefont {Laureijs}\ \emph {et~al.}(2011)\citenamefont
  {Laureijs} \emph {et~al.}}]{EUCLID:2011zbd}%
  \BibitemOpen
  \bibfield  {author} {\bibinfo {author} {\bibfnamefont {R.}~\bibnamefont
  {Laureijs}} \emph {et~al.} (\bibinfo {collaboration} {EUCLID}),\ }\href@noop
  {} {\bibfield  {journal} {\bibinfo  {journal} {1110.3193}\ } (\bibinfo {year}
  {2011})}\BibitemShut {NoStop}%
\bibitem [{\citenamefont {Ivezi\'c}\ \emph {et~al.}(2019)\citenamefont
  {Ivezi\'c} \emph {et~al.}}]{LSST:2008ijt}%
  \BibitemOpen
  \bibfield  {author} {\bibinfo {author} {\bibfnamefont {v.}~\bibnamefont
  {Ivezi\'c}} \emph {et~al.} (\bibinfo {collaboration} {LSST}),\ }\href
  {\doibase 10.3847/1538-4357/ab042c} {\bibfield  {journal} {\bibinfo
  {journal} {Astrophys. J.}\ }\textbf {\bibinfo {volume} {873}},\ \bibinfo
  {pages} {111} (\bibinfo {year} {2019})},\ \Eprint
  {http://arxiv.org/abs/0805.2366} {arXiv:0805.2366 [astro-ph]} \BibitemShut
  {NoStop}%
\bibitem [{\citenamefont {Spergel}\ \emph {et~al.}(2013)\citenamefont {Spergel}
  \emph {et~al.}}]{Spergel:2013tha}%
  \BibitemOpen
  \bibfield  {author} {\bibinfo {author} {\bibfnamefont {D.}~\bibnamefont
  {Spergel}} \emph {et~al.},\ }\href@noop {} {\bibfield  {journal} {\bibinfo
  {journal} {1305.5422}\ } (\bibinfo {year} {2013})}\BibitemShut {NoStop}%
\bibitem [{\citenamefont {Bacon}\ \emph {et~al.}(2020)\citenamefont {Bacon}
  \emph {et~al.}}]{SKA:2018ckk}%
  \BibitemOpen
  \bibfield  {author} {\bibinfo {author} {\bibfnamefont {D.~J.}\ \bibnamefont
  {Bacon}} \emph {et~al.} (\bibinfo {collaboration} {SKA}),\ }\href {\doibase
  10.1017/pasa.2019.51} {\bibfield  {journal} {\bibinfo  {journal} {Publ.
  Astron. Soc. Austral.}\ }\textbf {\bibinfo {volume} {37}},\ \bibinfo {pages}
  {e007} (\bibinfo {year} {2020})},\ \Eprint {http://arxiv.org/abs/1811.02743}
  {arXiv:1811.02743 [astro-ph.CO]} \BibitemShut {NoStop}%
\end{thebibliography}%
%-------------------------------------------------------------------------------
\section{Appendix-A: Reconstructions of $\widetilde{Q}(z)$ and $w_{\rm DM}^{\rm eff}$  for different values of the DE EoS}
\label{sec-appendix}

In this section we present the 
reconstructed graphs of the interaction function $\widetilde{Q} (z)$ and the effective EoS parameter for DM for different values of $w_{\rm DE}$ other than $-1$ considering both the data driven approaches (GP and ANN) and using CC+Pantheon+ dataset and only one value of $H_0$, namely, $H_0 = 73.04 \pm 1.04$ $\rm km\ s^{-1}\ Mpc^{-1}$ at 68\% CL \cite{Riess:2021jrx}. We have noticed that the other value of $H_0$ mentioned in the main text does not affect the reconstructions. Therefore, one can consider any of the two $H_0$ values for the purpose of reconstructions. 

Considering GP, in Fig. \ref{fig:GP-Q_diff_wde-Appendix-1} we show the reconstruction of  $\widetilde{Q} (z)$ for different values of $w_{\rm DE}$ and in Fig. \ref{fig:GP-eff-wdm_wde-Appendix-2} we present the reconstructed $w_{\rm DM}^{\rm eff}$. One can notice that when $w_{\rm DE}$ deviates from $-1$, the evidence of interaction is pronounced. Since the effective EoS parameter for DM involves $w_{\rm DE}$, therefore, the evidence of interaction corresponds to the deviation of  $w_{\rm DM}^{\rm eff}$ from its null value. 

On the other hand, considering ANN approach, in Figs. \ref{fig:ANN-Q_diff_wde-Appendix-1} and \ref{fig:ANN-eff-wdm_wde-Appendix-2} we respectively show the reconstructed graphs of the interaction function and the effective EoS parameter for DM. In both the plots, we use CC+Pantheon+ dataset and $H_0 = 73.04 \pm 1.04$ $\rm km\ s^{-1}\ Mpc^{-1}$ at 68\% CL \cite{Riess:2021jrx}. We again note that the choice of $H_0$ does not play any effective role in the reconstructions. Similar to the reconstructions using GP, here too, we notice that the evidence of interaction increases when $w_{\rm DE}$ deviates from $-1$. As a consequence of this, the effective EoS parameter for DM deviates from  zero. 
\appendix
\renewcommand{\thefigure}{A\arabic{figure}} 
\begin{widetext} 

\begin{figure*}[h]
  % \centering
\includegraphics[width=1.0\linewidth]{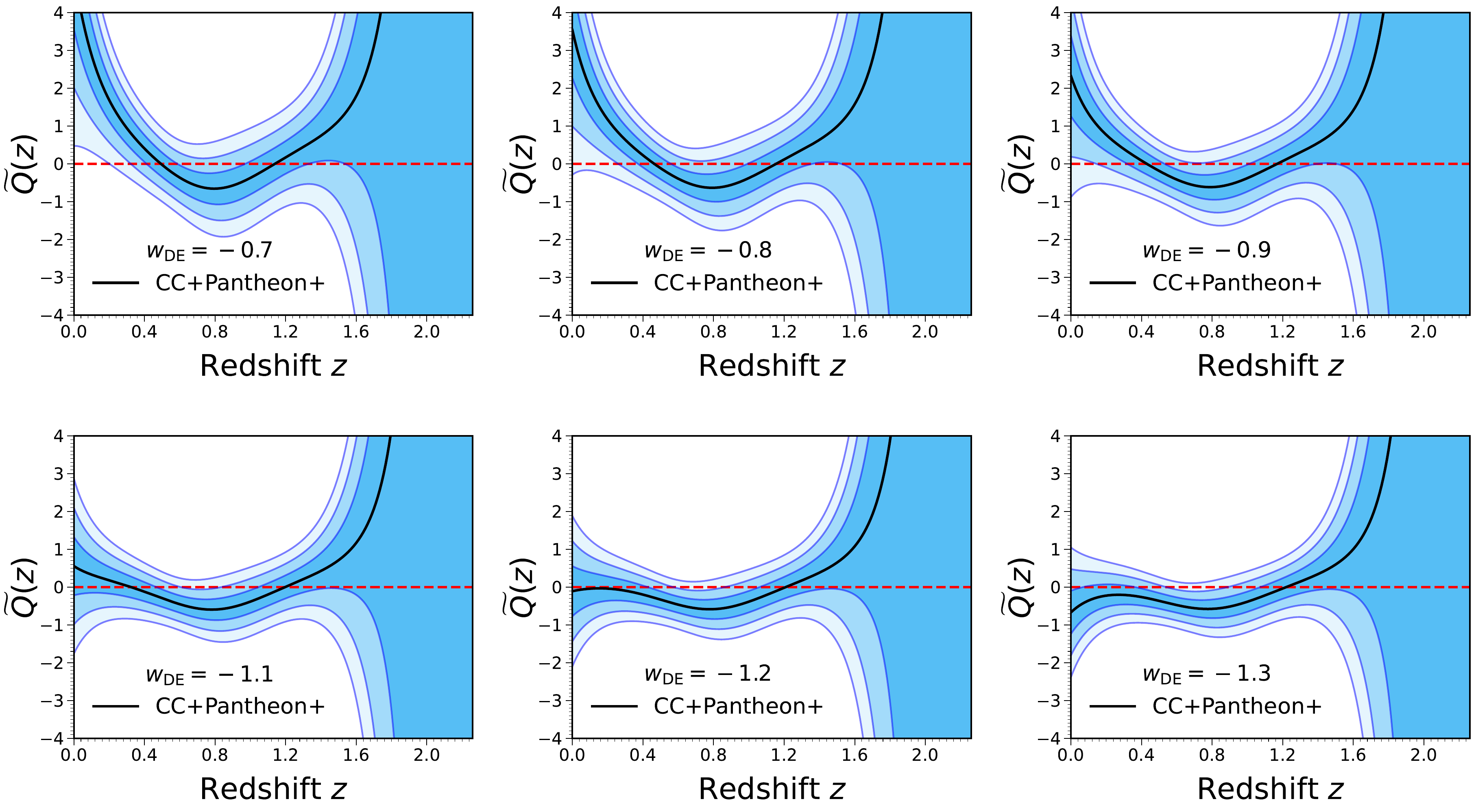}
  \caption{Reconstructed interaction function, $\widetilde{Q}(z)$ using GP for different values of $w_{\rm DE}$ considering CC+Pantheon+ and $H_0=73.04\pm 1.04$ $\rm km\ s^{-1}\ Mpc^{-1}$ at 68\% CL~\protect\cite{Riess:2021jrx}.}
  \label{fig:GP-Q_diff_wde-Appendix-1}
\end{figure*}

\end{widetext}
\begin{figure*}
    % \centering
    \includegraphics[width=1.0\linewidth]{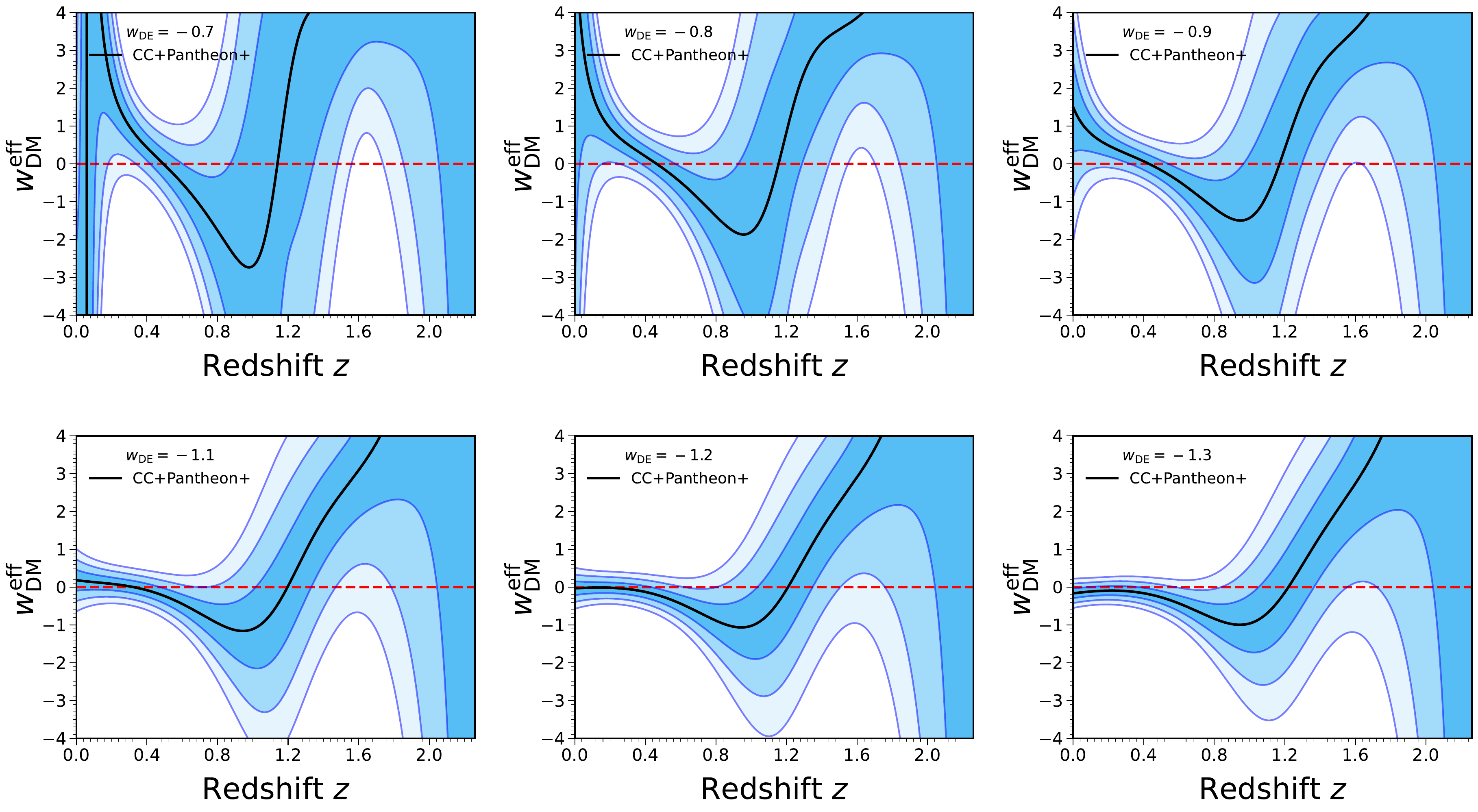}
    \caption{Reconstructed $w_{\rm DM}^{\rm eff}$ using GP for different value of $w_{\rm DE}$ considering CC+Pantheon+ and $H_0=73.04\pm 1.04$ $\rm km\ s^{-1}\ Mpc^{-1}$ at 68\% CL~\protect\cite{Riess:2021jrx}. }
    \label{fig:GP-eff-wdm_wde-Appendix-2}
\end{figure*}
\begin{figure*}
    % \centering
\includegraphics[width=0.3\linewidth]{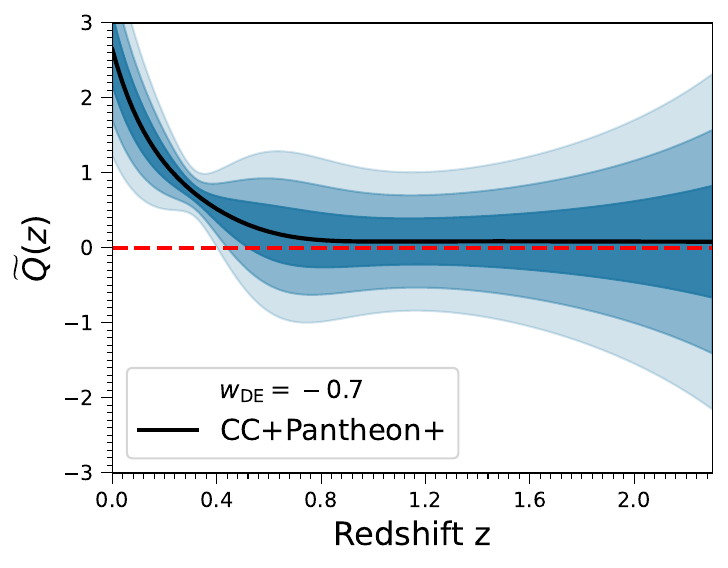}
\includegraphics[width=0.3\linewidth]{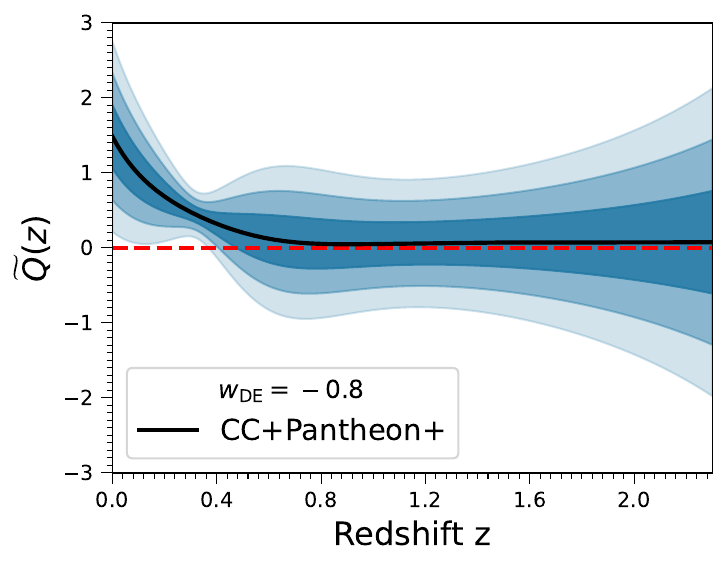}
\includegraphics[width=0.3\linewidth]{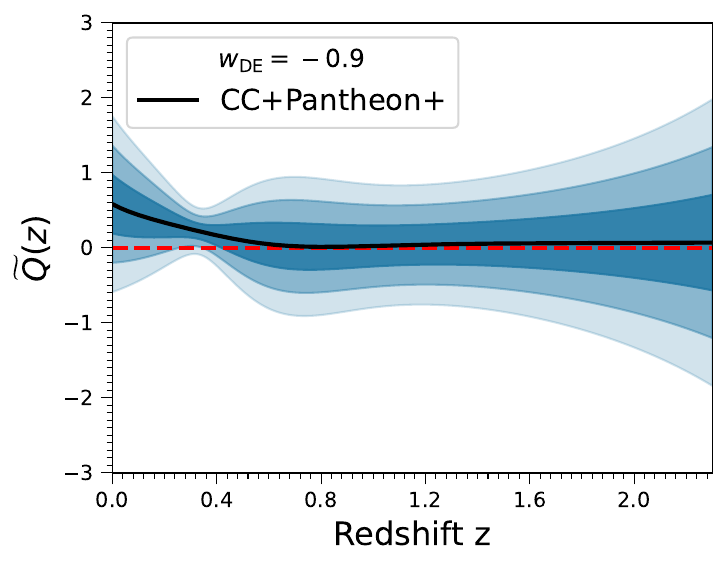}    \includegraphics[width=0.3\linewidth]{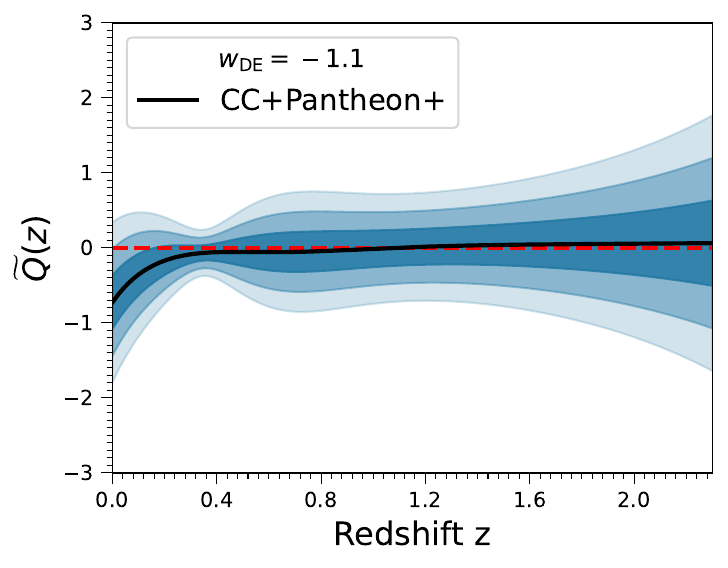}
    \includegraphics[width=0.3\linewidth]{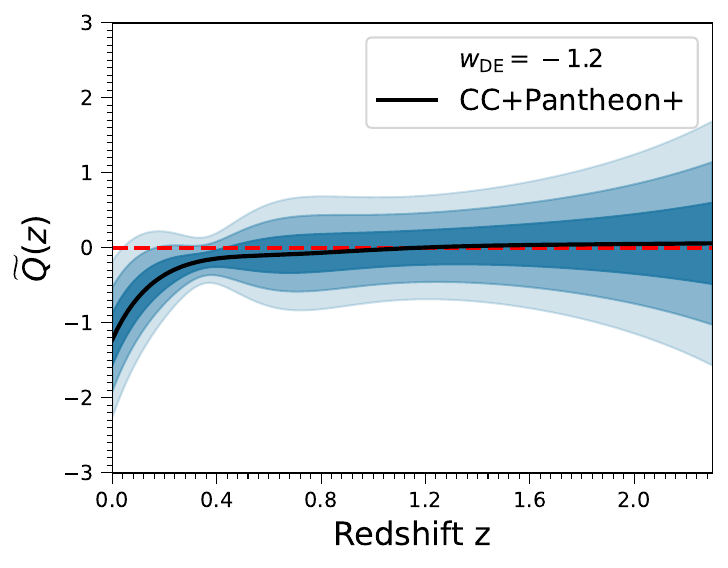}
    \includegraphics[width=0.3\linewidth]{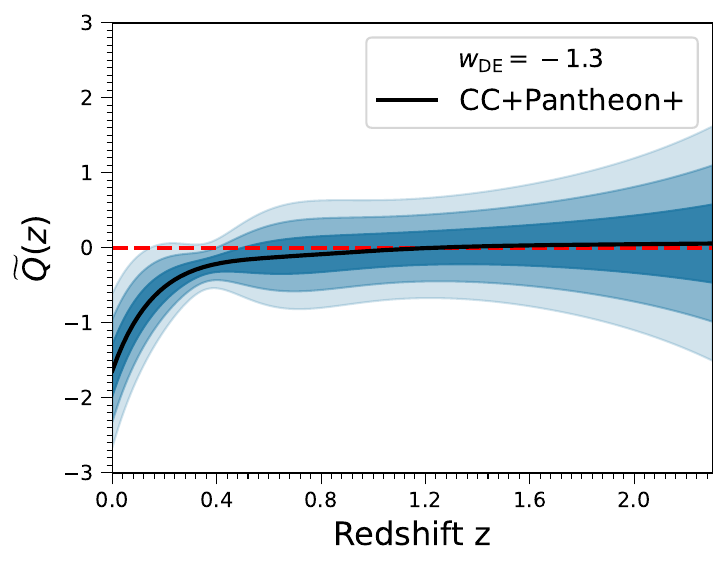}
    \caption{Reconstructed interaction function $\widetilde{Q}(z)$ using ANN for various values of $w_{\rm DE}$ considering CC+Pantheon+ and $H_0=73.04\pm 1.04$ $\rm km\ s^{-1}\ Mpc^{-1}$ at 68\% CL~\protect\cite{Riess:2021jrx}. }
    \label{fig:ANN-Q_diff_wde-Appendix-1}
\end{figure*}
\begin{figure*}
    % \centering
    \includegraphics[width=0.3\linewidth]{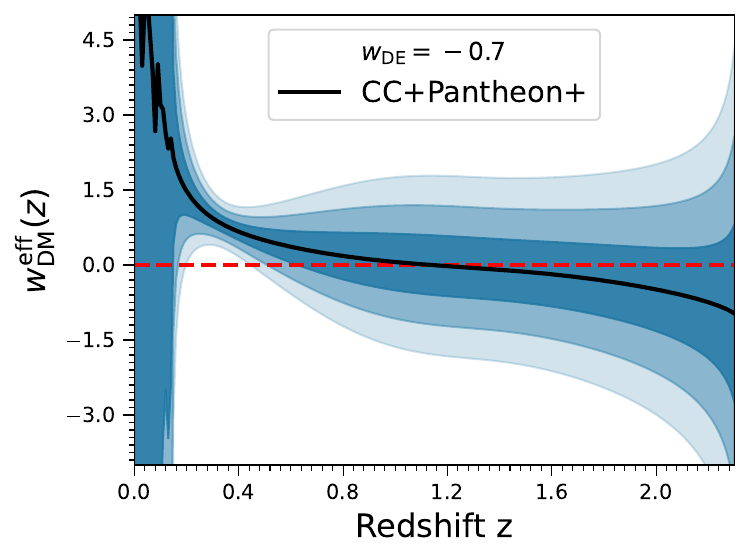}
    \includegraphics[width=0.3\linewidth]{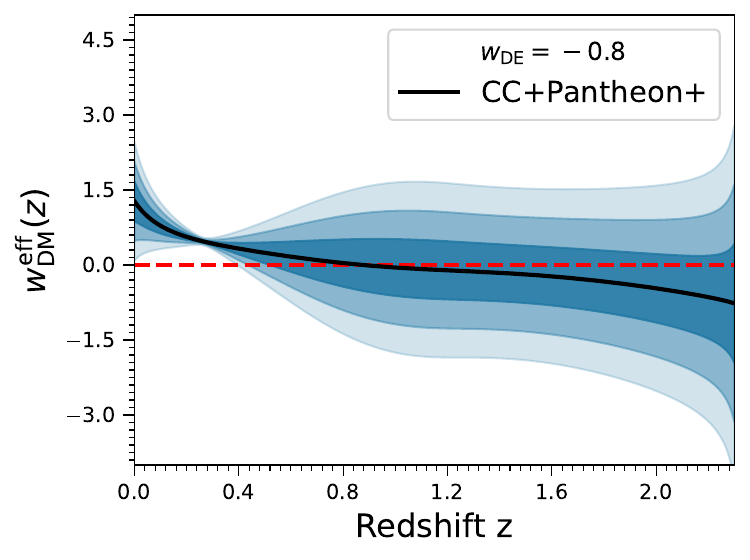}
    \includegraphics[width=0.3\linewidth]{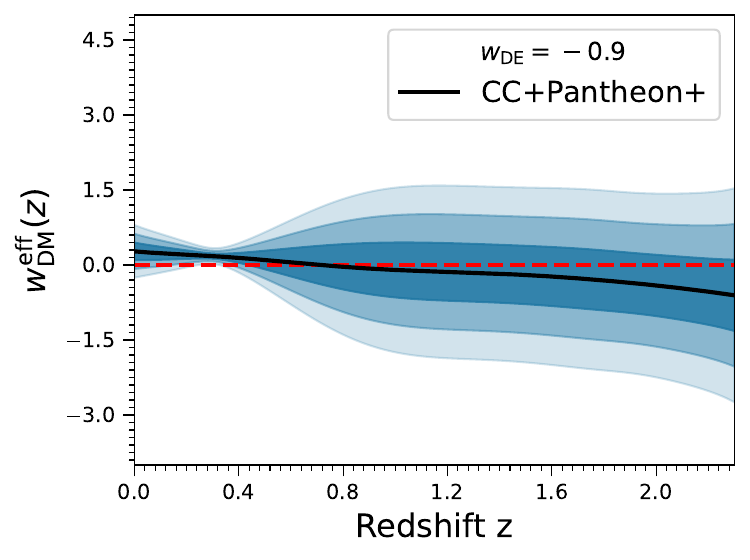}
    \includegraphics[width=0.3\linewidth]{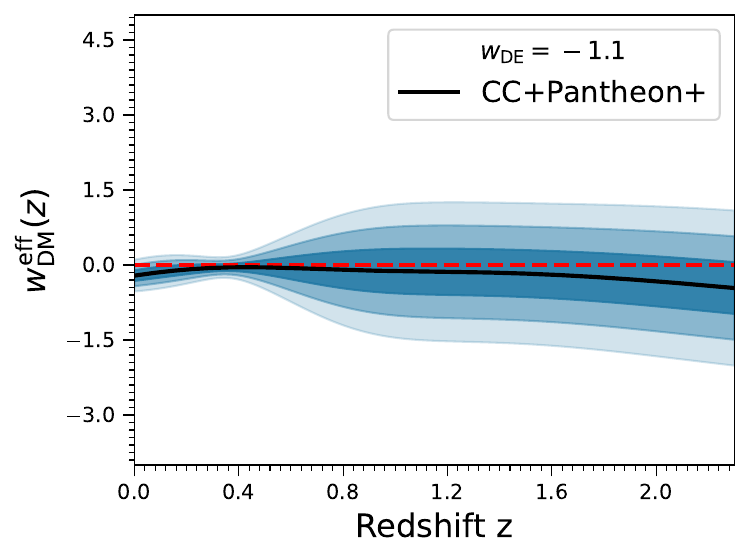}
    \includegraphics[width=0.3\linewidth]{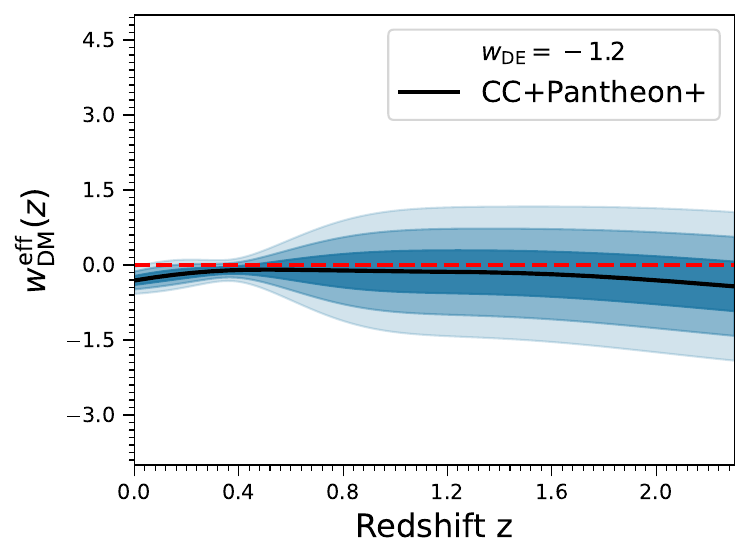}
    \includegraphics[width=0.3\linewidth]{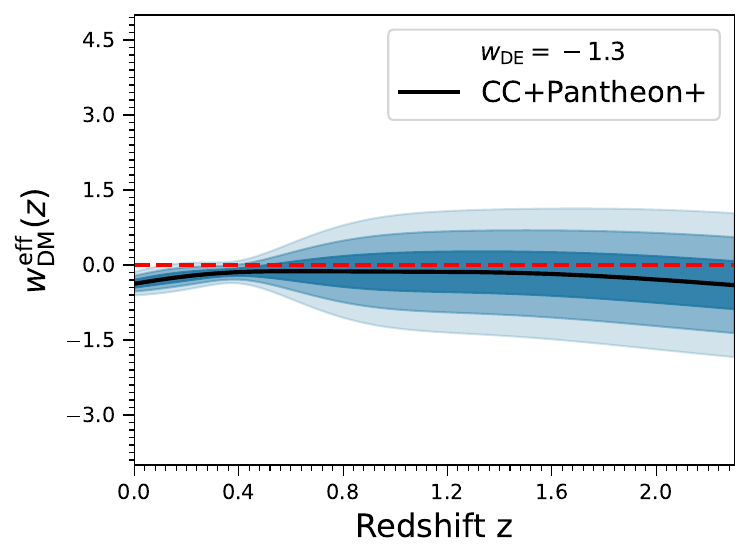}
    \caption{Reconstructed $w_{\rm DM}^{\rm eff}$ using ANN considering  CC+Pantheon+ and $H_0=73.04\pm 1.04$ $\rm km\ s^{-1}\ Mpc^{-1}$ at 68\% CL~\protect\cite{Riess:2021jrx}. }
    \label{fig:ANN-eff-wdm_wde-Appendix-2}
\end{figure*}

\end{document}